\documentclass[11pt]{article}
\usepackage[letterpaper,hmargin=0.75in,vmargin=0.7in,nohead,includefoot]{geometry}
\usepackage{mathtools}

\usepackage[T1]{fontenc}
\usepackage[utf8]{inputenc}
\usepackage{amsmath,amssymb,amsfonts}
\usepackage{amsthm}
\numberwithin{equation}{section}
\usepackage{bm}
\usepackage{xcolor}
\usepackage{braket}
\usepackage{hyperref}
\usepackage{tcolorbox}
\usepackage[
  backend=biber,
  style=numeric-comp,
  sorting=none,
  bibstyle=numeric,
  maxcitenames=3,
  natbib=true,
]{biblatex}
\hypersetup{
    colorlinks=true,
    linkcolor=blue!50!black,
    citecolor=blue!50!black,
    urlcolor=blue!50!black
}

\newtheorem{theorem}{Theorem}
\newtheorem{lemma}{Lemma}
\newtheorem{proposition}{Proposition}
\newtheorem{corollary}{Corollary}

\theoremstyle{definition}
\newtheorem{definition}{Definition}
\newtheorem{condition}{Condition}
\theoremstyle{plain}

\usepackage{mdframed}
\usepackage{enumitem}
\newenvironment{boxedthm}{%
  \begin{mdframed}[
    hidealllines=true,
    backgroundcolor=white,
    innertopmargin=10pt,
    innerbottommargin=10pt,
    innerleftmargin=10pt,
    innerrightmargin=10pt
  ]%
}{%
  \end{mdframed}%
}

\usepackage{subfig}
\usepackage{microtype}

\usepackage{authblk}
\usepackage{orcidlink}

\begin{document}
\pagenumbering{gobble}

\title{Strong hyperbolicity of puncture-gauge formulations of gravitational theories with second order equations of motion}
\author[1]{Shunhui Yao\,\orcidlink{0009-0003-8207-0335}}
\affil[1]{\small{\it Centre for Geometry, Analysis and Gravitation, School of Mathematical Sciences, Queen Mary University of London, Mile End Road, London E1 4NS, United Kingdom}}
\affil[ ]{\small\href{mailto:s.yao@qmul.ac.uk}{s.yao@qmul.ac.uk}}
\date{\today}

\maketitle

\begin{abstract}
We prove that the modified CCZ4 (mCCZ4) formulation, used in nonlinear puncture-gauge
simulations of $(d+1)$-dimensional Einstein-Gauss-Bonnet gravity and four-derivative
scalar-tensor theory in four dimensions, is strongly hyperbolic for both theories at
sufficiently weak coupling, provided that the gauge parameters are chosen to satisfy a
cone-separation condition. The proof applies to the full principal symbol of the
theories, extending previous analyses based on first-order perturbation theory about
the Einstein limit. We establish strong hyperbolicity by recasting the mCCZ4 equations
into a formally equivalent second-order system whose derivative structure resembles
that of the modified harmonic formulation of Kov\'acs and Reall. This allows us to adapt their covariant
characteristic analysis and continuity argument, without assuming that individual
eigenvalues or eigenvectors depend analytically on the coupling. We also show that standard CCZ4 in general
relativity is strongly hyperbolic in puncture gauge away from two defective
degeneracies.
\end{abstract}
\clearpage
\pagenumbering{arabic}
\tableofcontents

\section{Introduction}

\subsection{Gravity beyond general relativity with second order equations of motion}

Gravitational wave observations now probe the highly dynamical, strong field regime of
compact binary mergers, yet the analysis of these signals routinely assumes general
relativity (GR)~\cite{KAGRA:2021vkt,Yunes:2009ke}. Effective field theory (EFT)
provides a systematic and agnostic organising principle for parametrising deviations
from GR~\cite{Donoghue:1994dn,Figueras:2024bba}: GR is the leading term in a
derivative expansion, and higher curvature corrections are generically expected. The
leading corrections are quadratic in curvature, with coefficients of order $\ell^2$,
where $\ell$ is the ultraviolet (UV) length scale at which new physics is expected to
emerge. In vacuum, only the Gauss-Bonnet combination remains among the
parity-invariant terms after exploiting field redefinitions. This term is topological
in four spacetime dimensions unless coupled to a scalar field. The coupling, together
with the leading scalar self-interaction, gives the four-derivative scalar-tensor
theory ($4\partial\mathrm{ST}$), which contains Einstein-scalar-Gauss-Bonnet (EsGB)
gravity. In five or more dimensions, the Gauss-Bonnet term is dynamical and gives
Einstein-Gauss-Bonnet (EGB) gravity~\cite{Kovacs:2020pns,Kovacs:2020ywu}.

EGB gravity belongs to the Lovelock family, the most general theories in which the
gravitational field is described by a single metric tensor satisfying
diffeomorphism-invariant second order equations of motion~\cite{Lovelock:1971yv}.
Similarly, $4\partial\mathrm{ST}$ belongs to the Horndeski family~\cite{Kobayashi:2011nu},
the most general theories of a metric tensor coupled to a single scalar field, with
second-order field equations derived from a diffeomorphism-invariant action in four
spacetime dimensions~\cite{Horndeski:1974wa}. We give their field equations in
Sections~\ref{sec:egb-extension} and~\ref{sec:4dst-equations}.

Having second order equations of motion makes these families of particular interest. On the
one hand, theories in these families contain no Ostrogradsky ghosts, which generically
arise from higher-order equations of motion~\cite{Woodard:2015zca}. On the other hand,
at weak coupling, a suitable choice of formulation makes their equations
strongly hyperbolic and hence their initial value problem
well-posed~\cite{Kovacs:2020pns,Kovacs:2020ywu}. Their characteristic cones govern both the
hyperbolicity of the equations and the causal structure of these
theories~\cite{Reall:2014pwa,Reall:2021voz}, and can be monitored in
numerical evolutions~\cite{Ripley:2019hxt,Ripley:2019irj,Thaalba:2024crk,Figueras:2026evh,Figueras:2026epx}.
Generic higher derivative theories can also be evolved effectively, through
\textit{fixing-of-the-equations}~\cite{Cayuso:2023xbc} or
\textit{regularisation}~\cite{Figueras:2024bba}, but both approaches modify the UV
dynamics. Lovelock and Horndeski theories, by contrast, can be solved with
their UV dynamics intact, and therefore serve as benchmarks for these approaches.

The first formulation shown to be well-posed for general Lovelock and Horndeski
theories at weak coupling is the modified harmonic formulation of Kov\'acs and
Reall~\cite{Kovacs:2020pns,Kovacs:2020ywu}. It adds the same gauge-fixing term to the
metric equations $E^{\mu\nu}=0$ of both families, giving
\begin{equation}
E_{\mathrm{mhg}}^{\mu\nu}
=E^{\mu\nu}
+\hat P_\alpha{}^{\beta\mu\nu}\nabla_\beta\mathcal H^\alpha=0,
\label{eq:mHg}
\end{equation}
where
\begin{equation}
\hat P_\alpha{}^{\beta\mu\nu}
=\delta_\alpha^{(\mu}\hat g^{\nu)\beta}
-\tfrac12\delta_\alpha^\beta\hat g^{\mu\nu},
\qquad
\mathcal H^\mu
=-\tilde g^{\alpha\beta}\Gamma^\mu{}_{\alpha\beta}
=\tilde g^{\alpha\beta}\nabla_\alpha\nabla_\beta x^\mu.
\end{equation}
The modified harmonic gauge condition is $\mathcal H^\mu=0$.\footnote{Here we use the calligraphic symbol $\mathcal H^\mu$ for the gauge-constraint vector, denoted by $H^\mu$ in \cite{Kovacs:2020pns,Kovacs:2020ywu}, to avoid confusion with the gauge source used in \cite{AresteSalo:2023mmd}.}
With $n^\mu$ denoting the future-directed unit normal to the constant-$t$
hypersurfaces, the two auxiliary inverse metrics
\begin{equation}
\hat g^{\mu\nu}=g^{\mu\nu}-b(x)n^\mu n^\nu,
\qquad
\tilde g^{\alpha\beta}=g^{\alpha\beta}-a(x)n^\alpha n^\beta
\end{equation}
control the propagation of high-frequency solutions in the gauge-condition-violating
sector and the pure-gauge sector, respectively. The functions $a(x)$ and $b(x)$ are
chosen so that the null cones of $\tilde g^{\mu\nu}$, $\hat g^{\mu\nu}$, and $g^{\mu\nu}$
do not intersect, and the three sectors then propagate with different speeds in GR. In
standard harmonic gauge, by contrast, all three sectors propagate on the null cone of
$g^{\mu\nu}$, and strong hyperbolicity generically fails for Lovelock and Horndeski
theories~\cite{Papallo:2017qvl,Papallo:2017ddx}. Breaking this degeneracy keeps the
modified harmonic formulation strongly hyperbolic when the Lovelock or Horndeski terms
are turned on at weak coupling~\cite{Kovacs:2020pns,Kovacs:2020ywu}.

In numerical simulations, harmonic-type evolution requires excision of
black hole interiors~\cite{AresteSalo:2023mmd}, whereas many numerical-relativity codes
based on the BSSN, Z4c, or CCZ4
formulations~\cite{Shibata:1995we,Baumgarte:1998te,Bernuzzi:2009ex,Alic:2011gg} use
puncture gauge, which evolves black holes without
excision~\cite{Campanelli:2005dd,Baker:2005vv}. Arest\'e Sal\'o, Clough, and Figueras
therefore introduced the modified CCZ4 (mCCZ4) formulation, the puncture-gauge variant
of the modified harmonic formulation~\cite{AresteSalo:2022hua,AresteSalo:2023mmd}. It
is used in nonlinear simulations of Lovelock and Horndeski theories, from binary black hole mergers in
EsGB gravity~\cite{AresteSalo:2022hua,AresteSalo:2025sxc,Corman:2025wun} to the
Gregory-Laflamme instability of five-dimensional black strings in EGB
gravity~\cite{Figueras:2026epx,Figueras:2026evh}, as well as cosmological inflation and
preheating~\cite{Brady:2025zxp,Waeming:2026wrp}.
The strong hyperbolicity of mCCZ4 at weak coupling is the subject of this paper.

\subsection{The modified CCZ4 formulation}
\label{sec:mccz4-effective-mh}

The mCCZ4 formulation replaces the modified harmonic gauge with puncture
gauge~\cite{AresteSalo:2022hua,AresteSalo:2023mmd}, prescribing the
following modified $1+\log$ slicing for the lapse $\alpha$ and integrated
Gamma-driver for the shift $\beta^i$:\footnote{Latin indices denote spatial components and Greek indices spacetime components. The $3+1$ quantities in~\eqref{eq:puncture-gauge} follow the conventions of~\cite{AresteSalo:2023mmd}, and the conformal variables are defined in Section~\ref{sec:mccz4-application}.}
\begin{equation}
\partial_\perp\alpha=\frac{c_\alpha}{1+a(x)}\alpha(K-2\Theta),
\qquad
\partial_\perp\beta^i=\frac{c_\beta}{1+a(x)}\hat\Gamma^i-\frac{a(x)}{1+a(x)}\alpha D^i\alpha,
\label{eq:puncture-gauge}
\end{equation}
where $\partial_\perp:=\partial_t-\beta^i\partial_i$, $K$ is the trace of the extrinsic
curvature, and $D_i$ is the covariant derivative of the spatial metric. We keep
$c_\alpha<0$ and $c_\beta>0$ general, although the usual choices are $c_\alpha=-2$ and
$c_\beta=d/[2(d-1)]$. The lapse couples to the normal component of the Z4 vector
$Z^\mu$ through $\Theta:=-n_\mu Z^\mu$, and the shift couples to its spatial
components through $\hat\Gamma^i:=\tilde\Gamma^i+2\tilde\gamma^{ij}Z_j$, where
$\tilde\Gamma^i$ is the contracted Christoffel symbol of the conformal spatial metric
$\tilde\gamma_{ij}$.

The Z4 vector $Z^\mu$ enters the metric equation through a gauge-fixing term:
\begin{equation}
E^{\mu\nu}_{\mathrm{mCCZ4}}=E^{\mu\nu}
+2\hat P_\alpha{}^{\beta\mu\nu}\nabla_\beta Z^\alpha+\cdots=0,
\label{eq:eoms_lovelock}
\end{equation}
where the ellipsis denotes lower-order constraint-damping terms that do not enter the
principal symbol. Through its spatial and normal projections, equation~\eqref{eq:eoms_lovelock}
governs the evolution of the extrinsic curvature and the Z4 variables.
Equations~\eqref{eq:puncture-gauge}–\eqref{eq:eoms_lovelock} thus define the
mCCZ4 system.

As Arest\'e Sal\'o, Clough, and Figueras observed, the gauge-fixing operator
$\hat P_\alpha{}^{\beta\mu\nu}\nabla_\beta$ acts on $\mathcal H^\alpha$
in~\eqref{eq:mHg} and on $2Z^\alpha$ in~\eqref{eq:eoms_lovelock}, suggesting
the formal identification
\begin{equation}
Z^\mu=\frac12\mathcal H^\mu,
\qquad
\mathcal H^\mu:=-H^\mu
-\tilde g^{\rho\sigma}\Gamma^\mu{}_{\rho\sigma}.
\label{eq:eqZ2H}
\end{equation}

The gauge condition $\mathcal H^\mu=0$ corresponds to $Z^\mu=0$. Recovering the lapse
and shift equations~\eqref{eq:puncture-gauge} from the normal and spatial components
of~\eqref{eq:eqZ2H} requires a gauge source $H^\mu$ that depends on the ADM/BSSN
variables and their derivatives~\cite{AresteSalo:2023mmd}. Differentiating this source
in the gauge-fixing term changes the principal symbol, so strong hyperbolicity in
puncture gauge does not follow from the formal identification with the
modified harmonic system~\cite{Kovacs:2020ywu}.

Arest\'e Sal\'o, Clough, and Figueras therefore analysed the principal symbol of the
evolution system in ADM/BSSN form directly, treating the higher curvature corrections by
first-order degenerate perturbation theory about the Einstein limit and showing that
the first-order corrections preserve diagonalisability and the reality of the
eigenvalues~\cite{AresteSalo:2023mmd}. Whether this result holds to all orders in the
small couplings has remained an open question.

\subsection{Main result and proof strategy}

We prove that, at sufficiently weak coupling, the mCCZ4 system is strongly hyperbolic
for EGB gravity in $d+1$ spacetime dimensions and for four-derivative scalar-tensor
theory in four spacetime dimensions, provided that $c_\alpha,c_\beta,a(x),b(x)$ are
chosen to satisfy the cone-separation condition
(Condition~\ref{ass:mccz4-cone-separation}). The proof uses the full reduced principal
symbol at weak coupling and adapts the continuity argument of Kov\'acs and
Reall~\cite{Kovacs:2020ywu}, without assuming that individual eigenvalues or
eigenvectors depend analytically on the coupling. It proceeds in four steps.

In the first step, we recast the mCCZ4 evolution system, which is first order in
time, as a second-order equation for the spacetime metric so that we can apply the
Kov\'acs-Reall argument to its principal symbol. To eliminate $Z^\mu$ from the
gauge-fixing term in~\eqref{eq:eoms_lovelock}, we invert the lapse and shift
conditions~\eqref{eq:puncture-gauge}, which are linear in $\Theta$ and $Z^i$,
respectively. The resulting Z4 vector depends linearly on the first derivatives of
the spacetime metric (Appendix~\ref{app:gauge-driver-inversion}).
Substituting $Z^\mu(g,\partial g)$ into~\eqref{eq:eoms_lovelock} yields the
effective metric equation in generalised modified harmonic form:
\begin{equation}
E_{\textbf{mCCZ4-2nd}}^{\mu\nu}
=E^{\mu\nu}
+2\hat P_\alpha{}^{\beta\mu\nu}
\partial_\beta Z^\alpha(g,\partial g)+\cdots=0,
\label{eq:mCCZ4_2nd}
\end{equation}
where the ellipsis denotes terms of lower differential order in the metric.

In the second step, we prove that the original mCCZ4 system and~\eqref{eq:mCCZ4_2nd}
are equivalent with respect to strong hyperbolicity (Theorem~\ref{th:mccz4-equivalence}),
so it suffices to study the effective metric system in the remaining two steps.

In the third step, we establish strong hyperbolicity in the GR and Einstein-scalar
limits. With the cones separated by Condition~\ref{ass:mccz4-cone-separation}, we
construct eigenvectors in the physical sector, the pure-gauge sector, and the
gauge-condition-violating sector. Together they form a complete eigenbasis of the
reduced principal symbol, with real eigenvalues in each sector. In puncture gauge,
the temporal, longitudinal, and transverse cones replace the single pure-gauge cone
of the modified harmonic formulation.

In the fourth step, we extend this result to sufficiently weak coupling. The two
unphysical sectors keep their characteristic speeds and still admit a complete set of
eigenvectors. In the physical sector, degenerate eigenvalues may split but remain
grouped near their limiting values $\xi_0^\pm$. Following Kov\'acs and
Reall~\cite{Kovacs:2020ywu}, we construct a Hermitian form from the covariant
principal symbol and use its positivity to show that the eigenvalues in each group
remain real and the reduced principal symbol remains uniformly diagonalisable on the
corresponding subspace. Combining the three sectors and applying
Theorem~\ref{th:mccz4-equivalence}, we establish strong hyperbolicity of the original
mCCZ4 system at sufficiently weak coupling.

This result for EGB gravity and four-derivative scalar-tensor theory is based on
identities of the covariant principal symbol that follow from diffeomorphism
invariance and the variational principle. These identities hold throughout the
Lovelock and Horndeski families, and the puncture-gauge term has the same form in
both. Under Condition~\ref{ass:mccz4-cone-separation}, the argument therefore extends
to the full Lovelock and Horndeski families at sufficiently weak coupling about GR
and Einstein-scalar theory, respectively.

The rest of the paper is organised as follows. In
Section~\ref{sec:Equivalence_mccz4-effective-2nd}, we prove
Theorem~\ref{th:mccz4-equivalence}. In Sections~\ref{sec:proof-strong-hyperbolicity}
and~\ref{sec: Einstein-scalar-Gauss-Bonnet theory}, we prove strong hyperbolicity for
EGB gravity and four-derivative scalar-tensor theory, respectively. In
Section~\ref{sec:standard-ccz4}, we treat standard CCZ4 in GR, and in
Section~\ref{sec:discussion}, we summarise our results and discuss their limits and connections. In
Appendices~\ref{app:gauge-driver-inversion} and~\ref{app:moving-puncture}, we invert the
puncture-gauge conditions and extend the standard CCZ4 result to the original
form of the Gamma-driver condition~\eqref{eq:Gamma-driver} with auxiliary
variable $B^i$, respectively.

\section{Strong hyperbolicity in terms of effective second-order system}
\label{sec:Equivalence_mccz4-effective-2nd}

In this section, we shall prove that the effective second order form of modified CCZ4 is indeed equivalent to the original form in terms of the ADM/BSSN variables. We emphasise that the argument straightforwardly applies to the standard CCZ4 system when $a(x)=b(x)=0$. 
\subsection{Definitions of strong hyperbolicity for second‑order-in-space PDEs}
We begin by revisiting the definition of strong hyperbolicity for a generic first-order-in-time, second-order-in-space (FT2S) evolution system, as proposed in  \cite{Gundlach:2005eh,Hilditch:2013sba,Hilditch:2014naa}.

\begin{definition}[Generic FT2S system]
\label{def:generic-FT2S}
Consider a first-order-in-time, second-order-in-space system of the form
\begin{subequations}\label{eq:generic-FT2S}
\begin{align}
    \partial_t\widetilde u
    &=(A^{u}{}_{u})^{i}\partial_i\widetilde u
      +A^{u}{}_{v}\widetilde v+S_u,
      \\
    \partial_t\widetilde v
    &=(A^{v}{}_{u})^{ij}\partial_i\partial_j\widetilde u
      +(A^{v}{}_{v})^{i}\partial_i\widetilde v+S_v,
\end{align}
\end{subequations}
where $S_u$ and $S_v$ contain terms of lower differential order. For
every unit spatial covector $s_i$, the FT2S principal symbol is
\begin{equation}
    P_2^s:=
    \begin{pmatrix}
        (A^{u}{}_{u})^{i}s_i & A^{u}{}_{v} \\
        (A^{v}{}_{u})^{ij}s_i s_j & (A^{v}{}_{v})^{i}s_i
    \end{pmatrix}.
    \label{eq:generic-FT2S-symbol}
\end{equation}
The system is strongly hyperbolic if there exist a constant $K_2\geq1$
and a Hermitian matrix $H_2(s)$, depending smoothly on $s_i$ and on the
background field configuration, such that, for every unit $s_i$,
\begin{equation}
    H_2(s)P_2^s=(P_2^s)^\dagger H_2(s),
    \qquad
    K_2^{-1}I\leq H_2(s)\leq K_2 I.
    \label{eq:generic-FT2S-SH}
\end{equation}
The matrix is called the \textit{symmetrizer}. The inequalities are understood in the sense of matrix norm, and the
constant $K_2$ is independent of $s_i$.
\end{definition} 
A standard way to establish strong hyperbolicity is to show that $P_2^s$ is diagonalisable  with real eigenvalues and that the matrix $T$ formed with eigenvectors of $P_2^s$ as columns is smooth and bounded. In this case, the symmetrizer is given by 
\begin{equation}
    H_2(s)=T^{-1}(s)^\dagger T^{-1}(s).
\end{equation}

For a non-quasi-linear system, Definition~\ref{def:generic-FT2S} is applied to the linearisation  about an arbitrary smooth
background. The bounds in \eqref{eq:generic-FT2S-SH} are required to
hold locally uniformly in the background variables. This inspires the definition of strong hyperbolicity for a general class of 
partially reduced PDE systems as discussed in Sec.1.2.5  of \cite{Kovacs:2021vdk}.  
\begin{definition}[Partially reduced system] Consider the following u-v model:
\begin{subequations}\label{eq:u-v}
    \begin{align}
        \partial_t u&=M^k_{uu} (x,u)\partial_ku +M_{uv}(x,u)v, \label{eq:u-v1} \\
        \partial_t v&=X(x,u,v,\partial_iu,\partial_i\partial_ju,\partial_kv), \label{eq:u-v2}
    \end{align}
\end{subequations}
where $u$ is a n-component state vector and $v$ is a m-component state vector, with $M^{k}_{uu}$  $n\times n$  matrices, and $M_{uv}$ being  $n\times m$ matrix. Here, all the matrix elements as well as the $X$ function are smooth.\footnote{Widely used formulations in numerical GR belong to the
class~\eqref{eq:u-v}, except those reduced to fully first-order
form~\cite{Dumbser:2017okk,Lindblom:2005qh}.}

It follows from \cite{Kovacs:2021vdk}  that for the original u-v system \eqref{eq:u-v}, the strong hyperbolicity is defined in terms of its linearisation \textbf{about a generic background}, which is given by
\begin{subequations}
    \begin{align}
         \partial_t u&=M^k_{uu}u_k +M_{uv}v, \\
          \partial_t v&=M_{vv} ^i\partial_iv + M_{vu}^{ij}\partial_i\partial_ju,
    \end{align}
\end{subequations}
where we have defined the matrix-valued coefficients as
\begin{equation}
    M_{vv}^{i}:=\frac{\partial X}{\partial(\partial_i v)},\quad   M_{vu}^{ij}:=\frac{\partial X}{\partial(\partial_i\partial_j u)}.
\end{equation}
For every unit spatial covector $\xi_i$, define the principal symbol in analogy with \eqref{eq:generic-FT2S-symbol}:
\begin{equation}
    \tilde{M}(\xi_i):=\begin{pmatrix}
          M^{k}_{uu}\xi_k & M_{uv} \\
          M^{ij}_{vu}\xi_i\xi_j & M^{i}_{vv}\xi_i
    \end{pmatrix}. \label{eq:princ_u-v}
\end{equation}
The partially reduced system is strongly hyperbolic \eqref{eq:u-v} if and only if there exist a
constant $\tilde{K}\geq1$ and a Hermitian symmetrizer
$\tilde H(\xi_i)$, depending smoothly on $\xi_i$ and on the background field
configuration, satisfying
\begin{equation}
    \tilde H(\xi_i)\tilde M(\xi_i)
    =\tilde M(\xi_i)^\dagger\tilde H(\xi_i),
    \qquad
    \tilde K^{-1}I
    \leq\tilde H(\xi_i)\leq\tilde K I
    \label{eq:partially-reduced-SH}
\end{equation}
for every unit $\xi_i$.
\label{def:partially-reduced-SH}
\end{definition}
Just as in the previous case, the strong hyperbolicity can be established by showing that $M(\xi_i)$ is diagonalisable  with all eigenvalues being real,  and constructing a complete set of eigenvectors of $\tilde{M}(\xi_i)$ that depend smoothly on $\xi_i$ and on the background field configuration. 

Finally, in order to deduce the strong hyperbolicity of a generic fully second order system as studied in \cite{Kovacs:2020ywu,Kovacs:2020pns,Kovacs:2021vdk}, we need to adapt the Definition \ref{def:partially-reduced-SH} to fully second order PDEs.

\begin{definition}[Fully second-order system]
\label{def:fully-second-order-SH} 
Consider a generic $n$-component second order PDE system of the form $ \mathcal{F}(u,\partial_\mu u,\partial_\nu\partial_\alpha u)=0$. Linearising about a smooth background and retaining only the second derivative terms gives \begin{equation}
    \mathcal{P}^{\mu\nu}\partial_\mu\partial_\nu u=0,\quad  \mathcal{P}^{\mu\nu}\equiv \frac{\partial\mathcal{F}}{\partial(\partial_{\mu}\partial_\nu u)}. \label{eq:eqdefPu}
\end{equation}
Write the resulting linearised system to the first-order-in-time, second-order-in-space form:
\begin{equation}
\begin{gathered}
 \partial_t u = u_t, \\
 \partial_t u_t= -\mathcal{P}^{00}{}^{-1}(2\mathcal{P}^{0i}\partial_i u_t+\mathcal{P}^{ij}\partial_i\partial_j u).
\end{gathered} \label{eq:1stu}
\end{equation}
In line with \eqref{eq:generic-FT2S-symbol} as well as \eqref{eq:princ_u-v}, we define the principal symbol $M(\xi_i)$ of the above FT2S as 
\begin{equation}
    M(\xi_i)=\begin{pmatrix} 0 & I_{n\times n} \\-A^{-1}C & -A^{-1}B \end{pmatrix}, \label{eq:reducedM}
\end{equation}
where we have introduced the notations 
\begin{equation}
   A:= \mathcal{P}^{00} ,\quad B := 2\mathcal{P}^{0i}\xi_i ,\quad  C := \mathcal{P}^{ij}\xi_i\xi_j. \label{eq:matrixcoeff}
\end{equation}
 The fully second-order system $\mathcal{F}(u,\partial_\nu u,\partial_\nu\partial_\alpha u)=0$ is strongly hyperbolic if there exist a
constant $\lambda\geq1$ and a Hermitian symmetrizer $\mathcal{K}(\xi_i)$, depending
smoothly on $\xi_i$ and on the background field configuration, satisfying
\begin{equation}
    \mathcal{K}(\xi_i)M(\xi_i)=M(\xi_i)^\dagger\mathcal{K}(\xi_i),
    \qquad
    \lambda^{-1}I\leq \mathcal{K}(\xi_i)\leq \lambda I
    \label{eq:fully-second-order-SH}
\end{equation}
for every unit spatial covector $\xi_i$.
\end{definition}
Note that this definition coincides precisely with that introduced in \cite{Kovacs:2020ywu,Kovacs:2020pns,Kovacs:2021vdk}.
As in the preceding cases, the strong hyperbolicity of the fully second order system can also be deduced by showing the reality of the eigenvalues and constructing a complete set of bounded and smooth eigenvectors. 

It is useful to introduce a concept of equivalence for FT2S PDE systems, which can be formulated in the following proposition.
\begin{proposition}[Equivalence between two FT2S PDE systems]
Consider two FT2S systems $A,B$ whose principal symbols are denoted by $M_A(\xi_i)$ and $M_B(\xi_i)$. If there is a smooth similarity transformation generated by the matrix $S(\xi_i)$, then A is strongly hyperbolic if and only if B is strongly hyperbolic. \label{th:prop1} 
\end{proposition}
\begin{proof}
Suppose that $M_A(\xi_i)$ is related to $M_B(\xi_i)$ by a smooth similarity transformation:
\begin{equation}
    M_A(\xi_i)=S^{-1}(\xi_i)M_B(\xi_i)S(\xi_i),
\end{equation}
and that the symmetrizer of $B$ is given by $H_B$. 
Then, it is straightforward to check that the matrix 
\begin{equation}
    H_A:= S(\xi_i)^\dagger H_B S(\xi_i)
\end{equation}
gives the symmetrizer of $A$ and hence system $A$ is also strongly hyperbolic. The reverse implication is obtained immediately by interchanging $A$ with $B$ and $S^{-1}$ with $S$.
\end{proof}

\subsection{ Equivalence between a partially reduced system and a fully second-order system}
We return to the u-v model for the partially reduced system introduced in \eqref{eq:u-v}.
Suppose that $m=n$ and that the matrix $M_{uv}$ is invertible; one can eliminate the $v$-variables from \eqref{eq:u-v} by first solving \eqref{eq:u-v1} for $v$ in terms of the first derivatives of $u$, and then substituting these expressions for all occurrences of the $v$-variables in \eqref{eq:u-v2}, which leaves us with a second order PDE systems solely in $u$-variables, which can be written as
\begin{equation}
   M_{uv}^{-1} \partial_t^2u-  M_{uv}^{-1}M^{k}_{uu}\partial_t\partial_ku-X(x,u,v,\partial_iu,\partial_i\partial_ju,(M^{-1}_{uv})(\partial_t\partial_ku-M^{l}_{uu}\partial_k\partial_l u))\simeq0; \label{eq:usec}
\end{equation}
by $\simeq$ we mean that the equality holds only up to terms involving lower-order derivatives of $u$.

Below, we show that the strong hyperbolicity of the second order PDE system \eqref{eq:usec} is equivalent to that of the original system \eqref{eq:u-v}.

\begin{lemma}
A partially reduced system of the form \eqref{eq:u-v} with invertible $M_{uv}$ is, subject to definition \ref{def:partially-reduced-SH}, strongly hyperbolic if and only if the corresponding second-order PDE system \eqref{eq:usec} is strongly hyperbolic according to definition \ref{def:fully-second-order-SH}. \label{th:lem1}
\end{lemma} 
\begin{proof}
Based on the preceding discussion, we will show that the principal symbols $M(\xi_i)$  and $\tilde{M}(\xi_i)$ are connected by a similarity transformation.
According to \eqref{eq:usec}, the matrix-valued coefficients defined in \eqref{eq:matrixcoeff} can be expressed as 
\begin{equation}
\begin{aligned}
   \mathcal{P}^{00}&=M_{uv}^{-1},\\
   2\mathcal{P}^{0i}&= -M_{uv}^{-1}M^{i}_{uu}-\frac{\partial X}{\partial(\partial_i v)} M^{-1}_{uv}=-M_{uv}^{-1}M^{i}_{uu}-M_{vv}^i M_{uv}^{-1},\\
 \mathcal{P}^{ij}&=-\frac{\partial X}{\partial(\partial_i\partial_j u)} + \frac{\partial X}{\partial(\partial_i v)} M^{-1}_{uv} M^{j}_{uu}=-M_{vu}^{ij}+ M_{vv}^{i}M_{uv}^{-1}M_{uu}^j.
\end{aligned}
\end{equation}
Then the matrix $M(\xi_i)$ can be written in the block form as 
\begin{equation}
    M(\xi_i)=\begin{pmatrix}
        0 & I \\
        M_{uv} (M_{vu}^{ij}-M_{vv}^iM_{uv}^{-1}M_{uu}^j)\xi_i\xi_j  & (M_{uu}^i+M_{uv}M_{vv}^i M^{-1}_{uv})\xi_i
    \end{pmatrix}.
\end{equation}
Consider the transformation matrix
\begin{equation}
    S:=\begin{pmatrix}
        I & 0 \\ -M^{-1}_{uv}M^{k}_{uu}\xi_k & M_{uv}^{-1}
    \end{pmatrix},
\end{equation}
with the inverse matrix being 
\begin{equation}
    S^{-1}= \begin{pmatrix}
        I & 0\\
        M^{l}_{uu}\xi_l & M_{uv}
    \end{pmatrix}.
\end{equation}
Then, the similarity transformation of $\tilde{M}(\xi_i)$ yields
\begin{equation}
    \begin{aligned}
        S^{-1}\tilde{M}(\xi_i)S&=\begin{pmatrix}
        I & 0\\
        M^{l}_{uu}\xi_l & M_{uv}
    \end{pmatrix}
    \begin{pmatrix}
          M^{k}_{uu}\xi_k & M_{uv} \\
          M^{ij}_{vu}\xi_i\xi_j & M^{i}_{vv}\xi_i
    \end{pmatrix}
    \begin{pmatrix}
        I & 0 \\ -M^{-1}_{uv}M^{k}_{uu}\xi_k & M_{uv}^{-1}
    \end{pmatrix}\\
    &= \begin{pmatrix}
        I & 0\\
        M^{l}_{uu}\xi_l & M_{uv}
    \end{pmatrix} 
    \begin{pmatrix}
        0 & I\\
        (M^{ij}_{vu}-M^{i}_{vv}M_{uv}^{-1}M^{j}_{uu})\xi_i\xi_j & M^{i}_{vv}M^{-1}_{uv}\xi_i
    \end{pmatrix}\\
    &=\begin{pmatrix}
        0 &I \\
         M_{uv}(M^{ij}_{vu}-M^{i}_{vv}M_{uv}^{-1}M^{j}_{uu})\xi_i\xi_j & 
         (M_{uu}^i+M_{uv}M_{vv}^i M^{-1}_{uv})\xi_i
    \end{pmatrix}\\
    &\equiv M(\xi_i),
    \end{aligned}
\end{equation}
which concludes the proof.
\end{proof}

\subsection{Application to the (m)CCZ4 system}
\label{sec:mccz4-application}

We now write the (m)CCZ4 system in 3+1 form as a partially reduced system of the form~\eqref{eq:u-v}, with the ADM/BSSN variables grouped into the u-v sectors as
\begin{equation}
     u=(\chi,\tilde{\gamma}_{ij},\alpha,\beta^k)^T ,\quad
 v=(K,\tilde{A}_{ij},\Theta,\hat{\Gamma}^k)^T.
\end{equation}
Here $\gamma_{ij}$ is the induced metric on the constant-$t$ hypersurfaces and
$K_{ij}=-\frac12\mathcal L_n\gamma_{ij}$ is the extrinsic curvature. The conformal
variables are
\[
\chi:=(\det\gamma_{ij})^{-1/d},\qquad
\tilde\gamma_{ij}:=\chi\gamma_{ij},\qquad
\tilde A_{ij}:=\chi\Bigl(K_{ij}-\frac1d\gamma_{ij}K\Bigr),
\]
and $\tilde\Gamma^i:=\tilde\gamma^{jk}\tilde\Gamma^i{}_{jk}$ in $\hat\Gamma^i$ is the
contracted Christoffel symbol of $\tilde\gamma_{ij}$.

In order to apply Lemma~\ref{th:lem1}, we show that $M_{uv}$ is invertible by inverting the evolution equations of the $u$-variables and expressing the $v$-variables in terms of the normal and spatial derivatives of the $u$-variables as 
 \begin{subequations}\label{eq:u-vCCZ4}
     \begin{align}
K&=\frac{1}{\alpha}(\frac{d}{2\chi}\partial_\perp\chi+\partial_k\beta^k),\\
         \tilde{A}_{ij}&=-\frac{1}{2\alpha}(\partial_\perp\tilde{\gamma}_{ij}-2\tilde{\gamma
         }_{k(i}\partial_{j)}\beta^{k}+\frac{2}{d}\tilde{\gamma}_{ij}\partial_k\beta^k), \\
         \Theta& =-\frac{(1+a(x))}{2c_\alpha\alpha}\partial_\perp\alpha+\frac{1}{2\alpha}(\frac{d}{2\chi}\partial_\perp\chi+\partial_k\beta^k), \label{eq:eqTheta2}\\
       \hat{\Gamma}^i&=\frac{1+a(x)}{c_\beta} \partial_\perp \beta^i+ \frac{a(x)}{c_\beta}\alpha D^i\alpha \label{eq:eqGamma2},
     \end{align}
 \end{subequations}
 where equations \eqref{eq:eqTheta2} and \eqref{eq:eqGamma2}, combined with the definitions of $K$ and $\hat\Gamma^i$, give the expressions for $2\Theta$ and $2Z^i$ in~\eqref{eq:eqZofchris} of Appendix~\ref{app:gauge-driver-inversion}.
Inserting the expressions \eqref{eq:u-vCCZ4} for the $v$-variables into the evolution equations gives a manifestly second order PDE in $u=(\chi,\tilde{\gamma}_{ij},\alpha,\beta^k)^T$ of the form \eqref{eq:usec}, whose strong-hyperbolicity is equivalent to the original system according to the lemma \ref{th:lem1}.

Finally, we are able to prove that the (m)CCZ4 system written in terms of the metric variables, with the Z4-vectors traded for first derivatives of metric using \eqref{eq:eqZofchris},
is equivalent to the original 3+1 evolution system in the sense of strong hyperbolicity. 

\begin{boxedthm}
\begin{theorem}[The equivalence of 2nd order and 3+1 form of (m)CCZ4]
\label{th:mccz4-equivalence}
    The (modified) CCZ4 evolution system is strongly hyperbolic if and only if the associated second-order PDE in the metric variables, given by \eqref{eq:mCCZ4_2nd}, is itself strongly hyperbolic. 
\end{theorem}
\end{boxedthm}
\begin{proof}
    The proof amounts to showing that the second order PDE in $u=(\chi,\tilde{\gamma}_{ij},\alpha,\beta^k)^T$  variables, given by \eqref{eq:usec}, is equivalent to \eqref{eq:mCCZ4_2nd} in the sense of strong hyperbolicity.
The second-order system in the $u$-variables and the effective covariant
metric system \eqref{eq:mCCZ4_2nd} are two representations of the same
equations. Both are obtained from the covariant mCCZ4 equations after
eliminating the Z4 variables using the inverted gauge and kinematic
evolution equations. The two representations differ only by the smooth
algebraic change of variables $g_{\mu\nu}=g_{\mu\nu}(u)$ and by an
invertible recombination of the equations corresponding to the \(3+1\)
decomposition procedure. Consequently, their principal symbols are related by
    \begin{equation}
\mathcal{P}_u^{JI\mu\nu }\partial_\mu\partial_\nu u_I=\mathbf{L}^{J}_{\mu\nu}P_g^{\mu\nu\alpha\beta\rho\sigma}\partial_{\alpha}\partial_{\beta} g_{\rho\sigma} \simeq  \mathbf{L}^{J}_{\mu\nu}P_g^{\mu\nu\alpha\beta\rho\sigma} \mathbf{J}_{\rho\sigma}^I\partial_{\alpha}\partial_\beta u_I,
    \end{equation}
where we have omitted the first-order derivative terms in the last equality, and  \begin{equation}
    \mathbf{J}^{I}_{\mu\nu}:=\frac{\partial g_{\mu\nu
    }}{\partial u_I}
\end{equation}
is the Jacobian of the map from ADM variables to the metric;\footnote{The conformal variables are understood as coordinates on the
algebraic-constraint surface
$\det\tilde{\gamma}=1$. Hence admissible linearised perturbations
satisfy
$\tilde{\gamma}^{ij}\delta\tilde{\gamma}_{\langle ij\rangle}=0$, and the
Jacobian relevant to the principal-symbol transformation is its
restriction to
$(\delta\chi,\delta\tilde{\gamma}_{\langle ij\rangle})$.
The resulting map to $\delta\gamma_{ij}$ is nondegenerate for
$\chi>0$. The constrained and relaxed treatments of the algebraic
conformal constraints in CCZ4 are discussed in
Ref.~\cite{Alic:2011gg}.}
$\mathbf{L}^{J}_{\mu\nu}$ is the non-degenerate linear operator we use to project the covariant equations to the independent 3+1 equations. Both $\mathbf{L}$ and $\mathbf{J}$ are understood as invertible maps on the independent
components of the corresponding variables and equations.

The above relation between the second order principal symbols of the two PDEs can be written in an abstract form: \begin{equation}
    \mathcal{P}_u(\xi)= \mathbf{L}\cdot \mathcal{P}_g(\xi) \cdot \mathbf{J},
\end{equation}
and, accordingly, the matrix-valued coefficients $A,B,C$ associated with each PDE, as defined in \eqref{eq:matrixcoeff}, are also connected through  linear transformations as
\begin{equation}
    A_u= \mathbf{L}\cdot A_g \cdot \mathbf{J},\quad  B_u= \mathbf{L}\cdot B_g \cdot \mathbf{J},\quad C_u= \mathbf{L}\cdot C_g \cdot \mathbf{J}.
\end{equation}
Following the definition, we have
\begin{equation}
\begin{aligned}
     M_{u}(\xi_i)&= \begin{pmatrix}
        0& I \\
        -A_u^{-1}C_u & -A_u^{-1}B_u
    \end{pmatrix} =
    \begin{pmatrix}
        0& I \\
        -\mathbf{J}^{-1}A_g^{-1}\mathbf{L}^{-1}\cdot \mathbf{L}C_g \mathbf{J} &
        -\mathbf{J}^{-1}A_g^{-1}\mathbf{L}^{-1}\cdot \mathbf{L}B_g \mathbf{J}
    \end{pmatrix} \\
 &=\begin{pmatrix}
        \mathbf{J}^{-1}&0\\0&\mathbf{J}^{-1}
    \end{pmatrix}  M_g(\xi_i)
    \begin{pmatrix}
        \mathbf{J}&0\\0&\mathbf{J}
    \end{pmatrix} .
\end{aligned}
\end{equation}
According to Proposition~\ref{th:prop1}, the second order PDE system in $u$ variables is strongly hyperbolic if and only if its counterpart in metric variables is strongly hyperbolic. Combining this with lemma \ref{th:lem1}, we conclude the proof.
\end{proof}
The remaining task for the rest of the paper is then to prove the strong hyperbolicity of the modified CCZ4 evolution system in the second order form using the standard characteristic technique developed by \citet{Kovacs:2020ywu}.

\section{Proof of strong hyperbolicity for Einstein-Gauss-Bonnet gravity }
\label{sec:proof-strong-hyperbolicity}
A key lesson from the work of \citet{Kovacs:2020ywu} is that, to show the strong hyperbolicity, one does not need to explicitly diagonalise the principal symbol $M(\xi_i)$.  For a generic $N$-component second-order PDE of the form \eqref{eq:usec}, we first seek all polarisation vectors $t_I$ that solve the \textit{characteristic equation}:
\begin{equation}
    \mathcal{P}(\xi)^{IJ} t_J\equiv\mathcal{P}^{\mu\nu IJ}\xi_\mu\xi_\nu t_J=0 . \label{eq:character}
\end{equation}
Here, the covector $\xi_\mu$ is characteristic if the eigenpair $(\xi_0, t_I)$ solves the quadratic eigenvalue problem \eqref{eq:character}, namely, $\xi_0$ is a root of the characteristic polynomial $\det\mathcal{P}^{IJ}(\xi)$. For each eigenpair $(\xi_0, t_I)$, we can construct an eigenvector of the $2N \times 2N$ matrix $M(\xi_i)$ corresponding to the eigenvalue $\xi_0$, which is given by
\begin{equation}
    \mathcal{U} = \begin{pmatrix}
        t_I \\
        \xi_0 t_I \label{eq:eigenvec}
    \end{pmatrix}.
\end{equation}
In practice, if one can construct $2N$ linearly independent smooth eigenvectors in this manner—that is, if  $M(\xi_i)$ can be diagonalised via a smooth similarity transformation—and if all the $\xi_0$ are real, then the system is in fact strongly hyperbolic.

For GR in mCCZ4 formulation, we can explicitly construct all such eigenvectors following the prescriptions in \cite{Kovacs:2020ywu}, thereby proving the strong hyperbolicity. Similar to the modified harmonic case, the eigenvalues separate into three distinct sectors: \textbf{(i)} the pure-gauge sector, \textbf{(ii)} the gauge-condition-violating sector, and \textbf{(iii)} the physical sector. The required cone-separation conditions will be stated where they enter each construction. This exactly reproduces all eigenvalues obtained in \cite{AresteSalo:2023mmd} through direct diagonalisation, which confirms the validity of our approach.  In this situation, the symmetrizer of $M(\xi_i)$ is naturally expressed as $\mathcal{K} = (S^{-1})^\dagger S$, where $S$ is the matrix whose columns are the complete set of eigenvectors.

For EGB gravity in the weakly coupled regime, we can still construct the same number of eigenvectors associated with the pure-gauge sector and the gauge-condition-violating sector. Strong hyperbolicity then follows from the exclusion of Jordan block in the physical sector through a continuity argument and an alternative construction of symmetrizer as suggested in \cite{Kovacs:2020ywu}.

\subsection{Strong hyperbolicity of GR in mCCZ4 formulation}
\label{sec:gr-characteristic-problem}
Following \cite{Kovacs:2020ywu}, we derive the characteristic equation of the mCCZ4 system by substituting the second-order derivatives with wave vectors $\xi_\mu$ after linearising the non-linear mCCZ4  equation \eqref{eq:mCCZ4_2nd} around a generic background, namely $\partial_\mu\partial_\nu\delta g_{\rho\sigma}\to\xi_\mu\xi_\nu t_{\rho\sigma}$. We can split the full principal symbol into that obtained from the gauge-invariant EoMs and that obtained from the gauge-fixing term:
\begin{equation}
    0=\mathcal{P}(\xi)^{\mu\nu\rho\sigma} t_{\rho\sigma}=\mathcal{P}_{*}(\xi)^{\mu\nu\rho\sigma} t_{\rho\sigma}+ \mathcal{P}_{\mathrm{GF}}(\xi)^{\mu\nu\rho\sigma} t_{\rho\sigma}. \label{eq:Charactmunu}
\end{equation}
For the gauge-invariant part, we have
\begin{equation}
    \mathcal{P}_*(\xi)^{\mu\nu\rho\sigma}:=\mathcal{P}_*^{\mu\nu\alpha\beta\rho\sigma}\xi_\alpha\xi_\beta\equiv \frac{\partial E^{\mu\nu}}{\partial(\partial_\alpha\partial_\beta g_{\rho\sigma})}\xi_\alpha\xi_\beta.
\end{equation}
In GR, the expression is given by
\begin{equation}
         \mathcal{P}_*(\xi)^{\mu\nu\rho\sigma}=-\frac{1}{2} \xi^2P^{\mu\nu\rho\sigma}+ P_\gamma {}^{\alpha\mu\nu}\xi_{\alpha} P^{\gamma\beta\rho\sigma}\xi_{\beta}, \label{eq:Pstar}
\end{equation}
where $\xi^2 \equiv g^{\gamma\delta}\xi_\gamma \xi_\delta$, and the $P$- operator is defined by
\begin{equation}
    P_\mu {}^{\nu\rho\sigma}:=\delta_\mu^{(\rho}g^{\sigma)\nu}-\frac{1}{2}\delta_\mu^{\nu}g^{\rho\sigma}.
\end{equation}
For the gauge-fixing part, adopting a similar notation as  in \cite{Kovacs:2020ywu}, we express the principal symbol acting on the polarisation vector as 
\begin{equation}
     \mathcal{P}_{\mathrm{GF}}(\xi)^{\mu\nu\rho\sigma} t_{\rho\sigma}:=
     - \hat{P}_\gamma {}^{\alpha\mu\nu}\xi_{\alpha} \tilde{\mathbb{P}}^{\gamma\beta\rho\sigma}\xi_{\beta} t_{\rho\sigma}, \label{eq:PGauge}
\end{equation}
where
 \begin{equation}
\hat{P}_{\alpha}^{~\beta\mu\nu}=\delta_{\alpha}^{(\mu}\,\hat{g}^{\nu)\beta}-\tfrac{1}{2}\,\delta_{\alpha}^{\beta}\,\hat{g}^{\mu\nu}
\end{equation}
is the same tensor coefficient as in \eqref{eq:mHg} using the auxiliary metric $\hat{g}$, and
\begin{equation}
\tilde{\mathbb{P}}^{\gamma\beta\rho\sigma}:=-2\frac{\partial Z^{\gamma}}{\partial(\partial_\beta g_{\rho\sigma})}, \label{eq:deftildeP}
\end{equation}
with $Z^\mu$ treated as functions of the first derivatives of metric as given by \eqref{eq:eqZofchris}.\footnote{Here, the factor $-2$ is chosen such that, when it reduces to the harmonic gauge, we have $P=\hat{P}=\tilde{\mathbb{P}}$.} By using the explicit expression shown in \eqref{eq:eqZofg} of the appendix~\ref{app:gauge-driver-inversion}, we obtain
\begin{equation}
\begin{aligned}
      \tilde{\mathbb{P}}^{\mu\alpha\beta\lambda}\xi_\alpha t_{\beta\lambda}&\equiv-2\frac{\partial Z^{\mu}}{\partial(\partial_\alpha g_{\beta\lambda})}\xi_\alpha t_{\beta\lambda}\\
      &=
  (\gamma^{\mu\lambda}\mathcal{G}_{\bf T}^{\alpha\beta}-n^\mu n^\lambda \mathcal{G}_{\bf 0}^{\alpha\beta}) (\xi_{(\alpha} t_{\beta)\lambda}-\frac{1}{2}\xi_\lambda t_{\alpha\beta})+ \\ &\frac{1}{2}\gamma^{\alpha\mu} (\frac{d-2}{d}\gamma^{\beta\lambda}-\frac{\chi\alpha^2}{c_\beta}n^\lambda n^\beta )\xi_\alpha t_{\beta\lambda}.
\end{aligned} \label{eq:eq_gauge_fix_symbol}
\end{equation}

As noted in \cite{Kovacs:2020ywu,Reall:2021voz}, the principal symbol of the gauge-invariant EoMs possess two identities for arbitrary $\xi$:
\begin{equation}
    \xi_\mu\mathcal{P}_*(\xi)^{\mu\nu\rho\sigma}=0 ,\quad
   \mathcal{P}_*(\xi)^{\mu\nu\rho\sigma}  \xi_\rho=0, \label{eq:diffeom}
\end{equation}
which can be viewed as the frequency-domain manifestation of the gauge invariance and Bianchi identity of the Einstein Equation. Indeed, these identities are expected to persist for generic field equations of obtained from the variation of a diffeomorphism-invariant action with respect to the metric field \cite{Reall:2021voz}. Contracting now the full characteristic equation \eqref{eq:Charactmunu} with $\xi$ gives
\begin{equation}
    0=\xi_\mu \mathcal{P}_{\mathrm{GF}}(\xi)^{\mu\nu\rho\sigma} t_{\rho\sigma}= -\xi_\mu \hat{P}_\gamma {}^{\alpha\mu\nu}\xi_{\alpha} \tilde{\mathbb{P}}^{\gamma\beta\rho\sigma}\xi_{\beta} t_{\rho\sigma}=-\frac{1}{2} (\hat{g}^{\mu\alpha}\xi_\mu\xi_\alpha)(\tilde{\mathbb{P}}^{\nu\beta\rho\sigma}\xi_{\beta}t_{\rho\sigma}), \label{eq:Factor}
\end{equation}
which is exactly the factorisation identity in \cite{Kovacs:2020ywu}. We distinguish two cases according to whether $\hat{g}^{\mu\nu}\xi_\mu\xi_\nu$ vanishes, and state the required cone-separation conditions in the corresponding constructions below.

\subsubsection*{Case(i) : $\hat{g}^{\mu\alpha}\xi_\mu\xi_\alpha\neq 0$, while $\tilde{\mathbb{P}}^{\nu\beta\rho\sigma}\xi_{\beta}t_{\rho\sigma}=0$.}
    
    In this case, the characteristic equation \eqref{eq:Charactmunu} demands
    \begin{equation}
        0= \mathcal{P}_*(\xi)^{\mu\nu\rho\sigma}t_{\rho\sigma}=-\frac{1}{2} \xi^2P^{\mu\nu\rho\sigma} t_{\rho\sigma}+ P_\gamma {}^{\alpha\mu\nu}\xi_{\alpha} P^{\gamma\beta\rho\sigma}\xi_{\beta} t_{\rho\sigma}.
    \end{equation}
    In light of the symmetry of the $P$- operator, namely $P^{\gamma\alpha\mu\nu}=P^{\mu\nu\gamma\alpha}$ and the fact that it is invertible in the space of symmetric rank-two tensors, the outermost $P$- action in the above equation can be peeled off, yielding the trace-reversed version of the physical part of characteristic equation: 
    \begin{equation} 
    \xi^2 t_{\mu\nu}= 2\xi_{(\mu} P_{\nu)}{}^{\beta\rho\sigma}\xi_\beta t_{\rho\sigma}. \label{eq:onshell}
    \end{equation}
Then, according to whether $\xi$ is null w.r.t. the physical metric $g$, we have the following two subcases: 
\subsubsection*{Subcase (ia): The pure-gauge sector, characterised by $\xi^2 \neq 0$. } \label{sec:pure_gauge_mccz4}
   
From \eqref{eq:onshell} it immediately follows that the polarisation tensor $t$ should necessarily take the form
\begin{equation}
    t_{\mu\nu} = \xi_{(\mu} X_{\nu)}, \label{eq:puregauge}
\end{equation}
which is indeed a residual gauge transformation of the background metric in the high-frequency limit. Inserting the ansatz \eqref{eq:puregauge} to the gauge condition $\tilde{\mathbb{P}}^{\nu\beta\rho\sigma}\xi_{\beta}t_{\rho\sigma}=0$, we obtain a characteristic equation for the vector polarisation $X_\mu$
\begin{equation}
Q(\xi)^{\nu\sigma}X_\sigma\equiv\tilde{\mathbb{P}}^{\nu\beta\rho\sigma}\xi_{\beta}\xi_{\rho}X_\sigma=0. \label{eq:vectorEq}
\end{equation}

For \eqref{eq:vectorEq} to admit non-zero solution $X_\mu$, the eigenvalue $\xi_0$ must be a root of the characteristic polynomial $\det Q(\xi)=0$. Indeed, we shall see below that such $2(d+1)$ degree polynomial can further factorise into three quadratic factors with certain multiplicities.

To show this, we first introduce orthonormal basis 
\begin{equation}    
\{e^{a}_\mu\mid a=0,1,\ldots,d\} := \{e^0 := n_\mu,\ e^1 := k_\mu \equiv \tfrac{\gamma_{\mu}^{\nu}\xi_{\nu}}{\sqrt{\gamma^{\alpha\beta}\xi_{\alpha}\xi_{\beta}}},\ e^{\hat{i}}_{\mu}\mid \hat{i}=2,\ldots,d \}, \label{eq:tetrad1}
\end{equation}
 where $e^{\hat{i}}_\mu$ denotes the $A$-th covector transverse to both $\xi_\mu$ and $n_\mu$. The tetrad components of $X$ is defined such that 
\begin{equation}
    X_\mu=e_\mu^a X_a ,\quad X^a:= g^{\mu\nu}e_{\nu}^aX_\mu=e^{a}_\mu X^\mu,
\end{equation}
with $X_a$ ($X^a$)  raised (lowered) by the Minkowskian metric $\eta$.
 

Substituting the ansatz \eqref{eq:puregauge} to \eqref{eq:eq_gauge_fix_symbol} gives the expression of $Q$-operator acting on $X$:
\begin{equation}
    Q(\xi)^{\mu\sigma}X_\sigma= \frac{1}{2}(\gamma^{\mu\lambda}\mathcal{G}_{\bf T}^{\alpha\beta}-n^\mu n^\lambda \mathcal{G}_{\bf 0}^{\alpha\beta})  (\xi_\alpha \xi_\beta X_\lambda) + \frac{1}{2}\gamma^{\alpha\mu} (\frac{d-2}{d}\gamma^{\beta\lambda}-\frac{\chi\alpha^2}{c_\beta}n^\lambda n^\beta )\xi_\alpha \xi_{\beta}X_{\lambda}.
\end{equation}
Project it onto the temporal, longitudinal and transverse directions. We have
    \begin{subequations}
\begin{align}
     (Q(\xi)\cdot X)^0&:= n_\mu Q(\xi)^{\mu\sigma}X_\sigma=\frac{1}{2}(\mathcal{G}_{\bf 0}^{\alpha\beta}\xi_\alpha\xi_\beta) X^0, \\
     (Q(\xi)\cdot X)^1&:= k_\mu Q(\xi)^{\mu\sigma}X_\sigma= \frac{1}{2}(\mathcal{G}_{\bf T}^{\alpha\beta}\xi_\alpha\xi_\beta)X^1+\frac{d-2}{2d} (\gamma^{\alpha\beta}\xi_\alpha\xi_\beta) X^1 -\frac{\chi\alpha^2}{2 c_\beta} (k\cdot\xi)(n\cdot\xi)X^0 \nonumber \\
     &= \frac{d-1}{d}(\mathcal{G}_{\bf L}^{\alpha\beta}\xi_\alpha\xi_\beta) X^1-\frac{\chi\alpha^2}{2 c_\beta} (k\cdot\xi)(n\cdot\xi)X^0, \label{eq:Longi}\\
      (Q(\xi)\cdot X)^{\hat{i}}&:= e^{\hat{i}}_\mu Q^{\mu\sigma}X_\sigma=\frac{1}{2}(\mathcal{G}_{\bf T}^{\alpha\beta}\xi_\alpha\xi_\beta) X^{\hat{i}},
\end{align}        
    \end{subequations}
where, in the second equation \eqref{eq:Longi}  we have used the fact that $\xi_\mu \gamma^{\mu}_\beta=\sqrt{\gamma^{\rho\sigma}\xi_\rho\xi_\sigma} k_\beta $, and we have defined one additional auxiliary metric $\mathcal{G}_{\bf L}$ as 
\begin{equation}
    \mathcal{G}_{\bf L}^{\alpha\beta}:=\frac{d-2}{2(d-1)}\gamma^{\alpha\beta}+\frac{d}{2(d-1)}\mathcal{G}_{\bf T}^{\alpha\beta}.
\end{equation}
In summary, we have introduced three auxiliary metrics, with their explicit forms given by
\begin{subequations}
    \begin{align}
    &\mathcal{G}_{\bf 0}^{\alpha\beta}:=\gamma^{\alpha\beta}+ \frac{(1+a(x))\alpha}{c_\alpha} n^\alpha n^\beta, \\
        &\mathcal{G}_{\bf L}^{\alpha\beta}:=\gamma^{\alpha\beta}-\frac{d (1+a(x))\alpha^2\chi} {2(d-1)c_\beta} n^\alpha n^\beta,\\
         &\mathcal{G}_{\bf T}^{\alpha\beta}:=\gamma^{\alpha\beta}-\frac{(1+a(x))\alpha^2\chi}{c_\beta} n^\alpha n^\beta.
    \end{align}
\end{subequations}
Note that $c_\alpha<0<c_\beta,a(x)>0$ and hence $\mathcal{G}_{\bf 0,L,T}$ are all Lorentzian by construction. 

 We can then express the characteristic equation in matrix form as
\begin{equation}
    \begin{pmatrix}
        \frac{1}{2}\mathcal{G}_{\bf 0}^{\alpha\beta}\xi_\alpha\xi_\beta & 0 & 0 \\
         - \frac{\chi\alpha^2}{2c_\beta} (n\cdot\xi)(k\cdot\xi) & \frac{d-1}{d}\mathcal{G}_{\bf L}^{\alpha\beta}\xi_\alpha\xi_\beta & 0 \\
        0 & 0 & \frac{1}{2}(\mathcal{G}_{\bf T}^{\alpha\beta}\xi_\alpha\xi_\beta) \delta_{\hat{i}\hat{j}} \\ 
    \end{pmatrix} 
    \begin{pmatrix}
        X^0 \\ X^1 \\ X^{\hat{j}}
    \end{pmatrix}=0. \label{eq:mccz4-gauge-block}
\end{equation}
Hence, the characteristic polynomial factorises into three quadratic pieces as
\begin{equation}
    \det Q(\xi)^{ab}=\frac{d-1}{2^{d}d}\cdot( \mathcal{G}_{\bf 0}^{\alpha\beta}\xi_\alpha\xi_\beta) \cdot (\mathcal{G}_{\bf L}^{\alpha\beta}\xi_\alpha\xi_\beta)\cdot (\mathcal{G}_{\bf T}^{\alpha\beta}\xi_\alpha\xi_\beta)^{d-1}.
\end{equation}
For \eqref{eq:vectorEq} to admit a non-trivial solution, the characteristic covector $\xi_\mu$ must lie on one of the null cones defined by the three auxiliary metrics.

Suppose that $\mathcal{G}_{\bf 0}^{\mu\nu}\xi_\mu\xi_\nu=\mathcal{G}_{\bf L}^{\mu\nu}\xi_\mu\xi_\nu=0$. Equation~\eqref{eq:mccz4-gauge-block} then forces $X^0=0$, and hence we lose one independent polarisation associated with each branch of $\xi_0$ that solves the degenerate null cone equation. This defect is inherited from standard CCZ4, as will be analysed in detail in Sec.~\ref{sec:standard-ccz4}. To remove this defect, we therefore require $\mathcal{G}_{\bf 0}^{\mu\nu}\neq\mathcal{G}_{\bf L}^{\mu\nu}$, equivalently $-c_\alpha\alpha\chi\neq 2(d-1)c_\beta/d$. This can be achieved by a suitable choice of $c_\alpha,c_\beta,a(x),b(x)$.

With this degeneracy excluded, the corresponding eigenvalues and polarisation vectors are
\begin{align}
        \mathcal{G}_{\bf 0}^{\mu\nu}\xi_\mu\xi_\nu&=0:\quad \xi_0 =\beta^i\xi_i \pm \sqrt{\frac{-c_\alpha \alpha}{1+a(x)} },
        & X_\mu &= n_\mu+\frac{Q^{10}}{Q^{11}}k_\mu,\label{eq:mccz4-gauge-temporal}\\
        \mathcal{G}_{\bf L}^{\mu\nu}\xi_\mu\xi_\nu&=0:\quad \xi_0 =\beta^i\xi_i \pm \sqrt{\frac{2(d-1)}{d}\frac{c_\beta }{(1+a(x))\chi}},
        & X_\mu &= k_\mu,\label{eq:mccz4-gauge-longitudinal}\\
        \mathcal{G}_{\bf T}^{\mu\nu}\xi_\mu\xi_\nu&=0:\quad \xi_0 =\beta^i\xi_i \pm \sqrt{\frac{c_\beta }{(1+a(x))\chi}},
        & X_\mu &= e^{\hat{i}}_\mu,\label{eq:mccz4-gauge-transverse}
\end{align}
where we have normalised the spatial part of the wave vector as $\gamma^{ab}\xi_a\xi_b=1$. Note that the temporal and transverse polarisations remain linearly independent when $\mathcal G_{\mathbf0}^{\mu\nu}=\mathcal G_{\mathbf T}^{\mu\nu}$.
\textit{It is easy to verify that by setting $c_\beta=\frac{d}{2(d-1)},c_\alpha=-2$, we exactly reproduce all eigenvalues in this sector as obtained in \cite{AresteSalo:2023mmd} from a direct diagonalisation of the 3+1 principal symbol}.

Compared with the modified harmonic formulation \cite{Kovacs:2020ywu}, where the pure-gauge eigenvalues are governed by a single auxiliary metric $\tilde{g}^{\mu\nu}:=g^{\mu\nu}-a(x)n^\mu n^\nu$, the pure-gauge sector here decomposes into temporal, longitudinal and transverse sub-sectors.
Following the same reasoning as in the modified harmonic formulation, we shall require the null cones of $\mathcal{G}_{\bf 0,L,T}^{\mu\nu}$ to be disjoint from that of $g^{\mu\nu}$. This can be achieved by a suitable choice of $c_\alpha,c_\beta,a(x),b(x)$. Equivalently, $\det Q(\xi)\neq0$ for every covector $\xi_\mu$ null with respect to the physical metric $g^{\mu\nu}$.

\subsubsection*{Subcase (ib): The physical sector, defined by $\xi^2= 0$. }
The corresponding eigenvalues are $\xi_{0}^{\pm}=\beta^i\xi_i\pm\alpha$.
In this case, the trace-reversed characteristic equation \eqref{eq:onshell} implies that 
\begin{equation}
     P_\mu{}^{\beta\rho\sigma}\xi_\beta t_{\rho\sigma}=0=\xi^\sigma(t_{\mu\sigma}-\frac{1}{2} t^\alpha_\alpha g_{\mu\sigma}). \label{eq:transverse}
\end{equation}
    Recall that the gauge condition
    \begin{equation}
\tilde{\mathbb{P}}^{\nu\beta\rho\sigma}\xi_{\beta}t_{\rho\sigma}=0 \label{eq:gaugecond}
    \end{equation} 
    has to be satisfied as well.  Therefore, for each of $\xi^{\pm}_0$, there are supposed to be $\frac{(d+1)(d+2)}{2}-2(d+1)=\frac{(d+1)(d-2)}{2}$ physical polarisations, which would in turn constitute $(d+1)(d-2)$ independent eigenvectors of $M(\xi_i)$ of the form \eqref{eq:eigenvec}. The $\frac{(d+1)(d-2)}{2}$ linearly independent physical polarisations are constructed as follows. We first construct all $\frac{(d+1)(d-2)}{2}$ transverse-traceless polarisations, which automatically solve \eqref{eq:transverse}. Here, the transverse-traceless polarisations take the form
\begin{equation}
    \{t^{\rm TT}_{\mu\nu}=t_{\langle\hat{i}\hat{j}\rangle} e^{\hat{i}}_\mu e^{\hat{j}}_\nu \mid t_{\langle\hat{i}\hat{j}\rangle}\delta^{\hat{i}\hat{j}}=0,\; \hat{i},\hat{j}=2,\ldots,d\}, \label{eq:eqTT}
\end{equation}
where \(e^{\hat{i}}_\mu\) are the transverse tetrad vectors introduced in \eqref{eq:tetrad1}.

Observe that
\begin{equation}
    t_{\mu\nu}^{\rm TT} g^{\mu\nu}=0, \quad  t_{\mu\nu}^{\rm TT} n^{\mu}=0=t_{\mu\nu}^{\rm TT} k^{\mu}, \qquad t_{\mu\nu}^{\rm TT} \xi^\mu=0,
\end{equation}
and that
\begin{equation}
    \mathcal{G}_{\bf {0,L,T}}^{\alpha\beta} t_{\alpha\beta}^{\rm TT}=0= \gamma^{\alpha\beta} t_{\alpha\beta}^{\rm TT}.
\end{equation}
Consequently, plugging in a transverse-traceless polarisation into the expression for the gauge condition \eqref{eq:eq_gauge_fix_symbol} yields
\begin{equation}
\begin{aligned}
      \tilde{\mathbb{P}}^{\mu\alpha\beta\lambda}\xi_\alpha t_{\beta\lambda}^{\rm TT}&= (\gamma^{\mu\lambda}\mathcal{G}_{\bf T}^{\alpha\beta}-n^\mu n^\lambda \mathcal{G}_{\bf 0}^{\alpha\beta}) (\xi_{\alpha} t_{\beta\lambda}^{\rm TT}-\frac{1}{2}\xi_\lambda t_{\alpha\beta}^{\rm TT})+ \\ &\frac{1}{2}\gamma^{\alpha\mu} (\frac{d-2}{d}\gamma^{\beta\lambda}-\frac{\chi\alpha^2}{c_\beta}n^\lambda n^\beta )\xi_\alpha t_{\beta\lambda}^{\rm TT}=0. \label{eq:eqTTcondi}
\end{aligned}
\end{equation}
The transverse-traceless polarisations~\eqref{eq:eqTT} therefore satisfy the gauge condition~\eqref{eq:gaugecond}. As ${\det Q(\xi)\lvert_{\xi^2=0}\neq0}$, the gauge condition fixes the residual pure-gauge freedom, so these polarisations form a complete basis for the physical sector.
\subsubsection*{Case(ii) : $\hat{g}^{\mu\alpha}\xi_\mu\xi_\alpha= 0$, while $\tilde{\mathbb{P}}^{\nu\beta\rho\sigma}\xi_{\beta}t_{\rho\sigma}\neq 0$. This defines the gauge-condition-violating sector.}
The eigenvalues can be solved from the null cone equation of $\hat{g}$, namely  \begin{equation}
    \hat{\xi}_0^{\pm}=\beta^i\xi_i\pm \frac{ \alpha}{\sqrt{1+b(x)}}. \label{eq:gcv}
\end{equation}
For the construction in this sector, we require that the null cone of $\hat{g}^{\mu\nu}$ be disjoint from those of $g^{\mu\nu}$ and $\mathcal{G}_{\bf 0,L,T}^{\mu\nu}$. This can be achieved by a suitable choice of $c_\alpha,c_\beta,a(x),b(x)$. Thus, each characteristic covector $\hat{\xi}_\mu$ associated with~\eqref{eq:gcv} satisfies $\hat{\xi}^2\neq0$ and $\det Q(\hat{\xi})\neq0$.

We now construct the $d+1$-independent polarisations for each of the $\hat{\xi}^\pm_0$. 
Recall that in \cite{Kovacs:2020ywu}, the gauge-condition-violating polarisation is determined by a free vector constant $v^\mu$. Since the vector $v^\mu$ lies in a $(d+1)$-dimensional space, the polarisations constructed in this way span a $(d+1)$-dimensional space for each branch of the solutions \eqref{eq:gcv}, thereby producing $2(d+1)$ independent eigenvectors of the form \eqref{eq:eigenvec} for the matrix $M(\xi_i)$.  The procedure of constructing the polarisations for each $\hat{\xi}\in\{\hat\xi^{\pm}\mid \xi^{\pm}_\mu=(\xi_0^\pm,\xi_i)\}$ from $v^\mu$ amounts to solving the two equations at the same time:
\begin{align}
    \mathcal{P}_*(\hat\xi)^{\mu\nu\rho\sigma}t_{\rho\sigma}&=\hat{P}_\gamma {}^{\alpha\mu\nu}\hat{\xi}_{\alpha} v^\gamma,\label{eq:tensormode} \\
    \tilde{\mathbb{P}}^{\gamma\beta\rho\sigma}\hat{\xi}_{\beta} t_{\rho\sigma}&=v^\gamma.\label{eq:vecmode}
\end{align}
Note that the solution $t_{\mu\nu}(v)$ to the above equation system automatically solves the original characteristic equation \eqref{eq:Charactmunu} with the wave vector being $\hat{\xi}$.

Following the prescription in \cite{Kovacs:2020ywu}, we introduce a new set of orthonormal basis for each branch in \eqref{eq:gcv}:
\begin{equation}
   \{e^1_\mu=\tfrac{1}{\sqrt{\hat{\xi}^2}}\hat{\xi}_\mu,e^A_\mu\mid A=0,2,3,\cdots d \},\label{eq:tetrad2}
\end{equation}
with $\hat{\xi} \in \{\xi^\pm\}$, which are null w.r.t. the $\hat{g}^{\mu\nu}$ but normalisable w.r.t. the physical metric.  Then, the ansatz for the solution $t_{\mu\nu}(v)$ takes the form 
\begin{equation}
    t_{\mu\nu}(v)= t_{AB}(v)e^{A}_\mu e^B_\nu + \hat{\xi}_{(\mu} Y_{\nu)}.
    \label{eq:ansatzgcv}
\end{equation}
The property of the $\hat{P}$- operators yields
\begin{equation}
    \hat{\xi}_\mu\hat{P}_\gamma {}^{\alpha\mu\nu}\hat{\xi}_{\alpha} v^\gamma=\frac{1}{2}\hat{g}^{\mu\alpha}\hat{\xi}_\mu\hat{\xi}_\alpha v^\nu=0,
\end{equation}
and it follows from the diffeomorphism symmetry and Bianchi identities that 
\begin{equation}
    \hat{\xi}_\mu\mathcal{P}_*(\hat\xi)^{\mu\nu\rho\sigma}= 0=\mathcal{P}_*(\hat\xi)^{\mu\nu\rho\sigma}\hat{\xi}_\rho.
\end{equation}
Hence the tensor equation \eqref{eq:tensormode} is trivial when projected onto the $e^1_\mu\sim \xi_\mu$-direction and the second term in \eqref{eq:ansatzgcv} would not enter the tensor equation \eqref{eq:tensormode}, yielding a closed system for the transverse mode $t_{AB}$:
\begin{equation}
    \mathcal{P}_*(\hat\xi)^{ABCD}t_{CD}=S^{AB}(v)\equiv e^A_\mu e^B_\nu \hat{P}_\gamma {}^{\alpha\mu\nu}\hat{\xi}_{\alpha} v^\gamma. \label{eq:eqtransverse}
\end{equation}
The condition $\hat{\xi}^2\neq0$ ensures that $\mathcal{P}_*^{ABCD}(\hat\xi)$ is invertible on symmetric rank-two tensors transverse to $\hat{\xi}$. This can be seen by looking at the homogeneous part of \eqref{eq:tensormode}, namely $ \mathcal{P}^{\mu\nu\rho\sigma}_*(\hat\xi)t_{\rho\sigma}^{\rm hom}=0$, which is again equivalent to the trace-reversed equation \eqref{eq:onshell} with the $\xi$ replaced by $\hat{\xi}$:
\begin{equation}
\hat{\xi}^2 t_{\mu\nu}^{\rm hom}= 2\hat{\xi}_{(\mu} P_{\nu)}{}^{\beta\rho\sigma}\hat{\xi}_\beta t_{\rho\sigma} ^{\rm hom}.
\end{equation}
Note that $\hat{\xi}^2\neq 0$  by construction; it then follows that the homogeneous solution must  take form of a "pure-gauge" polarisation
\begin{equation}
    t^{\rm hom}_{\mu\nu}=\hat{\xi}_{(\mu} X_{\nu)},
\end{equation}
for some $X$. This implies that the transverse part of the homogeneous solution is indeed zero, and hence we have  
\begin{equation}
    \mathcal{P}_*(\hat\xi)^{ABCD}t_{CD}=0 \Longleftrightarrow t_{CD}=0,
\end{equation}
which guarantees the solvability  of the equation \eqref{eq:eqtransverse}, ensuring that $t_{AB}$ is a function of $v$.

Now we proceed with the remaining vector degree of freedom $Y_\mu$, which is determined by inserting the ansatz \eqref{eq:ansatzgcv} into the vector equation \eqref{eq:vecmode}, which gives 
\begin{equation}
   v^\gamma= \tilde{\mathbb{P}}^{\gamma\beta\rho\sigma}\hat{\xi}_{\beta} t_{\rho\sigma}= \tilde{\mathbb{P}}^{\gamma\beta\rho\sigma}\hat{\xi}_{\beta}e_\rho^A e_\sigma^B t_{AB}(v)+ Q^{\gamma\sigma}(\hat\xi)Y_{\sigma},
\end{equation}
where we have used the definition of $Q$-operator:
\begin{equation}
    Q(\hat\xi)^{\gamma\sigma}Y_{\sigma}\equiv \tilde{\mathbb{P}}^{\gamma\beta\rho\sigma}\hat{\xi}_\beta\hat{\xi}_\rho Y_{\sigma},
\label{eq:gauge-Q-hat}
\end{equation}
as given by \eqref{eq:vectorEq}. The condition $\det Q(\hat\xi)\neq0$ now gives
\begin{equation}
    Y_{\sigma}=Q^{-1}_{\gamma\sigma}\left(v^{\gamma}-\tilde{\mathbb{P}}^{\gamma\beta\rho\sigma}\hat{\xi}_{\beta}e_\rho^A e_\sigma^B t_{AB}(v)\right). \label{eq:v2Y}
\end{equation}
This is the counterpart of equation (43) in~\cite{Kovacs:2020ywu}.

\subsubsection*{A summary of the eigenvectors constructed }
Based on the above construction, we summarise the eigenvectors obtained for the principal symbol $M(\xi_i)$:
\begin{itemize}
    \item The pure-gauge sector: $\tilde{V}_{\bf 0}^{\pm}$ ,$\tilde{V}_{\bf L}^{\pm}$,$\tilde{V}_{\bf T}^{\pm}$. 
    
    The dimensions of the three sub-sectors are given by ${\rm dim}(\tilde{V}_{\bf 0}^{\pm})=1,\,{\rm dim}(\tilde{V}_{\bf L}^{\pm})=1, {\rm dim}(\tilde{V}_{\bf T}^{\pm})=d-1$, and thus we have $2(d+1)$ independent eigenvectors in this sector.
    \item The physical sector $V^{\pm}$ :
    
    Each of $V^\pm$ is of dimension $\frac{(d+1)(d-2)}{2}$, and hence we have $(d+1)(d-2)$ independent eigenvectors.
    \item The gauge-condition-violating sector: $\hat{V}^{\pm}$:
    
    For each of $\hat{V}^{\pm}$, we have constructed $d+1$ independent polarisations. Therefore we have $2(d+1)$ eigenvectors in this sector.
\end{itemize}

Recall that, for the construction to be admissible, we require the non-coincidence conditions below:
\begin{condition}[Separation of null cones]
\label{ass:mccz4-cone-separation}
In the GR or Einstein-scalar limit, the metrics satisfy, for $A\in\{\mathbf0,\mathbf L,\mathbf T\}$,
\[
\hat g^{\mu\nu}\neq g^{\mu\nu},\qquad
\mathcal G_A^{\mu\nu}\notin\{g^{\mu\nu},\hat g^{\mu\nu}\},\qquad
\mathcal G_{\mathbf0}^{\mu\nu}\neq\mathcal G_{\mathbf L}^{\mu\nu}.
\]
\end{condition}
These conditions separate the physical sector, pure-gauge sector, and gauge-condition-violating sector and exclude the temporal-longitudinal defect within the pure-gauge sector; they can be satisfied by a suitable choice of $c_\alpha,c_\beta,a(x),b(x)$.

\begin{figure}[htbp]
    \centering
    \includegraphics[width=0.4\linewidth]{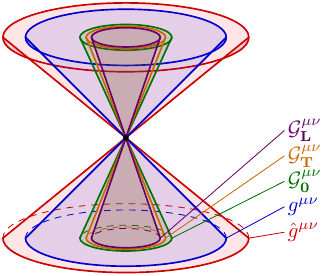}
    \caption{A representative ordering of the five null cones satisfying the cone-separation condition~\ref{ass:mccz4-cone-separation}.}
    \label{fig:lighcones}
\end{figure}

Figure~\ref{fig:lighcones} illustrates one possible ordering of the corresponding null cones.

To conclude, we have obtained $2(d+1) + (d+1)(d-2) + 2(d+1) = (d+1)(d+2)$ eigenvectors in total, which together form a complete basis to diagonalise $M(\xi_i)$. Note also that all the eigenvalues are determined by  particular null cone equations associated with some Lorentzian metric, which guarantees that all eigenvalues of $M(\xi_i)$ are real.

This completes the proof of strong hyperbolicity of GR in the mCCZ4 formulation.

\subsection{Weakly coupled Einstein-Gauss-Bonnet theory} 
\label{sec:egb-extension}
In $d+1$ spacetime dimensions, the vacuum Lovelock equations take the
form~\cite{Kovacs:2020ywu}
\begin{equation}
E^\mu{}_{\nu}
=G^\mu{}_{\nu}+\Lambda\delta^\mu_\nu
+\sum_{p=2}^{\lfloor d/2\rfloor}k_p
\delta_{\nu\rho_1\cdots\rho_{2p}}^{\mu\sigma_1\cdots\sigma_{2p}}
\prod_{r=1}^{p}
R_{\sigma_{2r-1}\sigma_{2r}}{}^{\rho_{2r-1}\rho_{2r}}=0,
\label{eq:lovelock-vacuum}
\end{equation}
where $\Lambda$ is the cosmological constant, $k_p$ are coupling
constants, and
$\delta^{\mu_1\cdots\mu_q}_{\nu_1\cdots\nu_q}
=q!\delta^{\mu_1}_{[\nu_1}\cdots\delta^{\mu_q}_{\nu_q]}$
is the generalised Kronecker delta. The antisymmetrisation makes the sum finite.
For EGB gravity we retain the quadratic-curvature correction, $k_2$, in the
Lovelock equations~\eqref{eq:lovelock-vacuum}. Comparing the convention of
Kov\'acs and Reall~\cite{Kovacs:2020ywu} with that of Arest\'e Sal\'o, Clough,
and Figueras~\cite{AresteSalo:2023mmd} gives $k_2=-\lambda^{\mathrm{GB}}/8$.
With $\Lambda=0$ and $k_{p>2}=0$, the field equations become
\begin{equation}
E^\mu{}_{\nu}
=G^\mu{}_{\nu}
-\frac{\lambda^{\mathrm{GB}}}{8}
\delta_{\nu\rho_1\rho_2\rho_3\rho_4}^{\mu\sigma_1\sigma_2\sigma_3\sigma_4}
R_{\sigma_1\sigma_2}{}^{\rho_1\rho_2}
R_{\sigma_3\sigma_4}{}^{\rho_3\rho_4}=0.
\label{eq:eomsEGB}
\end{equation}
The Gauss-Bonnet term contributes to the vacuum metric equations for
$d+1>4$.

We treat EGB gravity as a weakly coupled deformation of GR governed by the parameter $\lambda^{\rm GB}$. We impose the cone-separation condition~\ref{ass:mccz4-cone-separation} in the GR limit, following the motivation of the modified harmonic formulation~\cite{Kovacs:2020ywu}.

The principal symbol of the gauge-fixed EGB equations has the form~\eqref{eq:Charactmunu}. The contribution from the covariant equations~\eqref{eq:eomsEGB} is
\begin{equation}
    \mathcal{P}_*(\xi)^{\mu\nu\rho\sigma}=\mathcal{P}^{\rm GR}_*(\xi)^{\mu\nu\rho\sigma}+ \tfrac{\lambda^{\mathrm{GB}}}{2 }\bigl(\delta^{\mu\mu_1\mu_2\mu_3\mu_4}_{\nu_0\nu_1\nu_2\nu_3\nu_4} R^{\nu_3\nu_4}_{\mu_3\mu_4}\bigr)\xi^{\nu_1}\xi_{\mu_1}g^{\nu_2\rho}\delta_{\mu_2}^\sigma g^{\nu_0\nu}.
\end{equation}
Note that the EGB principal symbol satisfies the identities~\eqref{eq:diffeom}, while the unchanged gauge-fixing term preserves the identity~\eqref{eq:Factor}. These deformation-invariant properties allow us to construct the pure-gauge and gauge-condition-violating polarisations in the same way as in GR.

For the pure-gauge sector, the GR ansatz $t_{\mu\nu}=\xi_{(\mu}X_{\nu)}$ still satisfies $\mathcal P_*(\xi)t=0$. We therefore recover the same characteristic equation~\eqref{eq:vectorEq}, $Q(\xi)^{\mu\nu}X_\nu=0$, for $X_\mu$. We thus obtain $1,1,d-1$ independent polarisations for each branch in $\xi_0$ of $\mathcal G_A^{\mu\nu}\xi_\mu\xi_\nu=0$, with $A=\mathbf0,\mathbf L,\mathbf T$, respectively. Both the eigenvalues and the corresponding eigenvectors are unchanged under the Gauss-Bonnet deformation. Using the notations introduced in the GR proof in Sec.~\ref{sec:gr-characteristic-problem}, these eigenvectors span $\tilde V_{\mathbf0}^\pm$, $\tilde V_{\mathbf L}^\pm$, and $\tilde V_{\mathbf T}^\pm$, whose dimensions are unchanged from GR.

For the gauge-condition-violating sector, the defining conditions $\hat g^{\mu\nu}\hat\xi_\mu\hat\xi_\nu=0$ and $\tilde{\mathbb P}^{\nu\beta\rho\sigma}\hat\xi_\beta t_{\rho\sigma}\ne0$ remain unchanged under the Gauss-Bonnet deformation, since the identity~\eqref{eq:Factor} is preserved. The eigenvalues in this sector therefore remain the same as in GR. For each such characteristic covector, we use the ansatz~\eqref{eq:ansatzgcv}:
\[
t_{\mu\nu}(v)=t_{AB}(v)e^A_\mu e^B_\nu
+\hat\xi_{(\mu}Y_{\nu)}.
\]
As in the GR construction, the tensor equation reduces to its transverse projection~\eqref{eq:eqtransverse}, while the gauge equation gives the same solution~\eqref{eq:v2Y}. Under the cone-separation condition~\ref{ass:mccz4-cone-separation}, the transverse operator $\mathcal P_*^{ABCD}(\hat\xi)$ is invertible in GR and remains invertible for sufficiently weak coupling by continuity~\cite{Kovacs:2020ywu}. The components $t_{AB}(v)$ are therefore uniquely determined by equation~\eqref{eq:eqtransverse}. The $Q(\hat\xi)$ operator is still defined by~\eqref{eq:gauge-Q-hat} and remains invertible, so $Y_\mu(v)$ is given by the unique solution~\eqref{eq:v2Y}. The arbitrary vector $v^\mu$ supplies $d+1$ independent polarisations for each branch in $\hat\xi_0$ of the null cone equation. The corresponding eigenvectors span $\hat V^\pm$, whose dimensions are therefore unchanged from GR.

Under the cone-separation condition~\ref{ass:mccz4-cone-separation}, the GR physical eigenvalues $\xi_0^\pm$ of $M(\xi_i)$ are uniformly separated from the eigenvalues in the pure-gauge and gauge-condition-violating sectors. By continuity, this separation persists for sufficiently weak Gauss-Bonnet coupling. Following~\cite{Kovacs:2020ywu}, we call the eigenvalues that approach $\xi_0^\pm$ in the GR limit the $\xi_0^\pm$-group. The two physical groups remain separated from each other and from the unphysical eigenvalues, allowing each group to be isolated by a contour in the complex plane.

Following~\cite{Kovacs:2020ywu,kato1995perturbation}, we introduce the \textit{total generalised eigenspace} $V^\pm$ associated with each $\xi_0^\pm$-group using the projection operator defined as
\begin{equation}
\Pi(\xi_0^\pm)
=\frac{1}{2\pi i}\int_{\Gamma_{\xi_0^\pm}}
\frac{1}{z-M(\xi_i)}\,dz,
\qquad
V^\pm=\operatorname{im}\Pi(\xi_0^\pm).
\label{eq:egb-physical-projector}
\end{equation}
Here $\Gamma_{\xi_0^\pm}$ is a counterclockwise circle around the GR eigenvalue $\xi_0^\pm$, enclosing the corresponding $\xi_0^\pm$-group while all other eigenvalues of $M(\xi_i)$ lie outside. Equivalently,
\begin{equation}
\begin{gathered}
V^\pm
=\bigoplus_{\xi_0\in\,\xi_0^\pm\text{-group}}
V_{\xi_0},\\
V_{\xi_0}
=\operatorname{span}\{v\mid(\xi_0-M)^rv=0
\text{ for some }r\in\mathbb N\}.
\end{gathered}
\end{equation}

The dimensions of the physical spaces $V^\pm$ remain equal to their GR values by continuity of their projectors~\cite{Kovacs:2020ywu}. Since the dimensions of the physical and unphysical spaces sum to $\dim V$, we obtain
\begin{equation}
\begin{aligned}
V={}&\hat V^+\oplus\hat V^-
\oplus\tilde V_{\mathbf0}^+\oplus\tilde V_{\mathbf0}^-
\\&\oplus\tilde V_{\mathbf L}^+\oplus\tilde V_{\mathbf L}^-
\oplus\tilde V_{\mathbf T}^+\oplus\tilde V_{\mathbf T}^-
\oplus V^+\oplus V^-.
\end{aligned}
\label{eq:egb-spectral-decomposition}
\end{equation}

It remains to show that the restrictions of $M(\xi_i)$ to $V^\pm$ have real eigenvalues and are diagonalisable.

\subsubsection*{Diagonalisability within $V^\pm$}


In the spirit of \cite{Kovacs:2020ywu}, we introduce a hermitian matrix for each of the $V^{\pm}$, which takes the form
\begin{equation}
    H_*^{\pm}=\pm \begin{pmatrix}
        B_* & A_* \\ A_* &0
    \end{pmatrix},
\label{eq:physical-hermitian-matrix}
\end{equation}
where $A_*,B_*$ are matrix coefficients analogous to those defined in \eqref{eq:matrixcoeff}, which indeed can be read off from the $\xi_0$- expansion of the $\mathcal{P}_*(\xi)$-operator: 
\begin{equation}
    \mathcal{P}_*(\xi)=\xi_0^2 A_* +\xi_0 B_*+ C_*.
\end{equation}

The hermitian matrix $H_*^{\pm}$ thus defines an inner product $\braket{\cdot,\cdot}$ within $V^{\pm}$:
\begin{equation}
\braket{\mathcal{U}_1,\mathcal{U}_2}_\pm\equiv\mathcal{U}_1^\dagger H_*^\pm \mathcal{U}_2.
 \end{equation}

Taking the GR limit, we can write out the matrix coefficients directly:
\begin{equation}
\begin{aligned}
     A_*^{\mu\nu\rho\sigma}&= -\frac{1}{2}g^{00}P^{\mu\nu\rho\sigma}+P_\gamma {}^{0\mu\nu} P^{\gamma 0\rho\sigma}, \\
     B_*^{\mu\nu\rho\sigma}&= -(g^{0i}\xi_i)P^{\mu\nu\rho\sigma}+P_\gamma {}^{0\mu\nu} P^{\gamma i\rho\sigma}\xi_i+
     P_\gamma {}^{i\mu\nu} P^{\gamma 0\rho\sigma}\xi_i, \\
     C_*^{\mu\nu\rho\sigma}&= -\frac{1}{2}(g^{ij}\xi_i\xi_j)P^{\mu\nu\rho\sigma}+P_\gamma {}^{i\mu\nu} P^{\gamma j\rho\sigma}\xi_i \xi_j.
\end{aligned}
\end{equation}

Hence, for any two eigenvectors  of the form $\mathcal{U}_{1,2}=(t_{\mu\nu}^{1,2},\xi_0^{\pm} t_{\mu\nu}^{1,2})^T$ , with $t_{\mu\nu}^{1,2}$ being physical polarisations of the form \eqref{eq:eqTT}, we have 
\begin{equation}
\begin{aligned}
     \braket{\mathcal{U}_1,\mathcal{U}_2}_{\pm}=& \pm t^{1*}_{\mu\nu} (B_*+2\xi_0^{\pm}A_*)t^2_{\rho\sigma}\\
    =&\pm t^{1*}_{\mu\nu}(-(g^{0i}\xi_i)P^{\mu\nu\rho\sigma}+P_\gamma {}^{0\mu\nu} P^{\gamma i\rho\sigma}\xi_i+
     P_\gamma {}^{i\mu\nu} P^{\gamma 0\rho\sigma}\xi_i- \\ &\xi_0^{\pm}g^{00}P^{\mu\nu\rho\sigma}+2\xi_0 ^{\pm } P_\gamma {}^{0\mu\nu} P^{\gamma 0\rho\sigma})t^2_{\rho\sigma}\\
     =&\pm t^{1*}_{\mu\nu}(-\xi^{\pm 0}P^{\mu\nu\rho\sigma}+P_\gamma {}^{0\mu\nu}P^{\gamma \alpha \rho\sigma}\xi_\alpha^{\pm}+P_\gamma {}^{\alpha\mu\nu}P^{\gamma0 \rho\sigma}\xi_\alpha^{\pm})t_{\rho\sigma}^2\\
     =& \mp \xi^{\pm 0} t^{1*}_{\mu\nu} P^{\mu\nu\rho\sigma}t^2_{\rho\sigma},
\end{aligned}
\end{equation}
where in the last equality we have used the transversality \eqref{eq:transverse} of the physical polarisation, with $\xi^{\pm 0}\equiv g^{0\mu}\xi^{\pm}_\mu$.
Inserting the ansatz \eqref{eq:eqTT} into the expression, we get 
\begin{equation}
        \braket{\mathcal{U}_1,\mathcal{U}_2}_{\pm}=\mp \xi^{\pm 0} t_{\mu\nu}^{1*\rm TT} P^{\mu\nu\rho\sigma}t^{2\,\rm TT}_{\rho\sigma}
        = \mp \xi^{\pm 0} t_{\langle\hat{i}\hat{j}\rangle}^{1*} t_{\langle\hat{i}\hat{j}\rangle}^2,
\end{equation}
where we have used the traceless property of $t^{\rm TT}_{\mu\nu}$. Taking $\mathcal{U}_1=\mathcal{U}_2$ yields equation (54) of \cite{Kovacs:2020ywu}, and in the GR limit one readily sees that the Hermitian form induced by $H_*^{\pm}$ is positive definite on $V^{\pm}$. In the weakly coupled regime of the EGB theory, by continuity, the inner product $\braket{\cdot,\cdot}_\pm$ remains positive definite. 
Next, we prove that every physical eigenvalue $\xi_0$ of $V^+$ in EGB is real. This can be seen by inserting two eigenvectors in the $V^{+}$ space associated  to two eigenvalues $\xi_0^{(1,2)}$, and we have 
\begin{equation}
\begin{aligned}
   (\xi_0^{(1)}-\xi_0^{(2)}) \mathcal{U}_1^T H_*^{+} \mathcal{U}_2=&  (\xi_0^{(1)}-\xi_0^{(2)}) t^1_{\mu\nu} (B_*+\xi_0^{(1)}A_*+\xi_0^{(2)}A_*)^{\mu\nu\rho\sigma}t^2_{\rho\sigma}\\
   &= t^1_{\mu\nu}((\xi_0^{(1)}-\xi_0^{(2)}) B_*+(\xi_0^{(1)2}-\xi_0^{(2)2}) A_* ) ^{\mu\nu\rho\sigma} t^2_{\rho\sigma}\\
   &=t^1_{\mu\nu}(\mathcal{P}_*(\xi^{(1)})^{\mu\nu\rho\sigma}-\mathcal{P}_*(\xi^{(2)})^{\mu\nu\rho\sigma} )t_{\rho\sigma}^{2}=0, 
\end{aligned} \label{eq:diffeigen}
\end{equation}
where we have used the fact that $C_*$ is common for all  physical characteristics and that $\mathcal{P}_*(\xi)\cdot t=0$  for all physical polarisations $t$. Note that $M(\xi_i)$ is a real matrix and hence if $(\xi_0,\mathcal{U})$ is an eigenpair, so is $(\xi_0^*,\mathcal{U}^*)$. However, it follows from the identity~\eqref{eq:diffeigen} that 
\begin{equation}
    2\rm{Im}(\xi_0)\braket{\mathcal{U},\mathcal{U}}_{+}=0.
\end{equation}
The positivity of the hermitian form yields $\rm{Im} (\xi_0)=0$ for all eigenvalues in $V^+$.  The same argument also applies to the $V^-$ space. 

Before proceeding, it is useful to show the following property for the vectors in $V^{\pm}$.
\begin{lemma}
    Every generalised eigenvector $\mathcal{V}=(t_{\rho\sigma},t'_{\rho\sigma})\in V^{\pm}$ must satisfy the following constraint:
    \begin{equation}
       \tilde{\mathbb{P}}^{\mu i \rho\sigma}\xi_it_{\rho\sigma}+\tilde{\mathbb{P}}^{\mu0\rho\sigma}t'_{\rho\sigma}=0. \label{eq:generaldiv}
    \end{equation}
    In particular, for a genuine eigenvector in $V^{\pm}$ of the form $(t,\xi_0 t)^T$, the constraint \eqref{eq:generaldiv} reduces to the gauge condition \eqref{eq:gaugecond} for the physical polarisation.
    \label{th:lem2}
\end{lemma}
\begin{proof}
    Like in \cite{Kovacs:2020ywu}, we start from the orthogonality of 
    $\mathcal{V}$ and any \textit{left} eigenvector $\mathcal{W}$ associated with the gauge-condition-violating eigenvalue $\hat{\xi}_0^{\pm}$. Note that, by definition, any generalised eigenvector vector in the total generalised eigenspace $V^{\pm}$ must be annihilated by $(\xi_0-M(\xi_i))^r$ for some $r$. It then follows that 
    \begin{equation}
        0=\mathcal{W}(\xi_0-M(\xi_i))^r\mathcal{V}=(\xi_0-\hat{\xi}_0^{\pm})^r (\mathcal{W}\cdot \mathcal{V}).
    \end{equation}
     The gap between the physical and unphysical eigenvalues implies immediately that $\mathcal{W}\cdot \mathcal{V}=0$.

    It is straightforward to check that $\mathcal{W}$ can be  explicitly written out in terms of the left polarisation $s_{\mu\nu}$ following the prescription in \cite{Kovacs:2020ywu}:
\begin{equation}
    \mathcal{W}= (s_I,\hat{\xi}_0^{\pm} s_I)\begin{pmatrix}
        B & A\\
        A &0
    \end{pmatrix},
\end{equation}
where the left polarisation $s$ takes the form 
\begin{equation}
    s_{\mu\nu}=\hat{\xi}^{\pm}_{(\mu}X_{\nu)},
\end{equation}
for \textbf{any} $X$,
which automatically solves the left polarisation equation 
\begin{equation}
    s_{\mu\nu}\mathcal{P}(\hat{\xi}^{\pm})^{\mu\nu\rho\sigma}=0.
\end{equation}
Recall that the $A,B,C$ are again the matrix coefficients read off from the principal symbol:
\begin{equation}
    \mathcal{P}(\xi)=\xi_0^2 A+\xi_0 B+C.
\end{equation}
One could also define the associated coefficients for the gauge-fixing part of the principal symbol:
\begin{equation}
     \mathcal{P}_{\rm GF}(\xi)=\xi_0^2 A_{\rm GF}+\xi_0 B_{\rm GF}+C_{\rm GF}.
\end{equation}
Then the orthogonality condition $\mathcal{W}\cdot \mathcal{V}$ implies 
\begin{equation}
    0=\mathcal{W}\cdot \mathcal{V}=\begin{pmatrix}
        s_I &\hat{\xi}_0^{\pm} s_I
    \end{pmatrix}\begin{pmatrix}
        B^{IJ} & A^{IJ}\\
        A^{IJ} &0
    \end{pmatrix} \begin{pmatrix}
        t_J\\t'_J
    \end{pmatrix}=s_{\mu\nu}(B+\hat{\xi}_0^{\pm}A)^{\mu\nu\rho\sigma}t_{\rho\sigma}+s_{\mu\nu} A^{\mu\nu\rho\sigma}t'_{\rho\sigma},
\end{equation}
and the arbitrariness of $X$ yields 
\begin{equation}
\hat{\xi}^{\pm}_\nu(B+\hat{\xi}_0^{\pm}A)^{\mu\nu\rho\sigma}t_{\rho\sigma}+\hat{\xi}^{\pm}_\nu A^{\mu\nu\rho\sigma}t'_{\rho\sigma}=0. \label{eq:hatxieq}
\end{equation}
Expanding \eqref{eq:hatxieq} in powers of $\hat{\xi}_0^{\pm}$ while trading $\hat{\xi}_0^{\pm2}$ for lower powers of  $\hat{\xi}_0^{\pm}$  using the defining equation $\hat{g}^{\mu\nu}\hat{\xi}^{\pm}_{\mu}\hat{\xi}^{\pm}_{\nu}=0$, we have 
\begin{equation}
\begin{gathered}
    0=\hat{\xi}_0^{\pm}(-2 (\hat{g}^{00})^{-1}\hat{g}^{0i}\xi_i  A^{\mu 0\rho\sigma}t_{\rho\sigma}+A^{\mu i\rho\sigma}\xi_i t_{\rho\sigma}+B^{\mu0\rho\sigma}t_{\rho\sigma}+A^{\mu0\rho\sigma}t'_{\rho\sigma})+ \\ B^{\mu i\rho\sigma}\xi_i t_{\rho\sigma} -(\hat{g}^{00})^{-1}\hat{g}^{ij}\xi_i\xi_j A^{\mu 0\rho\sigma}t_{\rho\sigma}+A^{\mu i\rho\sigma}\xi_i t'_{\rho\sigma}.
\end{gathered}
\end{equation}
Note that $\hat{\xi}_0^{+}\neq \hat{\xi}_0^{-}$, which implies that, when viewed as a linear polynomial in $\hat{\xi}_0^{\pm}$, the expression on the r.h.s. above must vanish identically. Reading off the coefficient of $\xi_0^{\pm}$ thus gives 
\begin{equation}
    R^{\mu}\equiv-2 (\hat{g}^{00})^{-1}\hat{g}^{0i}\xi_i  A^{\mu 0\rho\sigma}t_{\rho\sigma}+A^{\mu i\rho\sigma}\xi_i t_{\rho\sigma}+B^{\mu0\rho\sigma}t_{\rho\sigma}+A^{\mu0\rho\sigma}t'_{\rho\sigma}=0.
\end{equation}
By looking at the $\xi_0$-expansion of the identity $\xi_\mu\mathcal{P}_*(\xi)^{\mu\nu\rho\sigma}=0$, one also gets
\begin{equation}
    A_*^{0\nu\rho\sigma}=0,\quad  \xi_iA_*^{i\nu\rho\sigma}+ B_*^{0\nu \rho\sigma}=0.
\end{equation}
Thus only the gauge-fixing part contributes to the expression for $R^\mu$, namely
\begin{equation}
    0=R^{\mu}=-2 (\hat{g}^{00})^{-1}\hat{g}^{0i}\xi_i  A_{\rm GF}^{\mu 0\rho\sigma}t_{\rho\sigma}+(A_{\rm GF}^{\mu i\rho\sigma}\xi_i + B_{\rm GF}^{\mu0\rho\sigma})t_{\rho\sigma}+A_{\rm GF}^{\mu0\rho\sigma}t'_{\rho\sigma}. \label{eq:R-equation1}
\end{equation}
Writing out the explicit expressions for the $A,B$ coefficients by reading off the $\xi_0$ coefficients in \begin{equation}
    \mathcal{P}_{\mathrm{GF}}(\xi)^{\mu\nu\rho\sigma}\equiv
     - \hat{P}_\gamma {}^{\alpha\mu\nu}\xi_{\alpha} \tilde{\mathbb{P}}^{\gamma\beta\rho\sigma}\xi_{\beta},
\end{equation} we have
\begin{subequations}
    \begin{align}
     A^{\mu 0\rho\sigma}_{\rm GF}=&  - \hat{P}_\gamma {}^{0\mu0} \tilde{\mathbb{P}}^{\gamma0\rho\sigma}=-\frac{1}{2}\hat{g}^{00}\tilde{\mathbb{P}}^{\mu0\rho\sigma},\\
     A_{\rm GF}^{\mu i\rho\sigma}\xi_i + B_{\rm GF}^{\mu0\rho\sigma}=&  - \xi_i\hat{P}_\gamma {}^{0\mu i} \tilde{\mathbb{P}}^{\gamma0\rho\sigma}  - \hat{P}_\gamma {}^{0\mu 0}\tilde{\mathbb{P}}^{\gamma i\rho\sigma}\xi_{i}-\hat{P}_\gamma {}^{i\mu 0}\xi_{i} \tilde{\mathbb{P}}^{\gamma 0\rho\sigma}\nonumber
     \\=&- \xi_i(\hat{P}_\gamma {}^{0 i\mu} +\hat{P}_\gamma {}^{i0\mu})\tilde{\mathbb{P}}^{\gamma0\rho\sigma} -\frac{1}{2}\hat{g}^{00}\tilde{\mathbb{P}}^{\mu i\rho\sigma} \xi_i\nonumber
     \\=& -\hat{g}^{0i}\xi_i \tilde{\mathbb{P}}^{\mu 0 \rho\sigma}-\frac{1}{2}\hat{g}^{00}\tilde{\mathbb{P}}^{\mu i \rho\sigma}\xi_i,
\end{align}
\end{subequations}
where we have repeatedly used the following property of the $\hat{P}$-operator:
\begin{equation}
    2\hat{P}_{\alpha}^{~(\beta\mu)\nu}
    =\frac{1}{2}\delta_{\alpha}^{\mu}\,\hat{g}^{\nu\beta}+\frac{1}{2}\delta_{\alpha}^{\nu}\,\hat{g}^{\mu\beta}-\tfrac{1}{2}\,\delta_{\alpha}^{\beta}\,\hat{g}^{\mu\nu}+(\mu\leftrightarrow\beta)=\delta_{\alpha}^{\nu}\,\hat{g}^{\mu\beta}.
\end{equation}
Inserting the expression for the $A_{\rm GF},B_{\rm GF}$ back to the $R^\mu$-equation gives 
\begin{equation}
    0=-\frac{1}{2}\hat{g}^{00}(\tilde{\mathbb{P}}^{\mu i \rho\sigma}\xi_it_{\rho\sigma}+\tilde{\mathbb{P}}^{\mu0\rho\sigma}t'_{\rho\sigma}).
\end{equation}
Recall that, by construction, we have $\hat{g}^{00}\neq 0$, which concludes the proof.
\end{proof}

Finally, we show by contradiction that $V^{\pm}$ is indeed a linear span of eigenvectors.
Assume that  $M(\xi_i)$ has a Jordan block within  $V^{\pm}$.Then there must exist a non-zero generalised eigenvector $\mathcal{V}$ such that
\begin{equation}
    (\xi_0 - M(\xi_i))\mathcal{V} = \mathcal{U} \neq 0 \,, \quad \textbf{while} \quad (\xi_0 - M(\xi_i))\mathcal{U} = 0. \label{eq:generalEigen}
\end{equation}
Writing explicitly $\mathcal{V}=(v_I,v'_I)^T$ and $\mathcal{U}=(u_I,\xi_0 u_I)^T$, and expressing $v_I'\equiv \xi_0 v_I+w_I$, we have  
\begin{equation}
   \begin{pmatrix}
       \xi_0 I & -I \\
       A^{-1}C & \xi_0 I +A^{-1} B 
   \end{pmatrix}
   \begin{pmatrix}
       v \\ \xi_0 v+w
   \end{pmatrix} =\begin{pmatrix}
       u \\ \xi_0 u
   \end{pmatrix}.
\end{equation}
It follows that  $w=-u$, and that
\begin{equation}
    A^{-1}C  v+  \xi_0^2 v -\xi_0 u +A^{-1} B(\xi_0 u-v)=\xi_0 u.
\end{equation}
A direct manipulation yields 
\begin{equation}
    \mathcal{P}(\xi)v=(2\xi_0A+B)u. 
\end{equation}
Splitting this into the gauge-invariant and gauge fixing part gives 
\begin{equation}
    \mathcal{P}_*(\xi) v =
    (2\xi_0A_*+B_*)u +   (2\xi_0A_{\rm GF}+B_{\rm GF})u-\mathcal{P}_{\rm GF}(\xi)v. \label{eq:P*v}
\end{equation}
Recall that $\mathcal{V}=(v,\xi_0u-v)^T$ must satisfy the constraint~\eqref{eq:generaldiv}. This gives  
\begin{equation}
    0=\tilde{\mathbb{P}}^{\mu i \rho\sigma}\xi_i v_{\rho\sigma}+\tilde{\mathbb{P}}^{\mu0\rho\sigma}(\xi_0 v_{\rho\sigma}-u_{\rho\sigma})= \tilde{\mathbb{P}}^{\mu \nu\rho\sigma}\xi_\nu v_{\rho\sigma}-\tilde{\mathbb{P}}^{\mu0\rho\sigma}u_{\rho\sigma} .\label{eq:u-vconstr}
\end{equation}
Then it can be shown that the gauge-fixing part of \eqref{eq:P*v} would eventually  cancel:
\begin{equation}
\begin{aligned}
    & ~ ~ ~((2\xi_0A_{\rm GF}+B_{\rm GF})u)^{\mu\nu}-(\mathcal{P}_{\rm GF}(\xi)v)^{\mu\nu} \\ 
     &=- 2\xi_0\hat{P}_\gamma {}^{0\mu\nu} \tilde{\mathbb{P}}^{\gamma0\rho\sigma}u_{\rho\sigma} -
     \hat{P}_\gamma {}^{i\mu\nu} \xi_i\tilde{\mathbb{P}}^{\gamma0\rho\sigma}u_{\rho\sigma}-
     \hat{P}_\gamma {}^{0\mu\nu} \xi_i\tilde{\mathbb{P}}^{\gamma i\rho\sigma}u_{\rho\sigma}+\hat{P}_\gamma {}^{\alpha\mu\nu}\xi_{\alpha} \tilde{\mathbb{P}}^{\gamma\beta\rho\sigma}\xi_{\beta} v_{\rho\sigma}\\
     &=-\xi_\alpha\hat{P}_\gamma {}^{\alpha\mu\nu} \tilde{\mathbb{P}}^{\gamma0\rho\sigma}u_{\rho\sigma}-
     \xi_\alpha\hat{P}_\gamma {}^{0\mu\nu} \tilde{\mathbb{P}}^{\gamma\alpha\rho\sigma}u_{\rho\sigma}+\hat{P}_\gamma {}^{\alpha\mu\nu}\xi_{\alpha} \tilde{\mathbb{P}}^{\gamma\beta\rho\sigma}\xi_{\beta}v_{\rho\sigma} \\
     &=-
     \hat{P}_\gamma {}^{0\mu\nu} \tilde{\mathbb{P}}^{\gamma\alpha\rho\sigma}\xi_\alpha u_{\rho\sigma}=0,
\end{aligned}
\end{equation}
where we have used the relation \eqref{eq:u-vconstr} and the gauge condition \eqref{eq:gaugecond} for the physical polarisation $u$.
The remaining part of the equation \eqref{eq:P*v} becomes 
\begin{equation}
    (\mathcal{P}_*(\xi) v)^{\mu\nu} =
    ((2\xi_0A_*+B_*)u)^{\mu\nu}.
\end{equation}
Contracting both sides with $u_{\mu\nu}^*$ and considering the symmetry of the $\mathcal{P}_*$-operator, we have 
\begin{equation}
    0=u^*_{\mu\nu}(\mathcal{P}_*(\xi) v)^{\mu\nu}= u^*_{\mu\nu}(2\xi_0A_*+B_*u)^{\mu\nu\rho\sigma}u_{\rho\sigma}= \pm\mathcal{U}^\dagger H_*^{\pm}\mathcal{U}=\pm\braket{\mathcal{U},\mathcal{U}}_{\pm},
\end{equation}
where the first equality comes from the reality and symmetry of $\mathcal{P}_*(\xi)$ and the fact that $\mathcal{P}_*(\xi)\cdot u=0$.
The positive definiteness of the Hermitian form $\braket{\cdot,\cdot}_{\pm}$ therefore forces $\mathcal{U}=0$, which contradicts the assumption \eqref{eq:generalEigen}. Hence, $V^{\pm}$ is indeed spanned by eigenvectors of $M(\xi_i)$. 
\subsubsection*{Construction of the symmetrizer}
In contrast to GR, even though one can prove that $M(\xi_i)$ is diagonalisable  for every $\xi_i$, the physical eigenvectors in $V^{\pm}$ need not vary smoothly with $\xi_i$. This lack of smoothness is due to the eigenvalue crossing, as already acknowledged in \cite{Reall:2021voz,Kovacs:2020ywu,Davies:2021frz}. Therefore, in general, the candidate symmetrizer constructed from the eigenvectors may fail to be valid. Nevertheless, one can bypass this difficulty by constructing an alternative form of symmetrizer as proposed in \cite{Kovacs:2020ywu}.

We begin by showing that $H_*^{\pm}$ is the symmetrizer within $V^{\pm}$.
The space $V^\pm$ is now spanned by eigenvectors. For any two
eigenvectors $\mathcal U_1,\mathcal U_2\in V^\pm$ with eigenvalues
$\xi_0^{(1)},\xi_0^{(2)}$, respectively, we have
\begin{equation}
   \mathcal{U}_1^T \left(M^\dagger(\xi_i) H_*^{\pm}-H_*^{\pm}M(\xi_i)\right) \mathcal{U}_2
   =(\xi_0^{(1)}-\xi_0^{(2)}) \mathcal{U}_1^T H_*^{\pm} \mathcal{U}_2=0,
\end{equation}
which follows directly from the identity~\eqref{eq:diffeigen}. 
The completeness of the eigenvector basis thus implies that $H^{\pm}_*(\xi_i)$ serves as a symmetrizer on the physical eigenspace and depends smoothly on $\xi_i$.

Recall that we already have $2(d+1)$ pure gauge eigenvectors $\{\tilde{v}_1,\tilde{v}_2,\ldots,\tilde{v}_{(2d+1)}\}$,and $2(d+1)$  gauge-condition-violating eigenvectors $\{\hat{v}_1,\hat{v}_2,\ldots,\hat{v}_{(2d+1)}\}$. For each of $V^{\pm}$, we can choose a collection of $\frac{(d+1)(d-2)}{2}$ smooth basis vectors, which we denote by $\{v_1^\pm, v_2^\pm, \ldots, v_{(d+1)(d-2)/2}^\pm\}$. Let $S$ be the matrix whose columns are the complete set of vectors obtained above. The smooth symmetrizer defined on the total space $V$ is then give by 
\begin{equation}
    \mathcal{K}(\xi_i)=(S^{-1})^\dagger \begin{pmatrix}
        I_{2(d+1)\times 2(d+1)} & 0 & 0 & 0 \\
        0 & I_{2(d+1)\times 2(d+1)} & 0 & 0 \\
        0 & 0 & \mathcal{H}_{*}^+ & 0 \\
        0 & 0 & 0 & \mathcal{H}_{*}^- \\   
    \end{pmatrix} S^{-1}, \label{eq:Kmatrix}
\end{equation}
where $\mathcal{H}_{*}^\pm$ are the $\frac{(d+1)(d-2)}{2}\times\frac{(d+1)(d-2)}{2}$ matrices with matrix elements of the form 
\begin{equation}
(\mathcal{H}_{*}^\pm){}_{AB}=\braket{v_A^\pm,v_B^\pm}_{\pm}\equiv (v_A^\pm)^T H_*^{\pm} v_B^\pm.
\end{equation}
It's clear that the matrix $\mathcal{K}(\xi_i)$ fulfils condition \eqref{eq:fully-second-order-SH} for a symmetrizer in Definition \ref{def:fully-second-order-SH}. This concludes the proof of strong hyperbolicity.
\section{Generalisation to Four-derivative Scalar-Tensor theory}
\label{sec: Einstein-scalar-Gauss-Bonnet theory}
We now extend the weak-coupling argument of Section~\ref{sec:egb-extension}
to four-derivative scalar-tensor theory and, in particular, to EsGB gravity.

\subsection{Principal symbol and effective metric }
\label{sec:4dst-equations}
After field redefinitions, the effective action of four-derivative scalar-tensor
theory through four derivatives can be written as~\cite{Kovacs:2020ywu}
\begin{equation}
S=\frac{1}{16\pi G}\int d^4x\sqrt{-g}\,
\left[R+X-V(\phi)+\alpha(\phi)X^2
+\beta(\phi)\mathcal L_{\mathrm{GB}}\right],
\label{eq:four-derivative-st-action}
\end{equation}
where
\begin{equation}
X=-\frac12\nabla_\mu\phi\nabla^\mu\phi,
\qquad
\mathcal L_{\mathrm{GB}}
=\frac14\delta_{\rho_1\rho_2\rho_3\rho_4}^{\sigma_1\sigma_2\sigma_3\sigma_4}
R_{\sigma_1\sigma_2}{}^{\rho_1\rho_2}
R_{\sigma_3\sigma_4}{}^{\rho_3\rho_4}.
\end{equation}
The functions $V(\phi)$, $\alpha(\phi)$, and $\beta(\phi)$ are arbitrary
in this action. We take $\alpha(\phi)=g_2$ and
$\beta(\phi)=\lambda^{\mathrm{GB}}f(\phi)$. The couplings $g_2$ and
$\lambda^{\mathrm{GB}}$ are small constants of order $\ell^2$, where $\ell$ is
the UV length scale. The choice $g_2=0$ gives
EsGB theory, while $g_2\neq0$ includes
the quartic kinetic interaction.
We set $V(\phi)=0$ here because the potential contributes only
lower-order terms to the field equations and hence does not affect
their principal symbol.

Varying the resulting action with respect to $g_{\mu\nu}$ and $\phi$
gives the equations
$E^{\mu\nu}=0$ and $E_\phi=0$, where
\begin{equation}
    E^{\mu\nu}=\frac{-16\pi G}{\sqrt{-g}}\frac{\delta S}{\delta g_{\mu\nu}}, \quad E_\phi=\frac{-16\pi G}{\sqrt{-g}}\frac{\delta S}{\delta \phi}.
\end{equation}
Their full expressions are given by 
\begin{align}
&\begin{aligned}
E^\mu{ }_\nu \equiv & G^\mu{ }_\nu-\frac{1}{2}\left(X+g_2 X^2\right) \delta_\nu^\mu-\frac{1}{2}\left(1+2 g_2 X\right) \nabla^\mu \phi \nabla_\nu \phi \\
& +\lambda^{\rm GB}\left(f^{\prime \prime}(\phi) \nabla_{\mu_1} \phi \nabla^{\nu_1} \phi+ f^{\prime}(\phi) \nabla_{\mu_1} \nabla^{\nu_1} \phi\right) \delta_{\nu \nu_1 \nu_2 \nu_3}^{\mu \mu_1 \mu_2 \mu_3} R_{\mu_2 \mu_3}^{\nu_2 \nu_3}=0,
\end{aligned}\\[1.5em]
&\begin{aligned}
E_\phi \equiv & -\left(1+6 g_2 X\right) \Box \phi-2 g_2 \delta_{\nu_1 \nu_2}^{\mu_1 \mu_2} \nabla_{\mu_1} \phi \nabla^{\nu_1} \phi \nabla_{\mu_2} \nabla^{\nu_2} \phi \\
& -\tfrac{1}{4} \lambda^{\rm GB} f^{\prime}(\phi) \delta_{\nu_1 \nu_2 \nu_3 \nu_4}^{\mu_1 \mu_2 \mu_3 \mu_4} R_{\mu_1 \mu_2}^{\nu_1 \nu_2} R_{\mu_3 \mu_4}^{\nu_3 \nu_4}=0.
\end{aligned}
\end{align}



As with any covariant metric theory based on the variational principle, diffeomorphism invariance results in the standard Noether-Bianchi identity:
\begin{equation}
    \nabla_\mu E^{\mu\nu} + E_\phi \nabla^\nu \phi = 0.
\end{equation}
Following the prescription for Einstein-Gauss-Bonnet (EGB) theory, we fix the tensor part of the equations with a $Z4$ term. By imposing the gauge conditions in~\eqref{eq:puncture-gauge}, we close the tensor evolution system. As in the EGB case,  the $Z4$ can be eliminated using \eqref{eq:eqZofchris},  which yields a fully second order tensor equation of motion of the same form of equation \eqref{eq:mCCZ4_2nd}. Combining both the tensor equation and the scalar equation, we express the principal symbol $\mathcal{P}(\xi)^{IJ}:=\mathcal{P}_*(\xi)^{IJ}+\mathcal{P}_{\rm GF}(\xi)^{IJ}$ , acting on the \textit{extended polarisation} vector $T_I=(t_{\mu\nu},\psi)$ , in  block form 
\begin{equation}
\begin{aligned}
    \mathcal{P}(\xi)^{IJ}T_J&:=\mathcal{P}_*(\xi)^{IJ}T_J+\mathcal{P}_{\rm GF}(\xi)^{IJ}T_J\\ &\equiv  \begin{pmatrix}
        \mathcal{P}_{gg*}(\xi)^{\mu\nu\rho\sigma} & \mathcal{P}_{g\phi*}(\xi)^{\mu\nu} \\
        \mathcal{P}_{\phi g*}(\xi)^{\rho\sigma}  & \mathcal{P}_{\phi \phi*}(\xi)
    \end{pmatrix} \begin{pmatrix}
        t_{\rho\sigma} \\ \psi
    \end{pmatrix}+ \begin{pmatrix}
         \mathcal{P}_{\mathrm{GF}}(\xi)^{\mu\nu\rho\sigma} t_{\rho\sigma}\\ 0
    \end{pmatrix},
\end{aligned} \label{eq:CharaEsGB}
\end{equation}
where $\mathcal{P}_{\mathrm{GF}}(\xi)^{\mu\nu\rho\sigma}$ is exactly the GR gauge-fixing term as defined in \eqref{eq:PGauge}, and the other terms in the symbol are given by 
\begin{equation}
\begin{aligned}
      &\mathcal{P}_{gg*}(\xi)^{\mu\nu\rho\sigma}:=\mathcal{P}_{gg*}^{\mu\nu\alpha\beta\rho\sigma}\xi_\alpha\xi_\beta\equiv \frac{\partial E^{\mu\nu}}{\partial(\partial_\alpha\partial_\beta g_{\rho\sigma})}\xi_\alpha\xi_\beta, \\
      &\mathcal{P}_{g\phi*}(\xi)^{\mu\nu}:=\mathcal{P}_{g\phi*}^{\mu\nu\alpha\beta}\xi_\alpha\xi_\beta\equiv 
      \frac{\partial E^{\mu\nu}}{\partial(\partial_\alpha\partial_\beta\phi)} \xi_\alpha\xi_\beta,  \\
      &\mathcal{P}_{\phi g*}(\xi)^{\rho\sigma}:=\mathcal{P}_{\phi g*}^{\alpha\beta\rho\sigma}\xi_\alpha\xi_\beta\equiv 
      \frac{\partial E_\phi}{\partial(\partial_\alpha\partial_\beta g_{\rho\sigma})} \xi_\alpha\xi_\beta , \\
      &\mathcal{P}_{\phi\phi*}(\xi):=\mathcal{P}_{\phi\phi*}^{\alpha\beta}\xi_\alpha\xi_\beta\equiv 
      \frac{\partial E_\phi}{\partial(\partial_\alpha\partial_\beta\phi)} \xi_\alpha\xi_\beta.
\end{aligned}
\end{equation}
It is easy to check that the matrix $\mathcal{P}_*(\xi)^{IJ}$ is symmetric, and in particular that
\begin{equation}
    \mathcal{P}_{g\phi*}(\xi)^{\mu\nu}= \mathcal{P}_{\phi g*}(\xi)^{\mu\nu}\,,
\end{equation}
which is indeed a consequence of variational principle \cite[see][]{Papallo:2017qvl,Reall:2021voz}.
It therefore follows that $\mathcal{P}_*^{IJ}$ is symmetric with respect to indices $I$ and $J$.

As is discussed in \cite{Reall:2021voz}, by exploiting the symmetry of the system in $d=4$ spacetime, one could write out the gauge invariant parts of the principal symbols in  compact form:
\begin{equation}
    \begin{aligned}
        &(\mathcal{P}_{gg*}(\xi)\cdot t)^\mu_\nu=
         \tfrac{1}{2} \delta_{\nu \nu_1 \nu_2 \nu_3}^{\mu \mu_1 \mu_2 \mu_3} C_{\mu_1}^{\nu_1}\xi^{\nu_2}\xi_{\mu_2}t^{\nu_3}_{\mu_3}, \\[0.5em]
        &(\mathcal{P}_{g\phi*}(\xi)\cdot \psi)^{\mu}_{\nu}=\lambda^{\rm GB} f'(\phi)\delta_{\nu \nu_1 \nu_2 \nu_3}^{\mu \mu_1 \mu_2 \mu_3} R_{\mu_2 \mu_3}^{\nu_2 \nu_3} \xi^{\mu_1} \xi_{\nu_1} \psi ,\\[0.5em]
        & (\mathcal{P}_{\phi g*}(\xi)\cdot t)= \lambda^{\rm GB} f^{\prime}(\phi) \delta_{\nu_1 \nu_2 \nu_3 \nu_4}^{\mu_1 \mu_2 \mu_3 \mu_4} R_{\mu_1 \mu_2}^{\nu_1 \nu_2} \xi^{\nu_3}\xi_{\mu_3}t^{\nu_4}_{\mu_4} ,\\[0.5em]
        &  (\mathcal{P}_{\phi \phi*}(\xi)\cdot \psi)=-\left((1+2g_2X)g^{\mu\nu} -2g_2\nabla^\mu\phi\nabla^\nu\phi\right)\xi_\mu\xi_\nu\psi,
    \end{aligned} 
\end{equation}
where
\begin{equation}
    C_{\mu\nu}:=g_{\mu\nu}-4\lambda^{\rm GB}\left(f^{\prime \prime}(\phi) \nabla_{\mu} \phi \nabla_{\nu} \phi+ f^{\prime}(\phi) \nabla_{\mu} \nabla_{\nu} \phi\right)
\end{equation}
is known as the \textit{effective metric}.

As pointed out in \cite{Reall:2021voz}, the diagonal tensor-sector block, $\mathcal{P}_{gg*}(\xi)^{\mu\nu\rho\sigma}$, can be obtained from the principal symbol of Einstein gravity \eqref{eq:Pstar} by performing the background-metric substitution $g_{\mu\nu}\to C_{\mu\nu},\, g^{\mu\nu}\to C^{-1}{}^{\mu\nu}$. This can be demonstrated as follows. We first expand the Kronecker delta into product of Levi-Civita tensors:
\begin{equation}
    \begin{aligned}
        (\mathcal{P}_{gg*}(\xi)\cdot t)^{\mu\nu}=&
         \tfrac{1}{2} g^{\alpha\nu}\delta_{\alpha \nu_1 \nu_2 \nu_3}^{\mu \mu_1 \mu_2 \mu_3} C_{\mu_1}^{\nu_1}\xi^{\nu_2}\xi_{\mu_2}t^{\nu_3}_{\mu_3} =
         -\tfrac{1}{2}\epsilon^{\mu\mu_1\mu_2\mu_3}\epsilon^{\nu\nu_1\nu_2\nu_3}
         C_{\mu_1\nu_1}\xi_{\mu_2}\xi_{\nu_2}t_{\mu_3\nu_3}
         \\
         =& -\frac{C}{2g} \epsilon_{\rm C}^{\mu\mu_1\mu_2\mu_3}\epsilon_{\rm C}^{\nu\nu_1\nu_2\nu_3}
         C_{\mu_1\nu_1}\xi_{\mu_2}\xi_{\nu_2}t_{\mu_3\nu_3},
    \end{aligned} \label{eq:effectsymbol}
\end{equation}
where $\epsilon^{\mu\nu\rho\sigma}:=\frac{-1}{\sqrt{-g}}\hat{\epsilon}^{\mu\nu\rho\sigma}$ is defined in terms of the physical metric $g$ and the standard Levi-Civita symbol $\hat{\epsilon}$, whilst we've also introduced the $C$-Levi-Civita tensor with respect to the effective metric:
\begin{equation}
\begin{aligned}
    \epsilon_{\rm C}^{\mu\nu\rho\sigma}\equiv &\frac{-1}{\sqrt{-C}}\hat{\epsilon}^{\mu\nu\rho\sigma}= C^{-1}{}^{\mu\mu_1}C^{-1}{}^{\nu\nu_1}C^{-1}{}^{\rho\rho_1}C^{-1}{}^{\sigma\sigma_1}\sqrt{-C}\hat{\epsilon}_{\mu_1\nu_1\rho_1\sigma_1} \\
\equiv & C^{-1}{}^{\mu\mu_1}C^{-1}{}^{\nu\nu_1}C^{-1}{}^{\rho\rho_1}C^{-1}{}^{\sigma\sigma_1}\epsilon^{\rm C}_{\mu_1\nu_1\rho_1\sigma_1}.
\end{aligned}
\end{equation}
Hence, in the following calculation, all the tensor indices shall be lowered and raised by the effective metric $C_{\mu\nu},C^{-1}{}^{\mu\nu}$ and the expression~\eqref{eq:effectsymbol} can thus be rewritten as 
\begin{equation}
    \begin{aligned}
        (\mathcal{P}_{gg*}(\xi)\cdot t)^{\alpha \beta}=&
        -\frac{C}{2g} \epsilon_{\rm C}^{\alpha\mu_1\mu_2\mu_3}\epsilon_{\rm C}^{\beta\nu_1\nu_2\nu_3}
         C_{\mu_1\nu_1}\xi_{\mu_2}\xi_{\nu_2}t_{\mu_3\nu_3}\\
         =&\frac{C}{g} \left(\mathcal{P}_*^{\rm GR}(\xi)^{\alpha\beta\rho\sigma}t_{\rho\sigma}\right)_{g\to C,g^{-1}\to C^{-1}}.
    \end{aligned}
\end{equation}

\citet{Reall:2021voz} explored the physical polarisations of the theory and showed that the characteristic polynomial corresponding to the gauge-invariant principal symbol, i.e., $\det \mathcal{P}_*(\xi)^{IJ}\propto p(\xi) $, factorises into a quadratic polynomial and a quartic polynomial:
\begin{equation}
    p(\xi)=( C^{-1}{}^{\mu\nu} \xi_\mu\xi_\nu)\mathcal{Q}(\xi),
\end{equation}
which implies that physical degrees of freedom in EsGB propagate along the union of the null cone defined by the effective metric and a \textit{quartic cone} defined by $\mathcal{Q}(\xi)=0$.  The (weak) hyperbolicity of the system amounts to the question of whether the propagation speeds of the physical modes are real, that is, whether the roots of the characteristic polynomial $p(\xi)$ are real for any wave vector $\xi_i$. For the quartic factor, finding an efficient diagnostic for this criterion is not straightforward. However, in practice, during the dynamical evolution,  a potential \textit{loss of hyperbolicity} can still be detected once $C^{-1}{}^{\mu\nu}$ ceases to have a Lorentzian signature. This phenomenon has been reported in full $3+1$ simulations of black hole  \cite{Doneva:2023oww} and cosmological spacetimes \cite{Waeming:2026wrp}.

Since the goal of this paper is to show the strong hyperbolicity in a weakly coupled regime,we will not proceed with a further review of the intricate characteristic analysis presented in \cite{Reall:2021voz}. In the following, we will prove the strong hyperbolicity of the weakly coupled EsGB theory by first considering the zero-coupling limit, namely the Einstein-Scalar system in the modified CCZ4 formulation—and then extending the result to EsGB by continuity, as done in the previous section.
\subsection{Proof of strong hyperbolicity}

Throughout the weak-coupling argument, we require the cone-separation condition~\ref{ass:mccz4-cone-separation} in the Einstein-scalar limit. For $g_2=\lambda^{\rm GB}=0$, the principal symbol acting on $T_I=(t_{\mu\nu},\psi)$ is
\begin{equation}
\mathcal P_{\rm Esf}(\xi)^{IJ}T_J
=\begin{pmatrix}
\mathcal P^{\rm GR}(\xi)^{\mu\nu\rho\sigma}&0\\
0&-\xi^2
\end{pmatrix}
\begin{pmatrix}t_{\rho\sigma}\\\psi\end{pmatrix}.
\label{eq:einstein-scalar-symbol}
\end{equation}

Since the principal symbol is block diagonal, the polarisation space is the direct sum of the mCCZ4-GR sector and the scalar sector. The GR polarisations embed into the full space as $(t_{\mu\nu},0)^T$, while the scalar polarisation associated with the physical characteristics $\xi^2=0$ takes the form $(0,1)^T$.

The scalar sector adds one physical polarisation for each branch of $\xi^2=0$, giving $22$ independent eigenvectors in total. The block-diagonal Einstein-scalar system inherits strong hyperbolicity from the GR and scalar blocks.

Note that the EsGB principal symbol satisfies the diffeomorphism identities
\begin{equation}
\xi_\mu\mathcal P_{gg*}(\xi)^{\mu\nu\rho\sigma}
=0=\mathcal P_{gg*}(\xi)^{\mu\nu\rho\sigma}\xi_\rho,
\qquad
\xi_\mu\mathcal P_{g\phi*}(\xi)^{\mu\nu}=0.
\label{eq:esgb-diffeom}
\end{equation}

The characteristic polarisations satisfy the same identity~\eqref{eq:Factor}, since the gauge-fixing term is unchanged:
\begin{equation}
-\frac12\bigl(\hat g^{\mu\alpha}\xi_\mu\xi_\alpha\bigr)
\bigl(\tilde{\mathbb P}^{\nu\beta\rho\sigma}\xi_\beta t_{\rho\sigma}\bigr)=0.
\label{eq:esgb-factorization}
\end{equation}

These deformation-invariant properties allow us to construct the pure-gauge and gauge-condition-violating polarisations in the same way as in the Einstein-scalar system.

For the pure-gauge sector, the ansatz from the Einstein-scalar limit,
\begin{equation}
T=(\xi_{(\mu}X_{\nu)},0)^T,
\label{eq:esgb-pure-gauge}
\end{equation}
still satisfies $\mathcal P_*(\xi)T=0$ and yields the same characteristic equation~\eqref{eq:vectorEq}, $Q(\xi)^{\mu\nu}X_\nu=0$, for $X_\mu$. The pure-gauge eigenvalues and the corresponding eigenvectors are unchanged by the deformation. Using the notations introduced in the GR proof in Sec.~\ref{sec:gr-characteristic-problem}, these eigenvectors span $\tilde V_{\mathbf0}^\pm$, $\tilde V_{\mathbf L}^\pm$, and $\tilde V_{\mathbf T}^\pm$, whose dimensions are the GR values at $d=3$.

For the gauge-condition-violating sector, the eigenvalues remain the same as in GR by the same argument as in the EGB proof in Sec.~\ref{sec:egb-extension}, using the identity~\eqref{eq:esgb-factorization}.

In line with the EGB construction, we parametrise the polarisations by an arbitrary vector $v^\mu$ and use the ansatz in the transverse basis~\eqref{eq:tetrad2}:
\begin{equation}
t_{\mu\nu}(v)=t_{AB}(v)e^A_\mu e^B_\nu+\hat\xi_{(\mu}Y_{\nu)}(v).
\label{eq:esgb-gcv-ansatz}
\end{equation}

We construct the polarisation $T(v)=(t_{\mu\nu}(v),\psi(v))^T$ by solving the following equations:
\begin{subequations}
\begin{gather}
\mathcal P_*(\hat\xi)^{IJ}T_J(v)
=\begin{pmatrix}
\hat P_\gamma{}^{\alpha\mu\nu}\hat\xi_\alpha v^\gamma\\0
\end{pmatrix},
\label{eq:EsGBTensor}\\
\tilde{\mathbb P}^{\gamma\beta\rho\sigma}\hat\xi_\beta t_{\rho\sigma}(v)
=v^\gamma,
\label{eq:EsGBVect}
\end{gather}
\end{subequations}
where the tensor sector of the system inherits the same prescription as in GR and EGB, whilst the scalar block has zero right-hand side because $\mathcal P_{\rm GF}$ has no scalar component in the principal symbol~\eqref{eq:CharaEsGB}.

With this ansatz, the components $(t_{AB}(v),\psi(v))$ satisfy the coupled transverse system
\begin{equation}
\begin{pmatrix}
\mathcal P_{gg*}(\hat\xi)^{ABCD}&\mathcal P_{g\phi*}(\hat\xi)^{AB}\\
\mathcal P_{g\phi*}(\hat\xi)^{CD}&\mathcal P_{\phi\phi*}(\hat\xi)
\end{pmatrix}
\begin{pmatrix}t_{CD}(v)\\\psi(v)\end{pmatrix}
=\begin{pmatrix}
e^A_\mu e^B_\nu\hat P_\gamma{}^{\alpha\mu\nu}\hat\xi_\alpha v^\gamma\\0
\end{pmatrix}.
\label{eq:esgb-transverse-system}
\end{equation}

In the Einstein-scalar limit, the coefficient matrix of the transverse system becomes
\begin{equation}
\begin{pmatrix}
\mathcal P_*^{\rm GR}(\hat\xi)^{ABCD}&0\\
0&-g^{\mu\nu}\hat\xi_\mu\hat\xi_\nu
\end{pmatrix}.
\label{eq:esgb-transverse-limit}
\end{equation}

Under the cone-separation condition~\ref{ass:mccz4-cone-separation}, both diagonal blocks are invertible. By continuity, the coefficient matrix of the transverse system remains invertible for sufficiently weak coupling. We therefore first solve the transverse system uniquely for $(t_{AB}(v),\psi(v))$. With these components fixed, we then solve the gauge equation for $Y_\mu(v)$:
\begin{equation}
Q(\hat\xi)^{\gamma\sigma}Y_\sigma(v)
=v^\gamma-\tilde{\mathbb P}^{\gamma\beta\rho\sigma}
\hat\xi_\beta e^A_\rho e^B_\sigma t_{AB}(v).
\label{eq:esgb-gauge-vector}
\end{equation}

The $Q(\hat\xi)$ operator is still defined by~\eqref{eq:gauge-Q-hat} and remains invertible, so $Y_\mu(v)$ is given by the unique solution~\eqref{eq:v2Y}. The arbitrary vector $v^\mu$ supplies four independent polarisations for each branch in $\hat\xi_0$ of $\hat g^{\mu\nu}\hat\xi_\mu\hat\xi_\nu=0$. The corresponding eigenvectors span $\hat V^\pm$, whose dimensions are unchanged from the Einstein-scalar limit.

Under the cone-separation condition~\ref{ass:mccz4-cone-separation}, the Einstein-scalar physical eigenvalues $\xi_0^\pm$ of $M(\xi_i)$ are uniformly separated from the eigenvalues in the pure-gauge and gauge-condition-violating sectors, as in Sec.~\ref{sec:egb-extension}. By continuity, this separation persists for sufficiently weak coupling. We use the $\xi_0^\pm$-group for the eigenvalues approaching $\xi_0^\pm$ in the Einstein-scalar limit. We define the \textit{total generalised eigenspaces} $V^\pm$ as in the EGB proof in Sec.~\ref{sec:egb-extension}, using the EsGB matrix $M(\xi_i)$. Their dimensions remain equal to their Einstein-scalar values, $\dim V^\pm=3$, by continuity of their projectors, as in the EGB construction~\cite{Kovacs:2020ywu,kato1995perturbation}.

The dimensions of the constructed unphysical eigenspaces and the physical spaces $V^\pm$ sum to $\dim V=22$. We therefore obtain a direct-sum decomposition of the same form as~\eqref{eq:egb-spectral-decomposition}. It remains to show that the restrictions of $M(\xi_i)$ to $V^\pm$ have real eigenvalues and are diagonalisable.

\subsubsection*{Diagonalisability within $V^\pm$}

We use the Hermitian matrices $H_*^\pm$ defined in~\eqref{eq:physical-hermitian-matrix}, with $A_*,B_*$ now obtained from the combined tensor-scalar symbol $\mathcal P_*(\xi)^{IJ}$. In the Einstein-scalar limit, the corresponding Hermitian form includes the additional scalar contribution:
\begin{equation}
\langle\mathcal U_1,\mathcal U_2\rangle_\pm
:=\mathcal U_1^\dagger H_*^\pm\mathcal U_2
=\mp\xi^{\pm0}
\left(t_{\langle\hat{i}\hat{j}\rangle}^{1*}t_{\langle\hat{i}\hat{j}\rangle}^{2}
+2\psi_1^*\psi_2\right),
\label{eq:esgb-physical-form}
\end{equation}
where $\mathcal U_a=(T_a,\xi_0^\pm T_a)^T$, $T_a=(t_{\mu\nu}^a,\psi_a)^T$, and $\xi^{\pm0}=g^{0\mu}\xi_\mu^\pm$. This form is positive definite on each Einstein-scalar physical space. By continuity of $H_*^\pm$ and the projectors defining $V^\pm$, it remains positive definite on $V^\pm$ for sufficiently weak coupling.

As in Sec.~\ref{sec:egb-extension}, the polarisation associated with a physical eigenvector satisfies the gauge condition $\tilde{\mathbb P}^{\nu\beta\rho\sigma}\xi_\beta t_{\rho\sigma}=0$, and hence $\mathcal P_*(\xi)T=0$. The eigenpairs $(\xi_0^{(a)},\mathcal U_a)$ in $V^+$ satisfy the same identity~\eqref{eq:diffeigen}:
\begin{equation}
(\xi_0^{(1)}-\xi_0^{(2)})
\mathcal U_1^T H_*^+\mathcal U_2=0.
\label{eq:esgb-two-eigenpair}
\end{equation}

Since $M(\xi_i)$ is real, the conjugate of an eigenpair in the $\xi_0^+$-group is also in that group. Taking $(\xi_0^{(2)},\mathcal U_2)=(\xi_0^{(1)*},\mathcal U_1^*)$ yields
\begin{equation}
2\operatorname{Im}(\xi_0^{(1)})
\langle\mathcal U_1,\mathcal U_1\rangle_+=0.
\label{eq:esgb-reality}
\end{equation}

The physical eigenvalue therefore satisfies $\operatorname{Im}(\xi_0^{(1)})=0$ by positivity of the Hermitian form. The same argument applies to $V^-$.

We now show that the constraint in Lemma~\ref{th:lem2} can be extended to the tensor-scalar system. Let $\mathcal V=(v_I,v'_I)^T$ be a generalised eigenvector at a physical eigenvalue, with $v_I=(t_{\mu\nu},\psi)$ and $v'_I=(t'_{\mu\nu},\psi')$. For either branch $\hat\xi_0^\pm$, the left polarisation and associated left eigenvector are again given by
\begin{equation}
s_I=(\hat\xi^\pm_{(\mu}X_{\nu)},0),
\qquad
\mathcal W=(s_I,\hat\xi_0^\pm s_I)
\begin{pmatrix}B&A\\A&0\end{pmatrix},
\label{eq:esgb-left-eigenvector}
\end{equation}
where $X_\mu$ is arbitrary and $A,B,C$ are the coefficients of the full symbol $\mathcal P(\xi)=\xi_0^2A+\xi_0B+C$. As in the EGB case, $\mathcal W\mathcal V=0$ gives
\begin{equation}
\hat\xi^\pm_\nu(B+\hat\xi_0^\pm A)^{\mu\nu I}v_I
+\hat\xi^\pm_\nu A^{\mu\nu I}v'_I=0.
\label{eq:esgb-left-orthogonality}
\end{equation}

The same reduction using the two branches $\hat\xi_0^\pm$ yields
\begin{equation}
\begin{aligned}
R^\mu:={}&-2(\hat g^{00})^{-1}\hat g^{0i}\xi_i
A^{\mu0I}v_I
+\bigl(A^{\mu iI}\xi_i+B^{\mu0I}\bigr)v_I
+A^{\mu0I}v'_I=0.
\end{aligned}
\label{eq:esgb-R-full}
\end{equation}

The $A_*,B_*$ coefficients of $\mathcal P_*$ satisfy the analogous relations from the identities~\eqref{eq:esgb-diffeom}, written as $\xi_\nu\mathcal P_*(\xi)^{\mu\nu I}=0$:
\begin{equation}
A_*^{\mu0I}=0,
\qquad
\xi_iA_*^{\mu iI}+B_*^{\mu0I}=0.
\label{eq:esgb-starred-coefficients}
\end{equation}
Thus $R^\mu$ again contains only gauge-fixing terms. With $A_{g\phi,\rm GF}^{\mu\nu}=B_{g\phi,\rm GF}^{\mu\nu}=0$, it reduces to
\begin{equation}
\begin{aligned}
0=R^\mu={}&-2(\hat g^{00})^{-1}\hat g^{0i}\xi_i
A_{\rm GF}^{\mu0\rho\sigma}t_{\rho\sigma}\\
&+\bigl(A_{\rm GF}^{\mu i\rho\sigma}\xi_i
+B_{\rm GF}^{\mu0\rho\sigma}\bigr)t_{\rho\sigma}
+A_{\rm GF}^{\mu0\rho\sigma}t'_{\rho\sigma}.
\end{aligned}
\label{eq:esgb-R-gauge}
\end{equation}

A calculation similar to that in the EGB proof yields the same constraints on the generalised eigenvector
\begin{equation}
\tilde{\mathbb P}^{\mu i\rho\sigma}\xi_i t_{\rho\sigma}
+\tilde{\mathbb P}^{\mu0\rho\sigma}t'_{\rho\sigma}=0.
\label{eq:generaldiv1}
\end{equation}

Suppose that $M(\xi_i)|_{V^\pm}$ has a nontrivial Jordan block. Then there exist $\mathcal V=(v_I,v'_I)^T$ and $\mathcal U=(u_I,\xi_0u_I)^T\ne0$ such that
\begin{equation}
(\xi_0-M)\mathcal V=\mathcal U,
\qquad
(\xi_0-M)\mathcal U=0.
\label{eq:generalEigen1}
\end{equation}

Expanding the above relation and using the constraint~\eqref{eq:generaldiv1} just established, we obtain
\begin{equation}
\begin{aligned}
v'&=\xi_0v-u,\\
\mathcal P(\xi)v&=(2\xi_0A+B)u,\\
\mathcal P_{\rm GF}(\xi)v&=(2\xi_0A_{\rm GF}+B_{\rm GF})u.
\end{aligned}
\label{eq:esgb-jordan-components}
\end{equation}

Subtracting the gauge-fixing part leaves $\mathcal P_*(\xi)v=(2\xi_0A_*+B_*)u$. Since $\mathcal P_*(\xi)u=0$ and the physical eigenvalue $\xi_0$ is real, contraction with $u_I^*$ yields
\begin{equation}
0=u_I^*\mathcal P_*(\xi)^{IJ}v_J
=u_I^*(2\xi_0A_*+B_*)^{IJ}u_J
=\pm\langle\mathcal U,\mathcal U\rangle_\pm.
\label{eq:esgb-jordan-contradiction}
\end{equation}

The eigenvector must therefore satisfy $\mathcal U=0$ by positivity of the Hermitian form, a contradiction. Thus $M(\xi_i)|_{V^\pm}$ has real eigenvalues and is diagonalisable.

\subsubsection*{Construction of the symmetrizer}

The eigenvectors now span $V^\pm$. Any pair of vectors in $V^\pm$ therefore satisfies
\begin{equation}
\langle\mathcal U_1,M\mathcal U_2\rangle_\pm
=\langle M\mathcal U_1,\mathcal U_2\rangle_\pm
\qquad\text{for all }\mathcal U_1,\mathcal U_2\in V^\pm.
\label{eq:esgb-physical-symmetrizer}
\end{equation}

The restriction of $H_*^\pm$ therefore symmetrizes $M|_{V^\pm}$. The form and the projectors defining $V^\pm$ depend smoothly on $\xi_i$, as in Sec.~\ref{sec:egb-extension}. The physical blocks of the full symmetrizer are therefore smooth even when eigenvalues within a $\xi_0^\pm$-group cross.

We use the construction in~\eqref{eq:Kmatrix}, with $S$ formed from bases of the constructed unphysical eigenspaces and smooth bases $\{v_A^\pm\}_{A=1}^3$ of $V^\pm$. On a region where these bases are chosen smoothly, set
\begin{equation}
\begin{gathered}
\mathcal K(\xi_i)
=(S^{-1})^\dagger
\operatorname{diag}(I_8,I_8,\mathcal H_*^+,\mathcal H_*^-)
S^{-1},
\\
(\mathcal H_*^\pm)_{AB}
=(v_A^\pm)^\dagger H_*^\pm v_B^\pm.
\end{gathered}
\label{eq:esgb-symmetrizer}
\end{equation}

The two identity blocks correspond to the pure-gauge and gauge-condition-violating sectors. Each physical block is $3\times3$. The resulting $\mathcal K$ is positive definite and satisfies $\mathcal K M=M^\dagger\mathcal K$. Following the symmetrizer construction in Sec.~\ref{sec:egb-extension}, the weakly coupled EsGB system is therefore strongly hyperbolic.

\section{Strong hyperbolicity of standard CCZ4 revisited}
\label{sec:standard-ccz4}

It is natural to examine how our method applies to standard CCZ4 in GR with puncture gauge, corresponding to $a(x)=b(x)=0$. Indeed, the eigenbasis constructed for the gauge-condition-violating sector becomes singular as $b(x)\to0$, because the operator $\mathcal{P}_{*}^{ABCD}(\hat{\xi})$ inverted in that construction loses invertibility when the null cones of $\hat g^{\mu\nu}$ and $g^{\mu\nu}$ merge. This singularity does not itself imply a failure of strong hyperbolicity, since standard CCZ4 in puncture gauge has been proved to be strongly hyperbolic except at certain known defective degeneracies~\cite{Dumbser:2017okk}. In this section, we address this issue by constructing an alternative eigenbasis directly for $a(x)=b(x)=0$.

The principal symbol of the second-order CCZ4 metric system is
\begin{equation}
(\mathcal P_{\mathrm{CCZ4}}(\xi)t)^{\mu\nu}
={}
(\mathcal P_\star(\xi)t)^{\mu\nu}
-
P_\lambda{}^{\beta\mu\nu}\xi_\beta
\tilde{\mathbb P}^{(0)}{}^{\lambda\delta\rho\sigma}
\xi_\delta t_{\rho\sigma},
\label{eq:standard-ccz4-symbol}
\end{equation}
obtained from equation~\eqref{eq:Charactmunu} by setting $a=b=0$. Here $\tilde{\mathbb P}^{(0)}{}^{\lambda\delta\rho\sigma}$ is the $a\to0$ limit of $\tilde{\mathbb P}{}^{\lambda\delta\rho\sigma}$ defined in~\eqref{eq:deftildeP}. Contracting equation~\eqref{eq:standard-ccz4-symbol} with $\xi_\mu$ and using the identity~\eqref{eq:diffeom} for $\mathcal{P}_*(\xi)$ gives
\begin{equation}
0= -\frac{1}{2}\xi^2 
\left(\tilde{\mathbb P}^{(0)}{}^{\lambda\delta\rho\sigma}
\xi_\delta t_{\rho\sigma}\right).
\label{eq:standard-ccz4-contracted}
\end{equation}
Following the procedure established for mCCZ4, we divide the analysis into Case I, the $\xi^2\neq0$ sector, and Case II, the $\xi^2=0$ sector.

\paragraph*{Case I: The $\xi^2\neq0$ sector.}

For a covector $\xi_\mu$ satisfying $\xi^2\neq0$, the identity~\eqref{eq:standard-ccz4-contracted} gives $\tilde{\mathbb P}^{(0)}{}^{\lambda\delta\rho\sigma}\xi_\delta t_{\rho\sigma}=0$. The trace-reversed Einstein characteristic equation then implies $t_{\mu\nu}=\xi_{(\mu}X_{\nu)}$, as in the GR mCCZ4 characteristic analysis in Sec.~\ref{sec:gr-characteristic-problem}. Thus every polarisation $t_{\mu\nu}$ associated with $\xi_\mu$ belongs to the pure-gauge sector. The gauge condition takes the form of equation~\eqref{eq:vectorEq}:
\begin{equation}
    Q^{(0)}(\xi)^{\mu\nu}X_{\nu}\equiv\tilde{\mathbb P}^{(0)}{}^{\mu\delta\rho\sigma}\xi_\delta\xi_\rho X_\sigma=0.
\end{equation}
Here $Q^{(0)}(\xi)$ is the $a\to0$ limit of $Q(\xi)$. In the temporal, longitudinal, and transverse components $(X^0,X^1,X^{\hat i})$, the equation reads
\begin{equation}
		\begin{pmatrix}
			q_{\bf0} & 0 & 0\\[1mm]
			s & q_{\bf L} & 0\\[1mm]
			0 & 0 & q_{\bf T}\,\delta_{\hat i\hat j}
		\end{pmatrix}
        \begin{pmatrix}
            X^0 \\ X ^1 \\ X^{\hat{j}}
        \end{pmatrix} =0,
		\label{eq:Q0-block_0}
	\end{equation}
where
	\begin{gather}
		q_{\mathbf 0} := \frac12\,\mathcal G_{\mathbf 0}^{\mu\nu}\,\xi_\mu\xi_\nu, \qquad q_{\mathbf L} := \frac{d-1}{d}\,\mathcal G_{\mathbf L}^{\mu\nu}\,\xi_\mu\xi_\nu, \qquad q_{\mathbf T} := \frac12\,\mathcal G_{\mathbf T}^{\mu\nu}\,\xi_\mu\xi_\nu, \label{eq:q0LT-def}\\
		s := -\frac{\chi\alpha^2}{2c_\beta}\,(n^\mu\xi_\mu)(k^\nu\xi_\nu). \label{eq:q0LT-offdiag}
	\end{gather}
The coefficients $q_{\mathbf0},q_{\mathbf L},q_{\mathbf T}$ defined in~\eqref{eq:q0LT-def} are evaluated at $a=0$, and $k_\mu$ is defined as in~\eqref{eq:tetrad1}. Under our sign assumptions, $\mathcal G_{\mathbf0}$, $\mathcal G_{\mathbf L}$, and $\mathcal G_{\mathbf T}$ are Lorentzian. For a fixed nonzero spatial covector $\xi_i$, each of the equations $q_{\mathbf0}=0$, $q_{\mathbf L}=0$, and $q_{\mathbf T}=0$ has two real roots in $\xi_0$, defining its two branches. The coefficient $s$ is nonzero for the corresponding characteristic covectors $\xi_\mu$. \textbf{In analogy with the cone-separation condition~\ref{ass:mccz4-cone-separation}, we initially work under the separation assumption that the four metrics $g^{\mu\nu}$, $\mathcal G_{\mathbf0}^{\mu\nu}$, $\mathcal G_{\mathbf L}^{\mu\nu}$, and $\mathcal G_{\mathbf T}^{\mu\nu}$ are pairwise distinct.} For $a=b=0$, this assumption can be satisfied by a suitable choice of $c_\alpha$ and $c_\beta$. Then equation~\eqref{eq:Q0-block_0} gives one temporal polarisation for each branch of $q_{\mathbf0}=0$, one longitudinal polarisation for each branch of $q_{\mathbf L}=0$, and $d-1$ transverse polarisations for each branch of $q_{\mathbf T}=0$ in the pure-gauge sector, hence $2(d+1)$ eigenvectors in total. We shall relax this assumption in Sec.~\ref{sec:ccz4-general-conditions}.

\paragraph*{Case II: The $\xi^2=0$ sector.}

With $\gamma^{ij}\xi_i\xi_j=1$, the null cone equation $\xi^2=0$ gives the eigenvalues $\xi_0^{\pm}=\beta^i\xi_i\pm \alpha$. For $\xi_\mu=(\xi_0^\pm,\xi_i)$, equation~\eqref{eq:standard-ccz4-symbol} reduces to
\begin{equation}
    P_\gamma {}^{\alpha\mu\nu}\xi_{\alpha} P^{\gamma\beta\rho\sigma}\xi_{\beta} t_{\rho\sigma}=
     P_\gamma {}^{\alpha\mu\nu}\xi_{\alpha} \tilde{\mathbb{P}}^{(0)}{}^{\gamma\beta\rho\sigma}\xi_{\beta} t_{\rho\sigma}.
\end{equation}
Trace reversal then gives the constraint
\begin{equation}
P^{\gamma\beta\rho\sigma}\xi_{\beta} t_{\rho\sigma}=\tilde{\mathbb{P}}^{(0)}{}^{\gamma\beta\rho\sigma}\xi_{\beta} t_{\rho\sigma}.\label{eq:phyiscal-constraint}
\end{equation}
To construct the polarisations for each branch of $\xi^2=0$, we introduce a free vector source $v^\mu$, as in the gauge-condition-violating sector of mCCZ4, and write the constraint~\eqref{eq:phyiscal-constraint} as
\begin{subequations}\label{eq:split}
    \begin{align}
    &P^{\gamma\beta\rho\sigma}\xi_{\beta} t_{\rho\sigma}=v^\gamma, \label{eq:split_tensor}\\
    &\tilde{\mathbb{P}}^{(0)}{}^{\gamma\beta\rho\sigma}\xi_{\beta} t_{\rho\sigma}=v^\gamma.
\end{align}
\end{subequations}
We use the $1+1+(d-1)$ orthonormal frame introduced in Sec.~\ref{sec:pure_gauge_mccz4}, choosing the {\it signed} longitudinal direction $k_\mu$ for $\xi_\mu$:
\begin{equation}
    \omega:=-n\cdot\xi\neq0,\qquad k_\mu:=\frac{\gamma_\mu{}^\nu\xi_\nu}{\omega},\qquad
    q_{\mu\nu}:=e^{\hat{i}}_\mu e^{\hat{j}}_\nu\delta_{\hat{i}\hat{j}}=\gamma_{\mu\nu}-k_\mu k_\nu.
\end{equation}
Then $k\cdot k=1$, $k\cdot\xi=\omega$, and $\xi_\mu=\omega(n_\mu+k_\mu)$ for covectors on both branches of $\xi^2=0$. Equation~\eqref{eq:split_tensor} is invariant under
\begin{equation}
    t_{\mu\nu}\to t_{\mu\nu}+t_{\langle \hat{i}\hat{j}\rangle}e^{\hat{i}}_\mu e^{\hat{j}}_{\nu}+ \xi_{(\mu} X_{\nu)},
\end{equation}
where $X_\mu$ and the transverse-traceless data $t_{\langle \hat{i}\hat{j}\rangle}$ are arbitrary, with $t_{\langle \hat{i}\hat{j}\rangle}\delta^{\hat{i}\hat{j}}\equiv0$. We therefore use the ansatz
\begin{equation}
    t_{\mu\nu}=t_{\langle \hat{i}\hat{j}\rangle}e^{\hat{i}}_\mu e^{\hat{j}}_{\nu}+ k_{(\mu}Y_{\nu)}+c~q_{\mu\nu}+\xi_{(\mu} X_{\nu)}, \label{eq:CCZ4ansatz}
\end{equation}
with $Y\cdot n=0$. Equation~\eqref{eq:split_tensor} determines $Y_\mu$ and $c$ from $v^\mu$, leaving $t_{\langle\hat{i}\hat{j}\rangle}$ and $X_\mu$ free.

Following the relation~\eqref{eq:eqTTcondi}, the transverse-traceless part $t_{\mu\nu}^{\rm TT}=t_{\langle\hat{i}\hat{j}\rangle}e^{\hat{i}}_\mu e^{\hat{j}}_\nu$ satisfies the gauge condition $\tilde{\mathbb{P}}^{(0)}{}^{\gamma\beta\rho\sigma}\xi_{\beta} t_{\rho\sigma}^{\rm TT}=0$. Substituting the ansatz for $t_{\rho\sigma}$ into equations~\eqref{eq:split} therefore gives
\begin{gather}
     \frac{\omega}{2}Y_\mu -\frac{\omega}{2} (Y\cdot k) n_\mu-c\frac{d-1}{2}\xi_\mu=v_\mu, \label{eq:Yceq}\\
       Q^{(0)}(\xi)^{\mu\nu}X_\nu=v^\mu- \tilde{\mathbb{P}}^{(0)}{}^{\mu\beta\rho\sigma}\xi_\beta(k_{(\rho}Y_{\sigma)}+c  ~q_{\rho\sigma}) \label{eq:QxEq}.
\end{gather}
Projecting equation~\eqref{eq:Yceq} along $k$ and $n$ gives the unique solution for $c$ and $Y_\mu$:
\begin{equation}
   c(v)=\frac{v\cdot n-v\cdot k}{\omega(d-1)},\quad  Y_\mu(v)=\frac{2}{\omega}(v_\mu+\frac{v\cdot n- v\cdot k}{2}(n_\mu+k_\mu)+\frac{v\cdot n+ v\cdot k}{2}n_\mu)\, . \label{eq:solu_cY}
\end{equation}

Before solving equation~\eqref{eq:QxEq} for $X_\mu$, we establish completeness of the ansatz independently of the separation assumption.

\begin{proposition}[Completeness of the polarisation ansatz]
\label{prop:ccz4-completeness}
For any nonzero covector $\xi_\mu$ satisfying $\xi^2=0$, every symmetric tensor $t_{\mu\nu}$ has a unique decomposition~\eqref{eq:CCZ4ansatz}, with $c(v)$ and $Y_\mu(v)$ given by the solution~\eqref{eq:solu_cY} and unrestricted data $(t_{\langle\hat{i}\hat{j}\rangle},v^\mu,X_\mu)$.
\end{proposition}
\begin{proof}
Suppose $t_{\mu\nu}=0$ after substituting the solution~\eqref{eq:solu_cY}. Applying $P^{\gamma\beta\rho\sigma}\xi_\beta$ gives $v^\gamma=0$, since this map annihilates $t_{\rho\sigma}^{\rm TT}$ and $\xi_{(\rho}X_{\sigma)}$ for $\xi^2=0$. Thus the solution~\eqref{eq:solu_cY} gives $Y_\mu=c=0$, and the transverse-traceless projection gives $t_{\langle\hat{i}\hat{j}\rangle}=0$. Finally, injectivity of $X_\mu\mapsto\xi_{(\mu}X_{\nu)}$ for $\xi\neq0$ gives $X_\mu=0$. The decomposition is injective and has
\begin{equation}
\frac{(d+1)(d-2)}2+(d+1)+(d+1)=N
\end{equation}
independent parameters, so it exhausts the symmetric-tensor space.
\end{proof}

\begin{corollary}[Restrictions on the free data]
\label{cor:ccz4-free-data}
Fix a nonzero covector $\xi_\mu$ satisfying $\xi^2=0$. With $c(v)$ and $Y_\mu(v)$ given by solution~\eqref{eq:solu_cY}, the ansatz~\eqref{eq:CCZ4ansatz} defines a linear isomorphism from the unrestricted triples $(t_{\langle\hat{i}\hat{j}\rangle},v^\mu,X_\mu)$ onto the space of symmetric rank-two tensors. Any proper restriction on these triples therefore prevents the resulting tensors from exhausting that space. In particular, $r$ independent homogeneous linear constraints on the free data yield a tensor subspace of dimension $N-r$.
\end{corollary}
\begin{proof}
The ansatz is linear in the triple, and Proposition~\ref{prop:ccz4-completeness} gives a unique triple for every symmetric tensor. Excluding a triple therefore excludes its associated tensor. Restricting the map to a linear subspace preserves its dimension.
\end{proof}

Using the expression~\eqref{eq:eq_gauge_fix_symbol} in the limit $a,b\to 0$, equation~\eqref{eq:QxEq} becomes
\begin{equation}
\begin{aligned}
 Q^{(0)}(\xi)^{\mu\nu}X_\nu&=v^\mu- \tilde{\mathbb{P}}^{(0)}{}^{\mu\beta\rho\sigma}\xi_\beta(k_{(\rho}Y_{\sigma)}+c  ~q_{\rho\sigma}) \\
    &=v^\mu- \frac{1}{2}(\gamma^{\mu\lambda}\mathcal{G}_{\bf T}^{\alpha\beta}-n^\mu n^\lambda \mathcal{G}_{\bf 0}^{\alpha\beta}) \bigl(\xi_{(\alpha} k_{\beta)}Y_{\lambda}+\xi_{(\alpha|} k_{\lambda}Y_{|\beta)}-\xi_\lambda k_{(\alpha} Y_{\beta)}+ \\ &2c~\xi_{(\alpha}q_{\beta)\lambda}-c~\xi_\lambda q_{\alpha\beta}\bigr)- \frac{1}{2}\gamma^{\alpha\mu} (\frac{d-2}{d}\gamma^{\beta\lambda}-\frac{\chi\alpha^2}{c_\beta}n^\lambda n^\beta )\xi_\alpha (k_{(\beta}Y_{\lambda)} +c~q_{\beta\lambda})\\
    &=v^\mu-\frac{1}{2}g^{\mu\lambda}\gamma^{\alpha\beta}\bigl(\xi_{(\alpha} k_{\beta)}Y_{\lambda}+\xi_{(\alpha|} k_{\lambda}Y_{|\beta)}-\xi_\lambda k_{(\alpha} Y_{\beta)}+ \\ &2c~\xi_{(\alpha}q_{\beta)\lambda}-c~\xi_\lambda q_{\alpha\beta}\bigr)-\frac{d-2}{2d}\gamma^{\alpha\mu}\gamma^{\beta\lambda} \xi_\alpha (k_{(\beta}Y_{\lambda)} +c~q_{\beta\lambda}) \\
    &= v^\mu-\frac{1}{2}(\omega Y^\mu+\omega (k\cdot Y) k^\mu-\xi^\mu(k\cdot Y)-c~\xi^\mu (d-1)) -\frac{d-2}{2d}\omega k^\mu(k\cdot Y+c~(d-1))\\
    &= -(v\cdot n)\frac{d-2}{d} k^\mu.
\end{aligned} \label{eq:QxEqSimp}
\end{equation}
The third equality uses the spatial character of $k_{(\rho}Y_{\sigma)}+c q_{\rho\sigma}$ and the form $\gamma^{\mu\nu}-\epsilon_{\bf0,L,T}n^\mu n^\nu$ of the auxiliary metrics $\mathcal{G}_{\bf 0,T,L}$. Substituting the solutions $c(v)$ and $Y_\mu(v)$ from~\eqref{eq:solu_cY} gives the final equality.

In components $(X^0,X^1,X^{\hat i})$, equation~\eqref{eq:QxEqSimp} reads
	\begin{equation}
		\begin{pmatrix}
			q_{\bf0} & 0 & 0\\[1mm]
			s & q_{\bf L} & 0\\[1mm]
			0 & 0 & q_{\bf T}\,\delta_{\hat i\hat j}
		\end{pmatrix}
        \begin{pmatrix}
            X^0 \\ X ^1 \\ X^{\hat{j}}
        \end{pmatrix} = \begin{pmatrix}
            0 \\ -\frac{d-2}{d}(v\cdot n) \\ 0
        \end{pmatrix}.
		\label{eq:Q0-block}
	\end{equation} 
By the separation assumption, $(q_{\mathbf0}q_{\mathbf L}q_{\mathbf T})\lvert_{\xi^2=0}\neq0$. Thus $Q^{(0)}(\xi)$ is invertible for $\xi^2=0$, and equation~\eqref{eq:QxEqSimp} has the unique solution
\begin{equation}
    X^{\mu}(v)= -\frac{(d-2)(v\cdot n)}{q_{\bf L}d} k^\mu. \label{eq:eqsolX}
\end{equation}
By Proposition~\ref{prop:ccz4-completeness}, the data $(t_{\langle\hat{i}\hat{j}\rangle},v^\mu)$ supply $d(d+1)/2$ independent polarisations for each branch of $\xi^2=0$. Together with the eigenvectors associated with $q_{\mathbf0}=0$, $q_{\mathbf L}=0$, and $q_{\mathbf T}=0$ in the pure-gauge sector, this yields $d(d+1)+2(d+1)=2N$ independent eigenvectors with real eigenvalues.

\subsection{General conditions for strong hyperbolicity}
\label{sec:ccz4-general-conditions}

The separation assumption is stronger than necessary for strong hyperbolicity. First-order CCZ4 in puncture gauge admits coincident characteristic speeds compatible with strong hyperbolicity~\cite{Dumbser:2017okk}. We determine the broader admissible domain by counting polarisations in the $\xi^2\neq0$ and $\xi^2=0$ sectors separately. We define
\begin{equation}
\delta_{Ag}:=\mathbf1_{\{\mathcal G_A=g\}},
\qquad A\in\{\mathbf0,\mathbf L,\mathbf T\},
\qquad
\delta_{\mathbf0\mathbf L}:=\mathbf1_{\{\mathcal G_{\mathbf0}=\mathcal G_{\mathbf L}\}},
\label{eq:ccz4-coincidence-indicators}
\end{equation}

where each indicator equals one when the corresponding metrics coincide and zero otherwise. All totals below include both branches of the relevant characteristic equations.

In the $\xi^2\neq0$ sector, Case I shows that every polarisation has the pure-gauge form $t_{\mu\nu}=\xi_{(\mu}X_{\nu)}$. Equation~\eqref{eq:Q0-block_0} gives the solution vectors $X^\mu$:
\begin{equation}
\begin{aligned}
q_{\mathbf0}=0:\quad &X^\mu=-q_{\mathbf L}n^\mu-sk^\mu,\\
q_{\mathbf L}=0:\quad &X^\mu=k^\mu,\\
q_{\mathbf T}=0:\quad &X^\mu=X_\perp^\mu,\qquad n\cdot X_\perp=k\cdot X_\perp=0,
\end{aligned}
\label{eq:ccz4-pure-gauge-vectors}
\end{equation}

with $X_\perp^\mu$ an arbitrary transverse vector. If $\mathcal G_{\mathbf0}=\mathcal G_{\mathbf L}$, the first two vectors are collinear because $q_{\mathbf L}=0$ and $s\neq0$, reducing the number of independent polarisations by one for each branch. The transverse vectors are orthogonal to $n^\mu$ and $k^\mu$ and hence independent of the scalar vectors. The $\xi^2\neq0$ sector therefore supplies exactly
\begin{equation}
2\left[
(1-\delta_{\mathbf0g})(1-\delta_{\mathbf0\mathbf L})+(1-\delta_{\mathbf Lg})
+(d-1)(1-\delta_{\mathbf Tg})
\right]
\label{eq:ccz4-nonnull-count}
\end{equation}

independent polarisations in the $\xi^2\neq0$ sector for both branches of each contributing characteristic equation. We introduce the factors $1-\delta_{Ag}$, $A\in\{\mathbf0,\mathbf L,\mathbf T\}$, to avoid double counting with the $\xi^2=0$ sector.

In the $\xi^2=0$ sector, set $v^\mu=0$ and $t_{\langle\hat{i}\hat{j}\rangle}=0$ in the ansatz~\eqref{eq:CCZ4ansatz}.  Note that the tensors $t_{\mu\nu}=\xi_{(\mu}X^{\mathrm{hom}}_{\nu)}$, with $Q^{(0)}(\xi)^{\mu\nu}X^{\mathrm{hom}}_\nu=0$, satisfy the constraint~\eqref{eq:phyiscal-constraint} and can be added to any solution without changing $v^\mu$ or $t_{\langle\hat{i}\hat{j}\rangle}$. We parametrise this residual gauge freedom using equation~\eqref{eq:Q0-block_0}:
\begin{equation}
\begin{aligned}
X_{\mathrm{hom}}^\mu
={}&-c_{\mathbf0}\delta_{\mathbf0g}(1-\delta_{\mathbf 0 \mathbf L})
\bigl(q_{\mathbf L}n^\mu+s k^\mu\bigr)\\
&+c_{\mathbf L}\delta_{\mathbf Lg}k^\mu
+\delta_{\mathbf Tg}X_\perp^\mu,
\qquad n\cdot X_\perp=k\cdot X_\perp=0.
\end{aligned}
\label{eq:ccz4-residual-gauge}
\end{equation}

Here $c_{\mathbf0}$ and $c_{\mathbf L}$ are arbitrary scalars, and $X_\perp^\mu$ is an arbitrary transverse vector. These give $\delta_{\mathbf0g}(1-\delta_{\mathbf0\mathbf L})+\delta_{\mathbf Lg}+(d-1)\delta_{\mathbf Tg}$ residual gauge degrees of freedom for each branch and parametrise the pure-gauge polarisations excluded from the $\xi^2\neq0$ sector.

For general source data $v^\mu$, equation~\eqref{eq:Q0-block} takes the form
\begin{equation}
q_{\mathbf0}X^0=0,
\qquad
q_{\mathbf T}X^{\hat i}=0,
\qquad
v\cdot n=-\frac{d}{d-2}\bigl(sX^0+q_{\mathbf L}X^1\bigr).
\label{eq:ccz4-null-source-system}
\end{equation}

The solvability of the above system requires $v\cdot n=0$ when $\mathcal G_{\mathbf L}=g\ne\mathcal G_{\mathbf0}$; otherwise $v^\mu$ is unrestricted. The admissible source data thus have $d+1-(1-\delta_{\mathbf0g})\delta_{\mathbf Lg}$ independent components. For every such $v^\mu$, equation~\eqref{eq:Q0-block} admits the particular solution
\begin{equation}
X_{\mathrm{part}}^\mu(v)=
\frac{(d-2)(v\cdot n)}
{d\bigl(q_{\mathbf L}+\delta_{\mathbf Lg}s\bigr)}
\left[\delta_{\mathbf Lg}n^\mu-(1-\delta_{\mathbf Lg})k^\mu\right].
\label{eq:ccz4-particular-solution}
\end{equation}

This expression is well-defined at $q_{\mathbf L}=0$ because $s\neq0$, unlike the solution in \eqref{eq:eqsolX}.

The admissible source data $v^\mu$, unrestricted transverse-traceless data $t_{\langle\hat{i}\hat{j}\rangle}$, and residual gauge freedom $X_{\mathrm{hom}}^\mu$ uniquely parametrise every polarisation by Corollary~\ref{cor:ccz4-free-data}. Their dimensions sum to exactly
\begin{equation}
\begin{aligned}
&2\Bigl[
\frac{(d+1)(d-2)}2
+\bigl(d+1-(1-\delta_{\mathbf0g})\delta_{\mathbf L g}\bigr)\\
&\qquad+\bigl(
\delta_{\mathbf0g}(1-\delta_{\mathbf 0\mathbf L})
+\delta_{\mathbf Lg}+(d-1)\delta_{\mathbf Tg}
\bigr)\Bigr]\\
&\qquad=d(d+1)+2\delta_{\mathbf0g}
+2(d-1)\delta_{\mathbf Tg}
\end{aligned}
\label{eq:ccz4-null-count}
\end{equation}

independent polarisations for both branches of $\xi^2=0$. Here we have used
\begin{equation}
\delta_{\mathbf0g}\delta_{\mathbf0\mathbf L}
=\delta_{\mathbf0g}\delta_{\mathbf Lg}
=\delta_{\mathbf0\mathbf L}\delta_{\mathbf Lg}.
\label{eq:ccz4-delta-identity}
\end{equation}

The $\xi^2\neq0$ and $\xi^2=0$ sectors together supply exactly $m$ independent eigenvectors of the principal symbol $M(\xi_i)$, where
\begin{equation}
\begin{aligned}
&m=2\left[
(1-\delta_{\mathbf0g})(1-\delta_{\mathbf0\mathbf L})+(1-\delta_{\mathbf Lg})
+(d-1)(1-\delta_{\mathbf Tg})
\right]\\
&\quad+d(d+1)+2\delta_{\mathbf0g}
+2(d-1)\delta_{\mathbf Tg}\\
&=2N-2\left[
\delta_{\mathbf0\mathbf L}+\delta_{\mathbf Lg}
-\delta_{\mathbf0\mathbf L}\delta_{\mathbf Lg}
\right].
\end{aligned}
\label{eq:ccz4-polarization-count}
\end{equation}

\subsubsection*{Final conditions for strong hyperbolicity}

When $\mathcal G_{\mathbf L}\neq\mathcal G_{\mathbf0}$ and $\mathcal G_{\mathbf L}\neq g$, the construction yields exactly $m=2N$ independent eigenvectors with real eigenvalues. The temporal polarisation and solution~\eqref{eq:eqsolX} remain regular because $q_{\mathbf L}\lvert_{q_{\mathbf0}=0}\neq0$ and $q_{\mathbf L}\lvert_{\xi^2=0}\neq0$. The eigenvector matrices $S(\xi_i)$ can be chosen to vary smoothly, with both $S(\xi_i)$ and $S(\xi_i)^{-1}$ uniformly bounded over normalised spatial directions for each admissible background. The criterion following Definition~\ref{def:fully-second-order-SH} therefore establishes strong hyperbolicity.

If $\mathcal G_{\mathbf0}=\mathcal G_{\mathbf L}$ or $\mathcal G_{\mathbf L}=g$, there are exactly $m=2N-2$ independent eigenvectors, so the system admits no complete eigenbasis. Under the stated assumptions, standard CCZ4 in GR is therefore strongly hyperbolic if and only if
\begin{equation}
\alpha^2\chi\neq\frac{2(d-1)c_\beta}{d},\qquad
-c_\alpha\alpha\chi\neq\frac{2(d-1)c_\beta}{d}.
\label{eq:ccz4-admissible}
\end{equation}

For the standard choice $c_\alpha=-2$ and $c_\beta=d/[2(d-1)]$, the conditions~\eqref{eq:ccz4-admissible} reduce to $\alpha^2\chi\neq1$ and $2\alpha\chi\neq1$. The defect at $\alpha^2\chi=1$ agrees with that reported for moving-puncture CCZ4 by Dumbser et al.~\cite{Dumbser:2017okk}. For $d=3$, the temporal-longitudinal degeneracy $2\alpha\chi=1$ was also identified in the earlier Z4 analysis of Bona and Palenzuela~\cite{BonaPalenzuela:2004}.

It shall be emphasised that the term \textit{moving puncture gauge} is commonly
used for $1+\log$ slicing combined with the original form of the Gamma-driver
condition with auxiliary variable $B^i$~\cite{Alic:2011gg}.
We consider here the fully advective version:
\begin{subequations}
\label{eq:Gamma-driver}
\begin{align}
\partial_\perp\beta^i&=c_\beta B^i,\\
\partial_\perp B^i&=\partial_\perp\hat\Gamma^i-\eta B^i,
\end{align}
\end{subequations}
where $\eta\geq0$ is a damping parameter.
Following~\cite{AresteSalo:2023mmd}, we have used the term moving puncture gauge
for the lapse and shift conditions~\eqref{eq:puncture-gauge}. Since the shift
satisfies an integrated Gamma-driver equation, we refer to this choice more
specifically as the \textit{integrated moving puncture gauge}.

For standard CCZ4 in GR, we keep the lapse condition
in~\eqref{eq:puncture-gauge} at $a=0$ and replace the integrated shift condition
by~\eqref{eq:Gamma-driver}. The resulting system is strongly hyperbolic under
the same admissibility conditions~\eqref{eq:ccz4-admissible}.
Appendix~\ref{app:moving-puncture} proves this extension by first bringing the
extended principal symbol to block triangular form with a spectral separation
of its diagonal blocks, and then constructing a smooth symmetrizer.

\section{Discussion}
\label{sec:discussion}
\subsection{Conclusions}

We have proved that, at sufficiently weak coupling, the mCCZ4 formulation is
strongly hyperbolic for $(d+1)$-dimensional EGB gravity and for four-derivative
scalar-tensor theory in four dimensions, provided that the gauge parameters
satisfy the cone-separation condition
(Condition~\ref{ass:mccz4-cone-separation}). The central step inverts the
puncture-gauge conditions, which expresses $Z^\mu$ in terms of first derivatives
of the metric and brings the mCCZ4 metric equation into the generalised modified
harmonic form~\eqref{eq:mCCZ4_2nd}, equivalent to the original $3+1$ system with
respect to strong hyperbolicity (Theorem~\ref{th:mccz4-equivalence}). The
covariant characteristic analysis of Kov\'acs and Reall then applies: the
polarisations again divide into the physical sector, the pure-gauge sector, and
the gauge-condition-violating sector, and in puncture gauge the temporal,
longitudinal, and transverse cones replace the single pure-gauge cone of the
modified harmonic formulation. We control each $\xi_0^\pm$-group using the
positive-definiteness of the Hermitian form used by Kov\'acs and Reall on
$V^\pm$, guaranteed by continuity at weak coupling.
The proof retains the full principal symbol
without assuming analytic dependence of individual eigenvalues or eigenvectors
on the coupling.
It thereby completes the first-order argument of
Arest\'e Sal\'o, Clough, and Figueras~\cite{AresteSalo:2023mmd}. The argument
rests on identities of the covariant principal symbol that hold throughout the
Lovelock and Horndeski families, and therefore extends to these families at
weak coupling about GR and Einstein-scalar theory, respectively.

Standard CCZ4, with $a=b=0$, requires a separate treatment: in this case, the
gauge-condition-violating cone merges with the physical cone, and the
eigenbasis constructed for mCCZ4 becomes singular. Constructing the eigenbasis
directly in Section~\ref{sec:standard-ccz4}, we showed that standard CCZ4 in GR
with integrated moving puncture gauge is strongly hyperbolic if and only if the
conditions~\eqref{eq:ccz4-admissible} hold.
The conditions exclude the two degeneracies involving the longitudinal pure-gauge cone,
$\mathcal G_{\mathbf0}=\mathcal G_{\mathbf L}$ and $\mathcal G_{\mathbf L}=g$,
whereas the remaining coincidences still admit a complete set of eigenvectors.
For the usual choice $c_\alpha=-2$ and $c_\beta=d/[2(d-1)]$, the
conditions~\eqref{eq:ccz4-admissible} reduce to $\alpha^2\chi\neq1$ and
$2\alpha\chi\neq1$. The first defect agrees with that reported by Dumbser et
al.~\cite{Dumbser:2017okk}, and the second was identified by Bona and
Palenzuela~\cite{BonaPalenzuela:2004}. The same conditions ensure strong
hyperbolicity of the system with the original Gamma-driver condition~\eqref{eq:Gamma-driver}
with auxiliary variable $B^i$ (Appendix~\ref{app:moving-puncture}).

With $c_\alpha$ and $c_\beta$ fixed to their usual values, mCCZ4 retains freedom
in $a$ and $b$ to adjust the pure-gauge and gauge-condition-violating cones,
respectively. As with the modified harmonic formulation~\cite{Kovacs:2020ywu},
our weak-coupling proof requires Condition~\ref{ass:mccz4-cone-separation}.
In contrast with $\tilde g^{\mu\nu}$ in the modified harmonic formulation,
$\mathcal G_A^{\mu\nu}$ have $n^\mu n^\nu$ offsets from $g^{\mu\nu}$ whose
coefficients depend on $\alpha$ and $\chi$. Constant choices of $a$ and $b$ therefore
may fail to maintain this condition during evolution. For example, with constant
$a>0$, the longitudinal and physical cones coincide
where $(1+a)\alpha^2\chi=1$. The temporal and longitudinal cones coincide
where $2\alpha\chi=1$, independently of $a$ and $b$. Typical moving-puncture
evolutions cross both surfaces as $\alpha,\chi$ range from nearly zero near
the puncture to unity far away. These degeneracies have caused no numerical
problems~\cite{AresteSalo:2022hua}, possibly because they occur on measure-zero
submanifolds. We discuss background-dependent choices of $a$ and $b$ in
Section~\ref{sec:discussion-outlook}.

\subsection{Connections and future directions}
\label{sec:discussion-outlook}

Simulations have so far used constant $a$ and
$b$~\cite{AresteSalo:2022hua,AresteSalo:2023mmd,AresteSalo:2025sxc,Shum:2025mBSSN,Figueras:2026epx,Figueras:2026evh},
although $\chi$-dependent choices have been tested~\cite{AresteSalo:2023mmd}.
For strongly varying backgrounds, gauge parameters can be made dynamical by prescribing the
relaxation drivers $\partial_\perp p=-\tau^{-1}(p-p_{\rm target})$, with
$p\in\{a,b,c_\alpha,c_\beta\}$ and $\tau>0$.
$c_\alpha$ and $c_\beta$ are not fixed because adjusting $a$ and $b$ alone
cannot remove the temporal-longitudinal coincidence.
For smooth targets depending on the metric and scalar fields through at most
first derivatives, the extended system remains
strongly hyperbolic at admissible weak-coupling backgrounds by the
block-triangular argument of Appendix~\ref{app:moving-puncture}.
Targets that maintain Condition~\ref{ass:mccz4-cone-separation} throughout
evolution remain to be found.

We note that our route to strong hyperbolicity, through the fully second-order
metric system, can be viewed as an instance of the matrix-pencil framework
recently introduced by Abalos and Hilditch~\cite{Abalos:2026kim}. They define
strong hyperbolicity of second-order systems through second-order matrix pencils
and show that this definition is equivalent to the standard one based on
first-order reductions. Lemma~\ref{th:lem1} expresses the same equivalence in
our setting. Their $Z$-extension of Maxwell's equations uses two auxiliary
Lorentzian metrics for the gauge-fixing term and the gauge condition,
paralleling the Z4-system for gravity studied here. They also examine strong
hyperbolicity when characteristic cones intersect, as we do for standard CCZ4
in GR in Section~\ref{sec:standard-ccz4}. The matrix-pencil framework thus offers
a natural setting for extending the present analysis to a broader class of
theories whose equations of motion are second order in spacetime derivatives.

We also notice that the modified BSSN (mBSSN) formulation recently proposed by
Shum et al.~\cite{Shum:2025mBSSN} is related to mCCZ4 through a
symmetry-breaking procedure analogous to the Z4-to-BSSN
construction~\cite{Bona:2003qn}. This involves redefining the extrinsic
curvature before removing $\Theta$ and its evolution equation.
In GR, they showed that mBSSN is strongly
hyperbolic, with the eigenvalues of mCCZ4 together with a vanishing eigenvalue
and one difference in multiplicity. For weakly coupled theories, they argued by
analogy with mCCZ4 that the coupling corrections affect only the physical modes,
and tested the formulation numerically. Given the close relation between mBSSN
and mCCZ4, the present approach, based on the full principal symbol, offers a
natural starting point for a corresponding analysis of mBSSN at weak coupling.

Finally, we suggest a further application of the standard CCZ4 analysis to
black hole evolution in asymptotically anti-de Sitter (AdS) spacetimes.
It is known that generalised harmonic gauge admits stable four-dimensional
AdS evolutions with prescribed sources consistent with the boundary
conditions~\cite{Bantilan:2021AdS}. We aim to adapt this prescription within
the conformal Einstein field equations, whose wave-equation formulation
uses fields regular at the conformal boundary
and a prescribed coordinate gauge source and conformal gauge
source~\cite{Carranza:2018AdS}.
We would connect integrated moving puncture gauge in the bulk to generalised
harmonic gauge near the boundary for singularity-avoiding evolution without
black hole excision~\cite{Campanelli:2005dd,Baker:2005vv}.
The outer source would enforce the desired Fefferman-Graham fall-off~\cite{deHaro:2000vlm}.
Compatible initial data and boundary conditions preserving the conformal
constraints would ensure equivalence to the Einstein equations for the
physical metric.
Appendix~\ref{app:interpolated-gauge} provides the gauge interpolation analysis
based on Section~\ref{sec:standard-ccz4}. Choosing the interpolation function
to ensure uniform strong hyperbolicity throughout the transition and testing
the gauge dynamics numerically will be addressed in our follow-up work.

\section*{Acknowledgements}
The author thanks Pau Figueras and \'Aron Kov\'acs for their help with the draft
of this paper, and Llibert Arest\'e Sal\'o for useful discussions. This work is supported by the China Scholarship
Council.

\appendix
\section{Inversion of the puncture-gauge conditions}
\label{app:gauge-driver-inversion}
Note that the derivatives of ADM variables are related to the spacetime Christoffel symbol through
\begin{subequations}
    \begin{align}
        \gamma^{k}_\lambda \gamma ^{\alpha}_i \gamma ^{\beta}_j \Gamma^\lambda_{\alpha\beta}&={}^{(\gamma)}\Gamma^k_{ij},\\
       n_\lambda\gamma ^{\alpha}_i \gamma ^{\beta}_j \Gamma^\lambda_{\alpha\beta}&=K_{ij}, \\
         n_\lambda n^{\alpha} n ^{\beta} \Gamma^\lambda_{\alpha\beta}&=-\frac{\partial_\perp\alpha}{\alpha^2},\\
         \gamma^{k}_\lambda n^{\alpha} n ^{\beta} \Gamma^\lambda_{\alpha\beta}&=\frac{\partial_\perp\beta^k}{\alpha^2}+\frac{1}{\alpha}D^k\alpha,\\
         n_\lambda \gamma^{k\alpha} n ^{\beta} \Gamma^\lambda_{\alpha\beta}&=-\frac{1}{\alpha}D^k\alpha,
    \end{align}
\end{subequations}
where ${}^{(\gamma)}\Gamma^k_{ij}$ denotes the Christoffel symbol associated with the induced metric $\gamma_{ij}$. In addition, the conformal Christoffel symbol can be expressed as 
\begin{equation}
    \tilde{\Gamma}^i=\tilde{\gamma}^{ab}\tilde{\Gamma}^i_{ab}=\frac{1}{\chi} {}^{(\gamma)}\Gamma^i+\frac{2-d}{2\chi^2}\gamma^{ai}\partial_a\chi=\frac{1}{\chi} {}^{(\gamma)}\Gamma^i+\frac{d-2}{d\chi}{}^{(\gamma)}\Gamma^{k}_{ak} \gamma^{ai}.
\end{equation}
Then the $Z4$ vector can be expressed in terms of linear combinations of the spacetime Christoffel symbol: 
\begin{subequations}\label{eq:eqZofchris}
    \begin{align}
       2\Theta& =-\frac{(1+a(x))}{c_\alpha\alpha}\partial_\perp\alpha+K=\frac{(1+a(x))\alpha}{c_\alpha} (n_\lambda n^{\alpha} n ^{\beta} \Gamma^\lambda_{\alpha\beta})+ n_\lambda\gamma ^{\alpha\beta} \Gamma^\lambda_{\alpha\beta}\equiv n_\lambda \mathcal{G}_{\bf 0}^{\alpha\beta}\Gamma^\lambda_{\alpha\beta},
       \\
       2Z^i&=\chi\left(\frac{1+a(x)}{c_\beta} \partial_\perp \beta^i+ \frac{a(x)}{c_\beta}\alpha D^i\alpha -\tilde{\Gamma}^i\right)\nonumber\\
       &=\chi\biggl(\frac{(1+a(x))\alpha^2}{c_\beta} (\gamma^{i}_\lambda n^{\alpha} n ^{\beta} \Gamma^\lambda_{\alpha\beta}+ n_\lambda \gamma^{i\alpha} n ^{\beta} \Gamma^\lambda_{\alpha\beta})-  \frac{a(x) \alpha^2}{c_\beta} n_\lambda \gamma^{i\alpha} n ^{\beta} \Gamma^\lambda_{\alpha\beta}) -\tilde{\Gamma}^i \biggr)\nonumber\\
       &=\chi\biggl(\frac{(1+a(x))\alpha^2}{c_\beta} \gamma^{i}_\lambda n^{\alpha} n ^{\beta} \Gamma^\lambda_{\alpha\beta}+\frac{\alpha^2}{c_\beta}n_\lambda \gamma^{i\alpha} n ^{\beta} \Gamma^\lambda_{\alpha\beta}\biggr)- \gamma^{i}_{\lambda}\Gamma^{\lambda}_{\alpha\beta}\gamma^{\alpha\beta}-\frac{d-2}{d}  \gamma^{\alpha i}\gamma^{\beta}_{\lambda}\Gamma^{\lambda}_{\alpha\beta} \nonumber\\
       &\equiv-\gamma^{i}_{\lambda}\mathcal{G}_{\bf T}^{\alpha\beta}\Gamma^{\lambda}_{\alpha\beta}-\gamma^{\alpha i}(\frac{d-2}{d}\gamma^{\beta}_{\lambda}-\frac{\chi\alpha^2}{c_\beta}n_\lambda n^\beta)\Gamma^{\lambda}_{\alpha\beta},
\end{align}
\end{subequations}
where we have defined two auxiliary metrics as \footnote{Note that $\Theta=0$ then yields an effective slicing condition: $\mathcal{G}_{\bf 0}^{\mu\nu}\nabla_{\mu}\nabla_{\nu}t=0$.}
\begin{subequations}
    \begin{gather}
    \mathcal{G}_{\bf 0}^{\alpha\beta}:=\gamma^{\alpha\beta}+ \frac{(1+a(x))\alpha}{c_\alpha} n^\alpha n^\beta, \\
    \mathcal{G}_{\bf T}^{\alpha\beta}:=\gamma^{\alpha\beta}-\frac{(1+a(x))\alpha^2\chi}{c_\beta} n^\alpha n^\beta.
\end{gather}
\end{subequations}
It is useful to further trade the spacetime Christoffel symbol in \eqref{eq:eqZofchris} for the first derivatives of the metric components. We thus have 
\begin{equation}
\begin{aligned}
    2Z^\mu&=2Z^i\gamma_{i}^\mu+2\Theta n^\mu\\
    &= -\gamma^{\mu}_{\lambda}\mathcal{G}_{\bf T}^{\alpha\beta}\Gamma^{\lambda}_{\alpha\beta}-\gamma^{\alpha \mu}(\frac{d-2}{d}\gamma^{\beta}_{\lambda}-\frac{\chi\alpha^2}{c_\beta}n_\lambda n^\beta)\Gamma^{\lambda}_{\alpha\beta}+n_\lambda \mathcal{G}_{\bf 0}^{\alpha\beta}\Gamma^\lambda_{\alpha\beta} n^\mu \\
    &=(-\gamma^{\mu\lambda}\mathcal{G}_{\bf T}^{\alpha\beta}-\frac{d-2}{d}\gamma^{\alpha\mu}\gamma^{\beta\lambda}+\frac{\chi\alpha^2}{c_\beta}n^\lambda n^\beta \gamma^{\alpha\mu})(\partial_{(\alpha} g_{\beta)\lambda}-\frac{1}{2}\partial_\lambda g_{\alpha\beta}) +\\
    & n^\mu n^\lambda \mathcal{G}_{\bf 0}^{\alpha\beta}(\partial_{(\alpha} g_{\beta)\lambda}-\frac{1}{2}\partial_\lambda g_{\alpha\beta}) \\
    &=(-\gamma^{\mu\lambda}\mathcal{G}_{\bf T}^{\alpha\beta}+n^\mu n^\lambda \mathcal{G}_{\bf 0}^{\alpha\beta}) (\partial_{(\alpha} g_{\beta)\lambda}-\frac{1}{2}\partial_\lambda g_{\alpha\beta})+ \frac{1}{2}\gamma^{\alpha\mu} (-\frac{d-2}{d}\gamma^{\beta\lambda}+\frac{\chi\alpha^2}{c_\beta}n^\lambda n^\beta )\partial_\alpha g_{\beta\lambda}.
\end{aligned}
    \label{eq:eqZofg}
\end{equation}

\section{Standard CCZ4 with the original form of the Gamma-driver condition}
\label{app:moving-puncture}
We consider standard CCZ4 in GR with $a=b=0$, $d>2$, and $c_\alpha<0<c_\beta$.
The lapse satisfies~\eqref{eq:puncture-gauge} at
$a=0$, and the shift satisfies~\eqref{eq:Gamma-driver}. We use the independent
conformal variables of Section~\ref{sec:mccz4-application} and assume the
admissibility conditions~\eqref{eq:ccz4-admissible}.

The algebraic change of variables $B'^i:=B^i-\hat\Gamma^i$ preserves strong
hyperbolicity by Proposition~\ref{th:prop1} and gives
\begin{equation}
\partial_\perp\beta^i=c_\beta\hat\Gamma^i+c_\beta B'^i,
\qquad
\partial_\perp B'^i=-\eta(B'^i+\hat\Gamma^i).
\end{equation}
The damping terms are lower order in the $B'^i$ equation. The extended system
therefore has the partially reduced form~\eqref{eq:u-v}, with
\begin{equation}
u=(\chi,\tilde\gamma_{ij},\alpha,\beta^k)^T,
\qquad
v=(K,\tilde A_{ij},\Theta,\hat\Gamma^k,B'^k)^T,
\end{equation}
and principal symbol
\begin{equation}
\tilde M_{\rm ext}(\xi_i)=
\begin{pmatrix}
\tilde M_{\rm CCZ4}(\xi_i)&L\\
0&(\beta^i\xi_i)I_d
\end{pmatrix}.
\label{eq:fullmccz4}
\end{equation}
Here $\tilde M_{\rm CCZ4}$ is the partially reduced symbol of CCZ4 in integrated
moving puncture gauge, and $L$ has entries $c_\beta I_d$ in the rows corresponding
to $\beta^i$ and zero entries elsewhere.

For $\gamma^{ij}\xi_i\xi_j=1$, the eigenvalues of $\tilde M_{\rm CCZ4}(\xi_i)$
are
\begin{equation}
\xi_0=\beta^i\xi_i\pm\alpha,\quad
\beta^i\xi_i\pm\sqrt{-c_\alpha\alpha},\quad
\beta^i\xi_i\pm\sqrt{\frac{2(d-1)c_\beta}{d\chi}},\quad
\beta^i\xi_i\pm\sqrt{\frac{c_\beta}{\chi}}.
\end{equation}
They are uniformly separated from $\beta^i\xi_i$ over normalised spatial
directions, locally uniformly in the admissible background variables. Thus
$\tilde M_{\rm CCZ4}(\xi_i)-(\beta^i\xi_i)I$ has a smooth, uniformly bounded
inverse. Let $\tilde H_{\rm CCZ4}(\xi_i)$ be a smooth positive-definite
symmetrizer for $\tilde M_{\rm CCZ4}(\xi_i)$, whose existence follows from
Section~\ref{sec:standard-ccz4}. We seek a symmetrizer for~\eqref{eq:fullmccz4}
of the form
\begin{equation}
\tilde H(\xi_i)=
\begin{pmatrix}
\tilde H_{\rm CCZ4}(\xi_i)&F^\dagger\\
F&E I_d
\end{pmatrix},
\end{equation}
where $E>0$. The equation
$\tilde H\tilde M_{\rm ext}=\tilde M_{\rm ext}^\dagger\tilde H$ requires
\begin{equation}
L^\dagger\tilde H_{\rm CCZ4}
=F\bigl(\tilde M_{\rm CCZ4}-(\beta^i\xi_i)I\bigr),
\qquad
FL=L^\dagger F^\dagger.
\end{equation}
The first relation is solved by
\begin{equation}
F=L^\dagger\tilde H_{\rm CCZ4}
\bigl(\tilde M_{\rm CCZ4}-(\beta^i\xi_i)I\bigr)^{-1}.
\end{equation}
Since $\tilde H_{\rm CCZ4}\tilde M_{\rm CCZ4}
=\tilde M_{\rm CCZ4}^\dagger\tilde H_{\rm CCZ4}$, the matrix
$\tilde H_{\rm CCZ4}\bigl(\tilde M_{\rm CCZ4}-(\beta^i\xi_i)I\bigr)^{-1}$
is Hermitian, so $FL=L^\dagger F^\dagger$ also holds. To establish positivity,
we follow the argument of Kov\'acs and Reall~\cite{Kovacs:2020ywu}.
For $T=(T_1,T_2)^T$, we have
\begin{equation}
T^\dagger\tilde H T
=T_1^\dagger\tilde H_{\rm CCZ4}T_1
+T_1^\dagger F^\dagger T_2+T_2^\dagger FT_1+E T_2^\dagger T_2.
\end{equation}
The matrices $\tilde H_{\rm CCZ4}$, $\tilde H_{\rm CCZ4}^{-1}$, and $F$ are
uniformly bounded over normalised spatial directions, locally uniformly in
the background variables. Taking $E$ sufficiently large therefore makes
$\tilde H$ positive definite with uniform upper and lower bounds. Hence
$\tilde H$ is a smooth symmetrizer for the extended system. Standard
CCZ4 with~\eqref{eq:Gamma-driver} is therefore strongly hyperbolic under the
conditions~\eqref{eq:ccz4-admissible}, for arbitrary initial $B^i$.

\section{Interpolation between puncture and generalised harmonic gauges}
\label{app:interpolated-gauge}
We illustrate an application of the standard-CCZ4 analysis in
Section~\ref{sec:standard-ccz4}. Consider vacuum GR in $d+1$ spacetime
dimensions, with $d>2$, $a=b=0$, $c_\alpha=-2$ and
$c_\beta=d/[2(d-1)]$. We assume the
admissibility conditions~\eqref{eq:ccz4-admissible}, which reduce to
$\alpha^2\chi\neq1$ and $2\alpha\chi\neq1$. A cosmological constant
contributes only lower-order terms and does not change the bulk principal
symbol. Let $Z_{\mathrm p}^\mu(g,\partial g)$ denote the gauge functional
in~\eqref{eq:eqZofg}, obtained by inverting~\eqref{eq:puncture-gauge} with these parameters.
Choose a smooth generalised harmonic source $H_{\mathrm H}^\mu(x,g)$, depending
only on the coordinates and the metric. For a prescribed smooth function
$0\leq F(x)\leq1$, define
\begin{equation}
Z_{(F)}^\mu
=-\frac{1-F}{2}\left(g^{\rho\sigma}\Gamma^\mu{}_{\rho\sigma}
+H_{\mathrm H}^\mu\right)+FZ_{\mathrm p}^\mu.
\label{eq:discussion-interpolated-gauge}
\end{equation}
The endpoints $F=1$ and $F=0$ give integrated moving puncture gauge and
generalised harmonic gauge, respectively. The contributions from
$H_{\mathrm H}^\mu(x,g)$ and from derivatives of $F$ are lower order in the
effective second-order metric equations and do not alter the principal symbol.

The $\tilde{\mathbb P}$-tensor defined in~\eqref{eq:deftildeP} therefore becomes
\begin{equation}
\tilde{\mathbb P}_{(F)}{}^{\mu\delta\rho\sigma}
=(1-F)P^{\mu\delta\rho\sigma}
+F\tilde{\mathbb P}^{(0)}{}^{\mu\delta\rho\sigma},
\label{eq:interpolated-gauge-tensor}
\end{equation}
where the $P$-tensor corresponds to generalised harmonic gauge and
$\tilde{\mathbb P}^{(0)}{}^{\mu\delta\rho\sigma}$ is the tensor
in~\eqref{eq:standard-ccz4-symbol}. The principal symbol is given by
\eqref{eq:standard-ccz4-symbol} with $\tilde{\mathbb P}^{(0)}$ replaced by
$\tilde{\mathbb P}_{(F)}$. We follow the same division into two sectors as in
Section~\ref{sec:standard-ccz4}.

\paragraph*{Case I: The $\xi^2\neq0$ sector.}

As in Case I of Section~\ref{sec:standard-ccz4}, every polarisation in this
sector has the pure-gauge form $t_{\mu\nu}=\xi_{(\mu}X_{\nu)}$ and satisfies
$Q_{(F)}(\xi)^{\mu\nu}X_\nu=0$, with
\begin{equation}
Q_{(F)}(\xi)^{\mu\nu}
=FQ^{(0)}(\xi)^{\mu\nu}+\frac{1-F}{2}\xi^2g^{\mu\nu},
\label{eq:discussion-interpolated-Q}
\end{equation}
where $Q^{(0)}(\xi)$ is the $Q$-operator of standard CCZ4 in
Section~\ref{sec:standard-ccz4}. Its temporal, longitudinal and transverse block
structure, with dimensions $1,1,d-1$, is preserved. The corresponding gauge metrics interpolate between
$g^{\mu\nu}$ and the puncture-gauge metrics:
\begin{equation}
\begin{aligned}
\mathcal G_{A,F}^{\mu\nu}
&=(1-F)g^{\mu\nu}+F\mathcal G_A^{\mu\nu},
\qquad A\in\{\mathbf0,\mathbf T\},\\
\mathcal G_{\mathbf L,F}^{\mu\nu}
&=\frac{d(1-F)}{d+(d-2)F}g^{\mu\nu}
+\frac{2(d-1)F}{d+(d-2)F}\mathcal G_{\mathbf L}^{\mu\nu}.
\end{aligned}
\label{eq:discussion-interpolated-metrics}
\end{equation}
For fixed nonzero $\xi_i$, the pure-gauge eigenvalues $\xi_0$ are determined
by the null cone equations $\mathcal G_{A,F}^{\mu\nu}\xi_\mu\xi_\nu=0$,
$A\in\{\mathbf0,\mathbf L,\mathbf T\}$, as in standard CCZ4.

\paragraph*{Case II: The $\xi^2=0$ sector.}

For $\xi^2=0$, use ansatz~\eqref{eq:CCZ4ansatz} with $c(v)$ and $Y_\mu(v)$
given by~\eqref{eq:solu_cY}. Equation~\eqref{eq:Yceq} gives
$P^{\mu\beta\rho\sigma}\xi_\beta(k_{(\rho}Y_{\sigma)}+c q_{\rho\sigma})=v^\mu$,
and the $P$- and $\tilde{\mathbb P}^{(0)}$-tensors
in~\eqref{eq:interpolated-gauge-tensor} annihilate the
transverse-traceless part when contracted with $\xi_\beta$. The interpolated
version of equation~\eqref{eq:QxEq} thus reads
\begin{equation}
\begin{aligned}
Q_{(F)}(\xi)^{\mu\nu}X_\nu
&=v^\mu-\tilde{\mathbb P}_{(F)}{}^{\mu\beta\rho\sigma}
\xi_\beta(k_{(\rho}Y_{\sigma)}+c q_{\rho\sigma})\\
&=F\left[v^\mu-\tilde{\mathbb P}^{(0)}{}^{\mu\beta\rho\sigma}
\xi_\beta(k_{(\rho}Y_{\sigma)}+c q_{\rho\sigma})\right]\\
&=-F\frac{d-2}{d}(v\cdot n)k^\mu.
\end{aligned}
\label{eq:interpolated-source-equation}
\end{equation}
Here the second equality uses~\eqref{eq:interpolated-gauge-tensor} and
the fact that $k_{(\rho}Y_{\sigma)}+c q_{\rho\sigma}$ solves
equation~\eqref{eq:split_tensor}, while the last follows from~\eqref{eq:QxEqSimp}. Since
$Q_{(F)}(\xi)=FQ^{(0)}(\xi)$ for $\xi^2=0$, cancelling $F>0$ recovers
equation~\eqref{eq:QxEqSimp} exactly. The ansatz and its completeness result
therefore carry over unchanged.
The longitudinal coefficient on $\xi^2=0$ is nonzero under
conditions~\eqref{eq:ccz4-admissible}, so the transverse-traceless and
unrestricted vector-source data supply $d(d+1)/2$ independent polarisations
per physical branch. The temporal and transverse gauge metrics may coincide
with $g$, with the additional polarisations accounted for by the residual
gauge freedom as in Section~\ref{sec:ccz4-general-conditions}.

At $F=0$, the system reduces to the generalised harmonic formulation, whose
characteristic equation is equivalent to $\xi^2t_{\mu\nu}=0$. For
$\xi^2=0$, this is an identity for arbitrary symmetric $t_{\mu\nu}$,
giving $(d+1)(d+2)/2$ independent polarisations for each branch.

\paragraph*{Conditions for strong hyperbolicity.}

For $F>0$, the metric $\mathcal G_{\mathbf L,F}$ differs from $g$ under the
standing admissibility conditions~\eqref{eq:ccz4-admissible}, namely
$\alpha^2\chi\neq1$ and $2\alpha\chi\neq1$. Combining the two sectors as in
Section~\ref{sec:ccz4-general-conditions} therefore gives a complete eigenbasis
precisely when $\mathcal G_{\mathbf L,F}\neq\mathcal G_{\mathbf0,F}$.
Using~\eqref{eq:discussion-interpolated-metrics}, the remaining condition is
\begin{equation}
(d-2)(1-F)(2-\alpha)
+2(d-1)\alpha(1-2\alpha\chi)\neq0.
\label{eq:interpolated-admissible}
\end{equation}
The strong-hyperbolicity argument of
Section~\ref{sec:ccz4-general-conditions} therefore applies at fixed $F>0$
under~\eqref{eq:interpolated-admissible}. This condition holds at $F=1$ by
\eqref{eq:ccz4-admissible}, and hence for $F$ sufficiently close to $1$, but
must also be checked along the transition.

Choosing $F(x)$ to ensure uniform strong hyperbolicity throughout the
interpolation, including the endpoint $F=0$, and assessing the numerical gauge
dynamics remain future work.

\printbibliography

\end{document}